\documentclass[preprint,12pt,authoryear,5p,twocolumn]{elsarticle}

\usepackage{multirow}
\usepackage{amssymb}
\usepackage{amsmath}

\newcommand{\ps}{{\it PolSTAR }}
\newcommand{\psc}{{\it PolSTAR}}
\newcommand{\psg}{{\it PolSTAR's }}

\newcommand{\ns}{{\it NuSTAR }}
\newcommand{\nsc}{{\it NuSTAR}}
\newcommand{\nsg}{{\it NuSTAR's }}

\def\polstar{{\it PolSTAR}}
\def\nustar{{\it NuSTAR}}

\def\ie{{i.e.}}
\def\eg{{e.g.}}

\newcommand{\ika}{ Fe K-$\alpha$ }
\def\arcsec{\hbox{$^{\prime\prime}$}}
\def\deg{\hbox{$^{\circ}$} }

\newcommand{\QQ}{{\cal Q}}
\newcommand{\UU}{{\cal U}}

\journal{Astroparticle Physics}

\begin{document}

\begin{frontmatter}


\title{X-Ray Polarimetry with the Polarization Spectroscopic Telescope Array {\it (PolSTAR)}}
\author[WUSTL]{Henric S.\ Krawczynski}
\author[JPL]{Daniel Stern}
\author[Caltech]{Fiona A.\ Harrison}
\author[WUSTL]{Fabian F.\ Kislat}
\author[WUSTL]{Anna Zajczyk}
\author[WUSTL]{Matthias Beilicke}
\author[WUSTL]{Janie Hoormann}
\author[WUSTL]{Qingzhen Guo}
\author[WUSTL]{Ryan Endsley}
\author[Amsterdam]{Adam R.\ Ingram}
\author[Caltech]{Hiromasa Miyasaka}
\author[Caltech]{Kristin K.\ Madsen}
\author[JPL]{Kim M. Aaron}
\author[WUSTL,JPL]{Rashied Amini}
\author[Rice]{Matthew G.\ Baring}
\author[WUSTL]{Banafsheh Beheshtipour}
\author[Giorgia]{Arash Bodaghee}
\author[JPL]{Jeffrey Booth}
\author[JPL]{Chester Borden}
\author[NWU]{Markus B\"ottcher}
\author[DTU]{Finn E.\ Christensen}
\author[Yale]{Paolo S.\ Coppi}
\author[WUSTL]{Ramanath Cowsik}
\author[Virginia]{Shane Davis}
\author[MPIE]{Jason Dexter}
\author[Durham]{Chris Done}
\author[JPL]{Luis A. Dominguez}
\author[NCSU]{Don Ellison}
\author[JPL]{Robin J.\ English}
\author[Cambridge]{Andrew C.\ Fabian}
\author[PSU]{Abe Falcone}
\author[JPL]{Jeffrey A.\ Favretto}
\author[UCB,UCB2]{Rodrigo Fern\'andez}
\author[ASI]{Paolo Giommi}
\author[Caltech]{Brian W.\ Grefenstette}
\author[Cambridge]{Erin Kara}
\author[JPL]{Chung H.\ Lee}
\author[Purdue]{Maxim Lyutikov}
\author[Texas]{Thomas Maccarone}
\author[Nagoya]{Hironori Matsumoto}
\author[UMD]{Jonathan McKinney}
\author[Riken]{Tatehiro Mihara}
\author[UMICH]{Jon M.\ Miller}
\author[Harvard]{Ramesh Narayan}
\author[INAF]{Lorenzo Natalucci}
\author[Arizona]{Feryal \"Ozel}
\author[LLNL]{Michael J.\ Pivovaroff}
\author[JPL]{Steven Pravdo}
\author[Arizona]{Dimitrios Psaltis}
\author[GSFC]{Takashi Okajima}
\author[Tohoku]{Kenji Toma}
\author[GSFC]{William W.\ Zhang}

\address[WUSTL]{Washington University in Saint Louis, Physics Department and McDonnell Center for the Space Sciences, Saint Louis, MO 63130, USA}
\address[JPL]{Jet Propulsion Laboratory, California Institute of Technology, Pasadena, CA 91109, USA}
\address[Caltech]{Cahill Center for Astronomy and Astrophysics, California Institute of Technology, Pasadena, CA 91125, USA}
\address[Amsterdam]{Anton Pannekoek Institute for Astronomy, Science Park 904, P.O. Box 94249, 1090 GE Amsterdam, The Netherlands}
\address[Rice]{Rice University, Department of Physics and Astronomy, P.O. Box 1892, Houston, TX 77251-1892, USA}
\address[Giorgia]{Department of Chemistry, Physics, and Astronomy, Georgia College, Milledgeville, GA 31061, USA}
\address[NWU]{Centre for Space Research, North-West University,  
Private Bag X6001, Potchefstroom, 2520, South Africa}
\address[DTU]{DTU Space, National Space Institute, Technical University of Denmark, Elektrovej 327, DK-2800 Lyngby, Denmark}
\address[Yale]{Department of Astronomy, Yale University, P.O. Box 208101, New Haven, CT 06520-8101, USA}
\address[Virginia]{Department of Astronomy, University of Virginia, P.O. Box 400325, Charlottesville, VA 22904-4325, USA}
\address[MPIE]{MPI for Extraterrestrial Physics, Giessenbachstr. 85748, Garching, Germany}
\address[Durham]{Centre for Extragalactic Astronomy, Department of Physics, Durham University, South Road, Durham, DH1 3LE} 
\address[NCSU]{Department of Physics, North Carolina State University, 400L Riddick Hall, Raleigh, NC 27695-8202, USA}
\address[Cambridge]{Institute of Astronomy, Madingley Road, Cambridge CB3 0HA, UK}
\address[PSU]{Penn State University, Department of Astronomy and Astrophysics, 516 Davey Lab, University Park, PA 16802, USA}
\address[UCB]{Department of Physics, University of California, Berkeley 94720, CA, USA}
\address[UCB2]{Department of Astronomy \& Theoretical Astrophysics Center, University of California, Berkeley CA 94720, CA, USA}
\address[ASI]{ASI Science Data Center,  Via del Politecnico s.n.c. I-00133 Rome Italy}
\address[Purdue]{Department of Physics and Astronomy, 525 Northwestern Avenue, West Lafayette, Indiana 47907-2036,USA}
\address[Texas]{Texas Tech University, Physics Department, Box 41051, Lubbock, TX 79409-1051, USA}
\address[Nagoya]{Nagoya University, Center for Experimental Studies, Kobayashi-Maskawa Institute for the Origin of Particles and the Universe, Furo-cho, Chikusa-ku, Nagoya 464-8602, Japan}
\address[UMD]{University of Maryland, Physics Department, College Park, MD 20742-4111, USA}
\address[Riken]{RIKEN, 2-1 Hirosawa, Wako, Saitama, Japan 351-0198}
\address[UMICH]{Univ.\ of Michigan in Ann Arbor, Astronomy Dept., 830 Dennison, 500 Church St., Ann Arbor, MI  48109-1042, USA}
\address[Harvard]{Harvard-Smithsonian Center for Astrophysics, 60 Garden Street, MS-51, Cambridge, MA 02138, USA}
\address[INAF]{Istituto di Astrofisica e Planetologia Spaziali, INAF, Via Fosso del Cavaliere 100, Roma I-00133, Italy}
\address[Arizona]{Department of Astronomy/Steward Observatory, 933 North Cherry Avenue, Tucson, AZ 85721-0065, USA}
\address[LLNL]{Lawrence Livermore National Laboratory, Livermore, CA 94550, USA}
\address[GSFC]{NASA Goddard Space Flight Center, Greenbelt, MD 20771, USA}
\address[Tohoku]{Astronomical Institute, Tohoku University, 6-3 Aramaki, Aoba-Ku, Sendai, Japan, 980-8578}

\end{frontmatter}
{\bf Abstract:} 
This paper describes the {\it Polarization Spectroscopic Telescope
Array} (\psc), a mission proposed to NASA's 
2014 Small Explorer (SMEX) announcement of opportunity.
\ps measures the linear polarization of 3-50 keV (requirement; goal: 2.5-70 keV) 
X-rays probing the behavior of matter, radiation and the very fabric of spacetime 
under the extreme conditions close to the event horizons of black holes, 
as well as in and around magnetars and neutron stars.
The \ps design is based on the technology 
developed for the {\it Nuclear Spectroscopic Telescope Array} (\nsc) mission
launched in June 2012.  In particular, it uses the same X-ray optics, 
extendable telescope boom, optical bench, and CdZnTe detectors as \nsc.
The mission has the sensitivity to measure $\sim$1\% linear polarization 
fractions for X-ray sources with fluxes down to $\sim$5 mCrab. 
This paper describes the \ps design as well as the science drivers 
and the potential science return.

\section{Introduction}
\label{Intro}
In the following, we describe the {\it Polarization Spectroscopic Telescope Array} (\psc),
a satellite-borne experiment measuring the linear polarization of X-rays 
in the energy range from 3-50 keV (requirement; goal: 2.5-70 keV).
The mission was proposed to NASA's 2014 Small Explorer (SMEX) announcement 
of opportunity \citep{SMEX2014}.  The mission concept builds on the highly 
successful {\it Nuclear Spectroscopic Telescope Array} (\nsc) hard
X-ray imaging mission \citep{2013ApJ...770..103H}. 
The main difference between \ps and \ns is the addition of a scattering element
in the focal plane of the X-ray telescope enabling the measurement of the
linear polarization properties. \ps measures the polarization fraction and angle, two properties
of photon beams characterizing the uniformity and orientation of
the electric field carried by the photons, respectively, as a
function of photon energy and arrival time.  
The two fundamentally new observables depend on
the emission mechanism, scattering angles, and the geometry and
properties of matter, electromagnetic fields and spacetime itself
of extreme objects such as black holes and neutron stars.
A mission like \ps gives geometric information about objects which are much 
too small to be imaged directly.  For example, consider the Galactic stellar 
mass black hole GRS~1915+105.  At a distance of $\sim$8.6~kpc \citep{2014ApJ...796....2R}, 
the gravitational radius $r_{\rm g}\,=\,G M/c^2$ (with gravitational constant $G$, 
black hole mass $M$, and speed of light $c$) measures 21~km, corresponding 
to an angle of 4.5 femto-degrees.  X-ray polarimetry allows us to measure angles 
in systems of such small angular extent.  

X-ray polarimetry is a largely unexplored field. NASA has so far
only launched one dedicated X-ray polarimetry mission, {\it OSO-8},
which was in orbit from 1975 to 1978 \citep{1975SSRv...18..389N}.
{\it OSO-8} measured the polarization fraction and angle of the
2.6~keV and 5.2~keV X-ray emission from the Crab Nebula
\citep{1978ApJ...220L.117W} and set upper limits on the polarization
fraction of the X-ray emission from 14 sources \citep{1979ApJ...232..248S,1984ApJ...280..255H}.

Two instruments on the International Gamma-Ray Astrophysics Laboratory {\it (INTEGRAL)}, 
the Spectrometer on {\it INTEGRAL} \citep[SPI; ][]{2003A&A...411L..63V} 
and the Imager on Board the {\it INTEGRAL} Satellite \citep[IBIS; ][]{2003A&A...411L.131U}, 
have revealed tentative evidence for highly polarized
$>$100 keV emission from the Crab Nebula
\citep{2008Sci...321.1183D,2008ApJ...688L..29F} and Cyg X-1
\citep{2011Sci...332..438L,2015ApJ...807...17R}, albeit with large systematic errors.
Several authors reported the detection of highly polarized X-ray
and/or gamma-ray emission from gamma-ray bursts \citep[GRBs;
e.g.,][and references
therein]{2005A&A...439..245W,2007A&A...466..895M,2007ApJS..169...75K,2011ApJ...743L..30Y,2012ApJ...758L...1Y,2013PhRvL.110t1601K},
but the evidence is not highly significant taking the statistical
and systematic errors into account.

\begin{table*}[t]
\centering
{\small
\begin{tabular}{|p{7.5cm}|p{3cm}|l|}
\hline
Parameter                                                                    & Requirement                         & Current Best Estimate \\ \hline
Telescope bandpass (keV)                                                     & 3-50                                & 2.5-70                      \\ \hline
Telescope effective area (effective \# of \nustar\ optics)                   & $\geq 0.9$                          & 1.1                         \\ \hline
Energy resolution (FWHM at 6 keV)                                            & $\leq 1$ keV                        & 0.45 keV                    \\ \hline
Absolute timing accuracy (msec)                                                   & $\leq 15$                           & 2                           \\ \hline
Angular resolution (half power diameter; arcsec)                                             & $\leq 80$                           & 60                          \\ \hline
Pointing, during science portion of orbits (99.7\% CL)                       & $\leq 62$\arcsec\ from stick center & 17\arcsec\ from stick center \\ \hline
Instrument reconstructed pointing knowledge (99.7\% CL)                      & $\leq 15$\arcsec                    & 8\arcsec                   \\ \hline
Minimum Detectable Polarization  (3-15 keV, 25 ks obs'n of 1 Crab source, 99\% CL)                               & $\leq 1\%$                          & 0.5\%                       \\ \hline
\begin{tabular}[c]{@{}l@{}}Polarization fraction systematic error\\ (3-15 keV; 99.7\% CL)\end{tabular} & $\leq 1.5\%$                        & 0.25\%                       \\ \hline
\begin{tabular}[c]{@{}l@{}}Polarization angle systematic error\\ ($\geq 6\%$ polarized source; 99.7\% CL)\end{tabular} & $\leq 20^\circ$                     & $2^\circ$                   \\ \hline
Bad pixel fraction                                                           & $\leq 2\%$                          & 1\%                         \\ \hline
Instrument mass (kg)                                                         & $\leq 170$                          & 131                         \\ \hline
Instrument power (W; orbital avg.)                                           & $\leq 45$                           & 28                          \\ \hline
\end{tabular}}
\caption{\polstar\ has significant margin on all primary instrument requirements, largely based on \nustar\ heritage. CL stands for confidence level.}
\label{table:requirements}
\end{table*}
\ps uses scattering off a lithium hydride (LiH) element to measure the linear polarization of X-rays.  
In the 2-10 keV energy band, a competing approach uses photoelectric effect interactions 
in a gas chamber read out by gas electron multipliers (GEMs). The proposed
{\it Imaging X-ray Polarimetry Explorer} \citep[{\it IXPE};][]{2014Ramsey}
and {\it X-ray Imaging Polarimetry Explorer} \citep[{\it
XIPE};][]{2013ExA....36..523S} 
missions use gas pixel detectors for the
readout, enabling spectropolarimetric imaging with an angular
resolution of $\sim 25\arcsec$. 
The proposed {\it Polarization from
Relativistic Astrophysical X-raY Sources} ({\it PRAXYS}) (former {\it GEMS}) mission
uses a gas chamber operated as a time projection chamber
\citep[TPC;][]{2015AAS...22533840J,2014SPIE.9144E..4ME,2014SPIE.9144E..1NH,2014SPIE.9144E..4NT,2014SPIE.9144E..4LK,2014SPIE.9144E..0NJ}.
The electron track perpendicular to the beam direction are measured
in two dimensions based on strip and drift time measurements.  
\ps is unique in offering a broad energy range, as that of {\it IXPE}, {\it XIPE} and {\it PRAXYS} 
is limited to 2-10 keV. \ps does not offer the imaging capabilities of {\it IXPE} and {\it XIPE}. 
At the time of writing this paper, NASA selected {\it IXPE} and {\it PRAXYS} for Phase A studies \citep{SMEXD}, 
and the European Space Agency (ESA) selected {\it XIPE}  as one of three M4 candidate missions \citep{m4}.
The NASA review classified \ps as a Category II proposal defined as a 
``well-conceived and scientifically or technically sound investigations which are 
recommended for acceptance, but at a lower priority than Category I'' 
attesting to the soundness of the approach.  
We are currently considering proposing an enlarged version of \ps to 
NASA's upcoming Medium Class Explorer (MIDEX) announcement of opportunity.

The paper is structured as follows.  After presenting the \ps
design in \S~\ref{S:Design} and the analysis methods and projected performance in \S~\ref{S:perf}, 
we discuss the \ps science program in \S~\ref{S:Science}. 
Section \S~\ref{S:Discussion} gives a summary.

Unless otherwise noted, all figures and sensitivity estimates assume
source fluxes normalized to the observed time-averaged 2-12 keV
fluxes measured from 1996-2011 with the All-Sky Monitor \citep[ASM;][]{1996ApJ...469L..33L}
on the {\it Rossi X-ray Timing Explorer} \citep[{\it RXTE};][]{1996SPIE.2808...59J} mission
(from R. Remillard, private communication).  All statistical errors are
given as 1$\sigma$ errors (68.7\% confidence level, CL).  As
systematic errors tend to exhibit non-Gaussian distributions, we
provide them as 99.7\% CL (equivalent to 3$\sigma$ for a Gaussian
distribution).  We follow the standard in the field to give fluxes
in units of mCrab (milli-Crab), equaling 1/1000$^{\rm th}$ of the
flux from the Crab Pulsar and Nebula.
See 
Meszaros et al.~\citeyearpar{1988ApJ...324.1056M},
Lei et al.~\citeyearpar{1997SSRv...82..309L},
Costa et al.~\citeyearpar{2006astro.ph..3399C},
Weisskopf et al.~\citeyearpar{2006astro.ph.11483W},
Bellazzini et al.~\citeyearpar{2010xrp..book.....B}, and 
Krawczynski et al.~\citeyearpar{2011APh....34..550K} 
for reviews of the science drivers and detection techniques of X-ray polarimetry.
\section{Design of \ps}
\label{S:Design}
\polstar\ measures the flux, polarization fraction, and polarization
angle of astrophysical sources as a function of energy. Table~\ref{table:requirements}
summarizes the primary \polstar\ instrument requirements and current
best estimate (CBE) of capabilities.

\begin{figure}[t]
\includegraphics[width=3.5in]{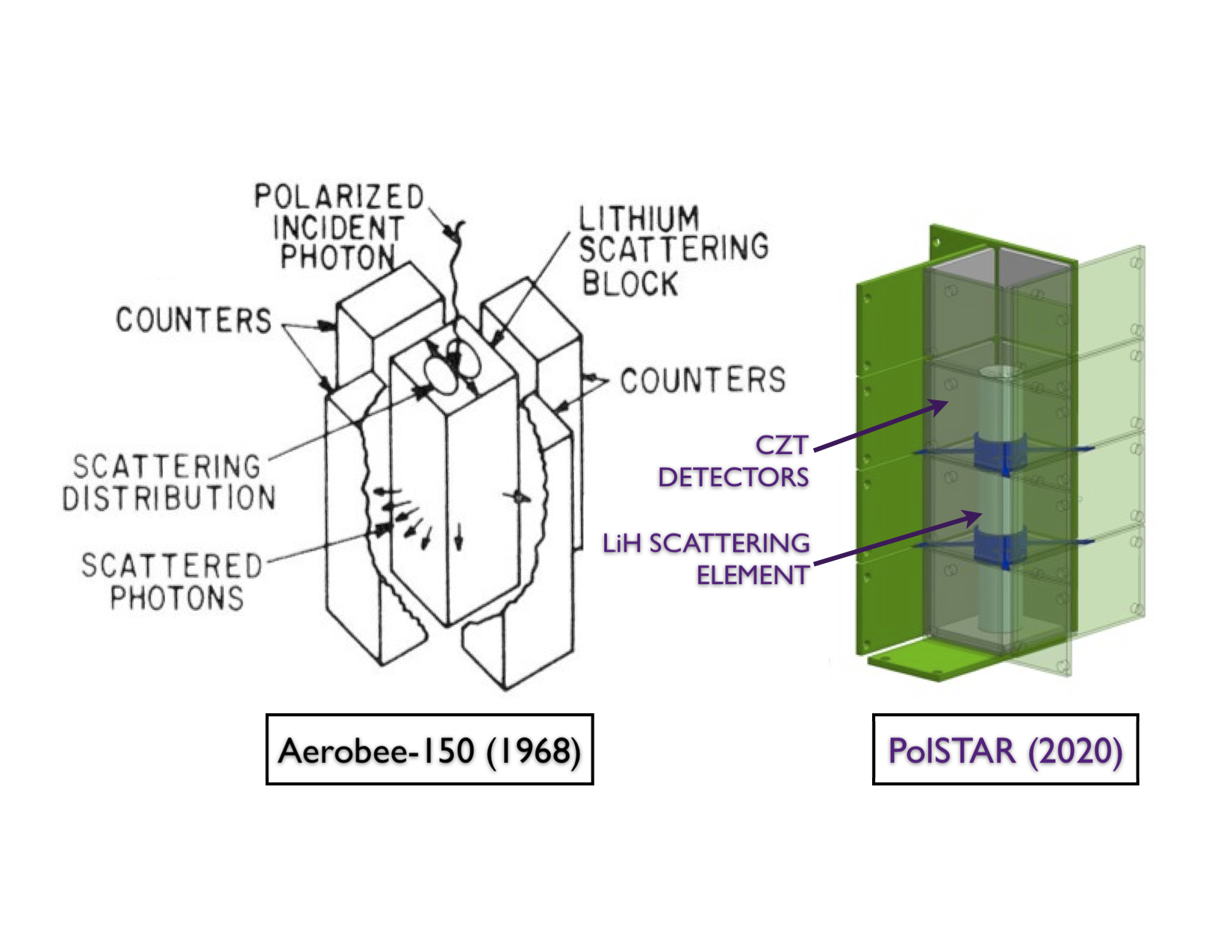}
\caption{\ps uses the same simple instrument design as the
first astronomical X-ray polarimeter, which flew in July 1968 on
an Aerobee-150 rocket \citep{1969ApJ...158..219A}. Fifty years later, \ps
can use a similar LiH scattering element design with modern CZT
detectors.
\label{fig:aerobee150}}
\end{figure}

\begin{table}[tb]
\centering
{\small
\begin{tabular}{|p{3cm}|l|l|}
\hline
Parameter                   & \nustar                       & \polstar                \\ \hline
\# telescope modules        & 2                             & 1                       \\ \hline
Effective focal length      & \multicolumn{2}{l|}{10.14 m}                            \\ \hline
Optics                      & \multicolumn{2}{l|}{\begin{tabular}[c]{@{}l@{}}Grazing incidence,\\ (conical approx.)\end{tabular}} \\ \hline
\# shells per optics module & \multicolumn{2}{l|}{133}                                \\ \hline
Multilayer coating          & W/Si and Pt/C                 & W/Si                    \\ \hline
Detectors                   & \multicolumn{2}{l|}{$32 \times 32$ pix CZT hybrids}   \\ \hline
\# detectors                & 8                             & 17                      \\ \hline
Shielding                   & \multicolumn{2}{l|}{CsI anti-coincidence}               \\ \hline
\end{tabular}}
\caption{Based largely on \nustar\ heritage, the \polstar\ instrument
is 82\% build-to-print by mass.}
\label{table:nustarpolstar}
\end{table}
As mentioned above, \ps has high heritage. Largely based on \ns
\citep[see][for a description of the flight hardware and its pre-flight and in-flight performance]{2013ApJ...770..103H}, 
\ps uses an identical extendable mast, structures,
metrology system, and cadmium zinc telluride (CZT) detectors (Table~\ref{table:nustarpolstar}).
The optics use the \ns design and assembly with simplified \nsc-heritage coatings.
There are three main differences from \nsc:  first, \ps
flies only one telescope rather than two. Second, \ps 
slowly rotates every 10 minutes to minimize systematic
errors on the polarization measurements.  Third, \ps inserts a passive
scattering element into the light path and arrays the CZT detectors
around this element, parallel to the incident photon path, to enable 
the measurement of the photon polarization; 
\ns uses the same CZT detectors perpendicular to the incident photon
path to provide focused images of the high-energy sky. The \ps
detection principle is very simple, essentially identical to that
of the very first astronomical X-ray polarimeter, an experiment
flown on an Aerobee-150 rocket in July 1968 (Figure~\ref{fig:aerobee150}).
Fifty years later, \ps can use modern X-ray optics and more capable
detectors to provide the first sensitive polarization measurements
of a representative sample of high-energy sources across a broad,
scientifically compelling energy range.
\subsection{Detection Principle}
\ps has a 3-50~keV energy range requirement driven by the
science objectives, and a 2.5-70~keV capability inherited from
\nustar. \nustar\ showed that many interesting and diagnostic
phenomena uniquely present themselves in this energy range, such
as the Compton reflection hump \citep[\eg,][]{2013Natur.494..449R},
and cyclotron absorption lines in neutron stars
\citep[\eg,][]{2013ApJ...779...69F,2014ApJ...784L..40F,2014ApJ...780..133F,2015ApJ...806L..24F,2014ApJ...795..154T}. 

\ps uses X-ray scattering off a passive element to measure the polarization of astrophysical
targets. Coherent scattering is the dominant process below 6~keV,
while at higher energies Thomson scattering and Compton scattering
are dominant. Thus, the effective energy range of a scattering
polarimeter can be broad, reaching from below a keV to greater than
a MeV.

\begin{figure*}[tbh]
\begin{center}
\includegraphics[width=5in]{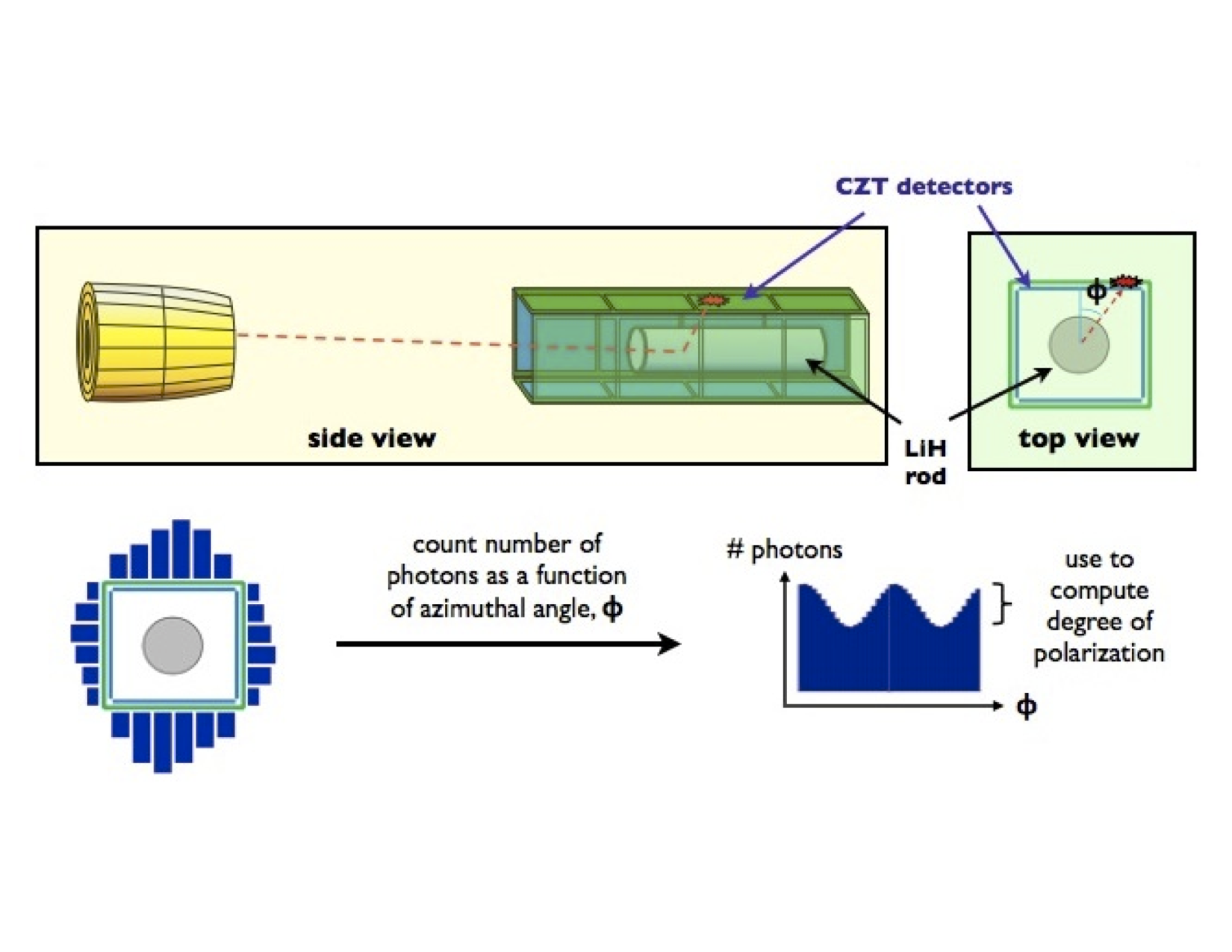}
\end{center}
\vspace*{-2cm}
\caption{How \ps works: photons are focused by the grazing
incidence optics onto the LiH scattering rod, where they preferentially
scatter perpendicular to the electric field vector.  Four columns of four
CZT detectors surround the rod to detect scattered
photons. A 17$^{\rm th}$ detector at the far end of the scattering
element provides a source image.  
\ps measures the polarization fraction and angle by measuring the 
azimuthal distribution of scattered photons.
\label{fig:detectionprinciple}}
\end{figure*}

\begin{figure*}[htb]
\begin{center}
\vspace*{-2cm}
\includegraphics[width=4in]{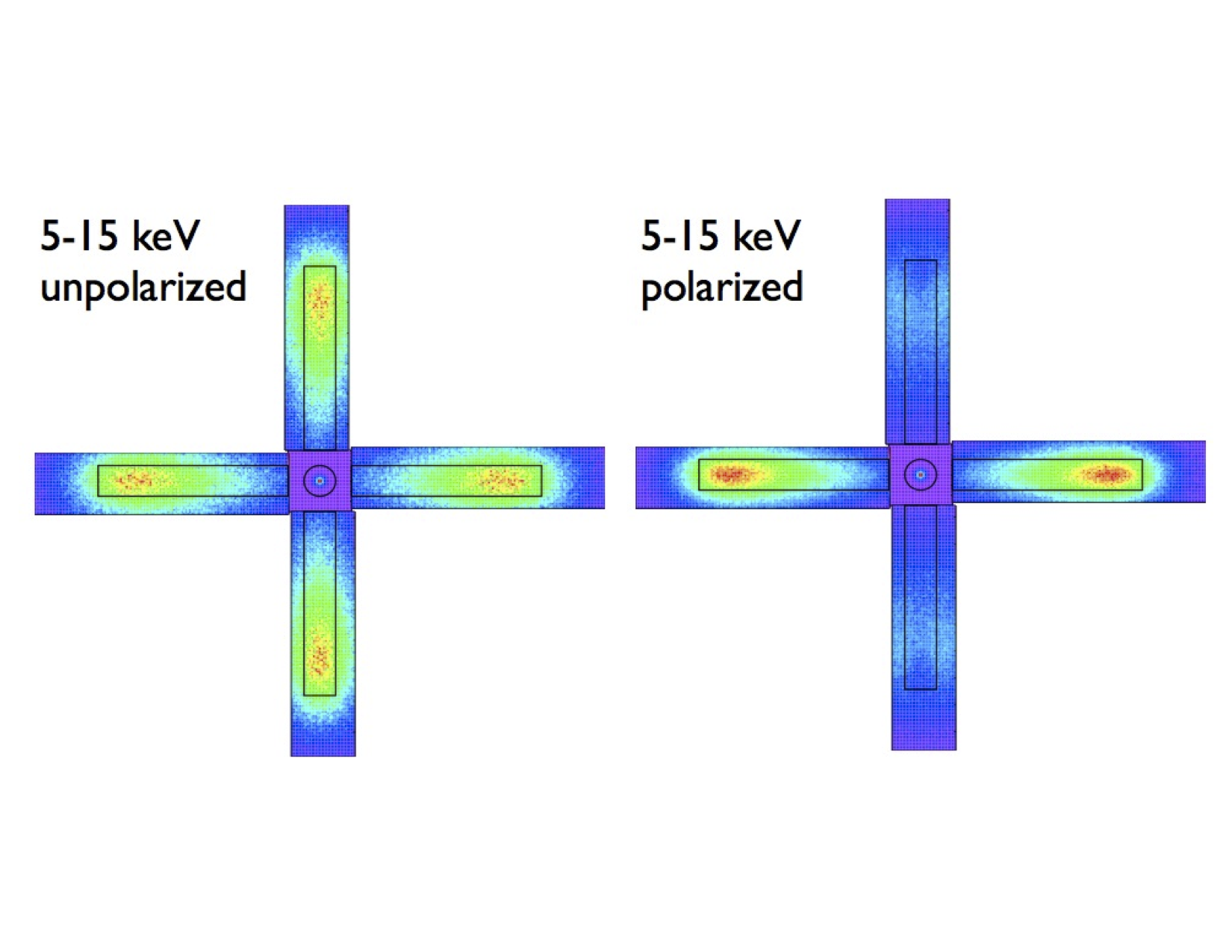}
\vspace*{-2cm}
\end{center}
\caption{End-to-end Monte Carlo simulations of an unpolarized (left) and
polarized (right) source in the 5-15~keV energy range, showing the four
unfolded CZT detector modules surrounding the LiH stick, with
Detector 17 in the center (see \S~\ref{S:perf} for details about the simulations).  
The inset black rectangles and circles indicate the LiH stick
size and location. Color scale encodes the number of hits, with red
indicating the highest flux.  Polarized photons preferentially
scatter perpendicular to the electric field vector, creating an
azimuthally asymmetric photon distribution.
\label{fig:simulation}}
\end{figure*}

Figures~\ref{fig:detectionprinciple} and \ref{fig:simulation} illustrate
how the instrument works.  
The mirror assembly focuses X-rays onto
the passive scattering element, a 6~cm long, 1~cm diameter lithium hydride (LiH) cylinder.  
As photons scatter preferentially perpendicular to the electric field vector, 
the azimuthal scattering angle distribution can be used to measure 
the polarization degree and angle (Figure~\ref{fig:simulation}). The scattered
photons are photo-absorbed in the assembly of CZT
detectors that surround the LiH. These detectors, identical to those
on \nustar, register each incident photon with a time, energy, and
location. The intensity as a function of azimuthal scattering angle
constrains the polarization fraction and angle, and this is done
as a function of energy.
%
\begin{table}[t]
\centering
{\small
\begin{tabular}{|l|l|l|ll}
\cline{1-3}
\multicolumn{2}{|l|}{Component}         & Heritage    &  &  \\ \cline{1-3}
Optics       & Segmented design         & \nustar     &  &  \\ \cline{1-3}
             & Glass substrates         & \nustar     &  &  \\ \cline{1-3}
             & Multilayers              & \nustar     &  &  \\ \cline{1-3}
             & Mounting/assembly        & \nustar     &  &  \\ \cline{1-3}
\multicolumn{2}{|l|}{Extendable mast}   & \nustar     &  &  \\ \cline{1-3}
\multicolumn{2}{|l|}{Metrology systems} & \nustar     &  &  \\ \cline{1-3}
Focal Plane  & CZT material             & \nustar     &  &  \\ \cline{1-3}
             & ASIC                     & \nustar     &  &  \\ \cline{1-3}
             & Hybrid sensor            & \nustar     &  &  \\ \cline{1-3}
             & CsI shield               & \nustar     &  &  \\ \cline{1-3}
Polarimeter  & LiH scatterer            & Aerobee-150 &  &  \\ \cline{1-3}
             & Design                   & {\it X-Calibur}   &  &  \\ \cline{1-3}
\end{tabular}}
\caption{Strong heritage pervades the \polstar\ design.}
\label{table:heritage}
\end{table}

It is instructive to compare the design of \ps to that of other scattering polarimeters, i.e.\ 
NASA's balloon-borne {\it X-Calibur} hard X-ray polarimeter experiment  
\citep{2011APh....34..550K,2013APh....41...63G,2014JAI.....340008B,2014JAI.....340008B} 
and the soft gamma-ray telescope (SGT) of JAXA's {\it ASTRO-H}  mission \citep{2014SPIE.9144E..2CF}. 
Whereas \ps uses a passive low-atomic number (low-$Z$) scattering element, 
{\it X-Calibur} and the SGT use active scattering elements made of heavier elements. 
\psg passive LiH scattering element enables polarimetric measurements in the key 
2.5-30 keV energy range. The active scattering elements of {\it X-Calibur} and the SGT 
result in a much higher energy thresholds ($\sim$30 keV for {\it X-Calibur} and 
$\sim$50 keV for the SGT), but enable additional background suppression 
capabilities through the coincident detection of the Compton scattered photon and the
Compton electron. In the case of the SGT, the active scattering elements enables 
furthermore an improved energy resolution by measuring the energy given to the Compton electron.
One reason for the higher energy threshold of active scattering elements is the higher $Z$
of active detector elements. {\it X-Calibur} uses the scintillator EJ-200 which contains roughly
equal amounts of H ($Z$=1) and C ($Z$=6), and the SGT uses Si ($Z$=14) pad detectors.
The heavier elements exhibit a much lower scattering efficiency than LiH owing to the prevalence of
photoelectric effect absorption over scattering interactions.  More quantitatively, the energy at 
which the scattering cross section starts to dominate above the photoelectric absorption cross section is 
9 keV, 20 keV, and 80 keV for LiH, C and Si, respectively.
The requirement to trigger the active scattering elements elevate the energy threshold for polarimetric 
studies even more. Using the standard Compton equations, and assuming a trigger threshold of 2 keV 
for the {\it X-Calibur} scintillator and 5.4 keV for the SGT Si pad detectors, we infer effective 
energy thresholds of 33 keV and 55 keV for {\it X-Calibur} and the SGT, respectively. 
Note that {\it X-Calibur} is optimized for operation on a balloon. As the residual atmosphere 
at a flight altitude of 125,000 feet anyhow absorbs $<$30 keV photons, an active 
scintillator scattering element is a good choice for {\it X-Calibur}. 

\ps achieves excellent energy resolutions although it does not measure the energy of the 
Compton electrons. The main reason is that 2.5-70 keV photons loose only a small fraction
of their energy when Compton scattering. For example, a $E_{\gamma}\,=$ 10~keV photon 
scattering by $\theta\,=$ 90$^{\circ}$ gives only 
$E_{\rm e}\,=\,(1-(1+E_{\gamma}/m_e c^2)^{-1})E_{\gamma}\,\approx$ 0.2 keV 
to the Compton electron (see \S \ref{S:perfX} and Fig.\ \ref{fig:energyRes}). 
\subsection{Instrument Subsystems}
Table~\ref{table:heritage} summarizes the heritage of the \ps
instrument components. \ps can re-use a large fraction of the
\ns hardware and software. The most significant changes with respect to
\nustar\ are the introduction of the passive LiH scattering element
and the slow rotation of the satellite. The scattering 
element has heritage from the Aerobee-150 rocket experiment 
\citep{1969ApJ...158..219A} and the more recent {\it X-Calibur} balloon experiment.
Below we discuss key instrument subsystems in more detail.

{\bf Grazing incidence optics:} The \polstar\ optics are a
simplified version of the \nustar\ grazing incidence optics,
fabricated by the same personnel using the same equipment.  
The reflecting surface of each glass substrate is coated with a
depth-graded multilayer consisting of up to several hundred alternating
thin layers of high and low index of refraction material.  The small
reflections from each layer add in phase, achieving a broad bandpass
over a relatively large field-of-view.  \polstar\ uses
\ns bilayer thickness recipes, deposited
using the same custom, high-throughput planar magnetron sputtering
facility at DTU-Space (Denmark's national space institute) as used by \nustar.
As shown by \citet{1993ApOpt..32.4231S}, grazing incidence optics produce
negligible instrumental polarization ($<0.1\%$).  The \polstar\
grazing incidence optics contain 133 nested multilayer-coated
shells in a conical approximation to a Wolter-I geometry
\citep{1952AnP...445...94W}.  

The only change relative to \nustar\ is that
\nustar\ used two mirror coating recipes: the inner 89~shells are
coated with depth-graded Pt/C multilayers which provide sensitivity
up to 79 keV and the outer 44~shells are coated with depth-graded
W/Si multilayers which provide sensitivity to 70~keV.  \polstar\
only requires an energy range of 3-50~keV, and therefore uses the less
expensive and easier to apply W/Si multilayers on all shells.
Eliminating the Pt/C multilayers also provides a 20\% larger effective 
area below 50 keV relative to the \ns optics.

\begin{figure}
\includegraphics[width=3.5in]{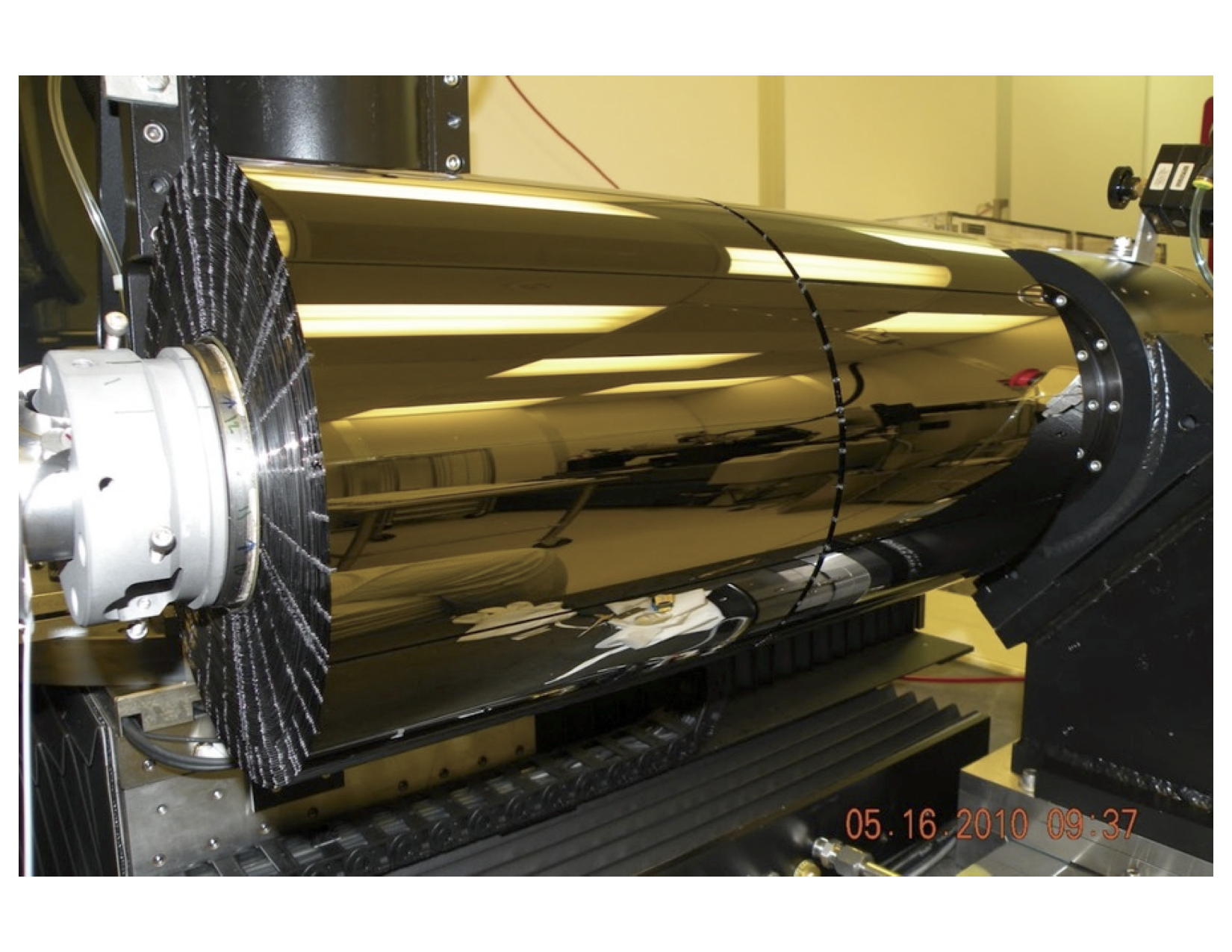}
\vspace*{-1.0cm}
\caption{Assembly of the first \nustar\ optics module (FM0), which
is essentially identical to the \ps optics module.  The optics modules
are built up from 133 layers of grazing incidence optics, built up
using epoxy and graphite spacers on a computed, numerically controlled
(CNC) lathe assembly machine.  The optics module is 47.2~cm long, 
19.1~cm in diameter and weighs 31~kg.  This picture, taken on May 16, 
2010, shows 82 layers.
\label{fig:assemblymachine}}
\end{figure}

The final step in fabricating the nested optic is to mount the glass
segments.  Alternating layers of precision-milled graphite spacers
and glass segments are epoxied together using one of the lathe-like
assembly machines procured to build the \nustar\ optics
(Figure~\ref{fig:assemblymachine}).  The final optics module produces
an azimuthally symmetric point-spread function (PSF) with a tight
18\arcsec\ full-width at half maximum (FWHM) core and a 58\arcsec\
half-power diameter (HPD) \citep{2013ApJ...770..103H}.  The \nustar\ PSF
varies by $<5\%$ as a function of energy \citep{2015arXiv150401672M}.

\citet{2010SPIE.7732E..0TH} give a detailed description of the \nustar\ optics,
\citet{2009SPIE.7437E..0NZ} describes the substrate production, 
\citet{2011SPIE.8147E..0UC} summarizes the coatings, and 
the overall optics fabrication is detailed in \citet{2011SPIE.8147E..0HC}.

\begin{figure}[t]
\vspace*{-3cm}
\includegraphics[width=3.5in]{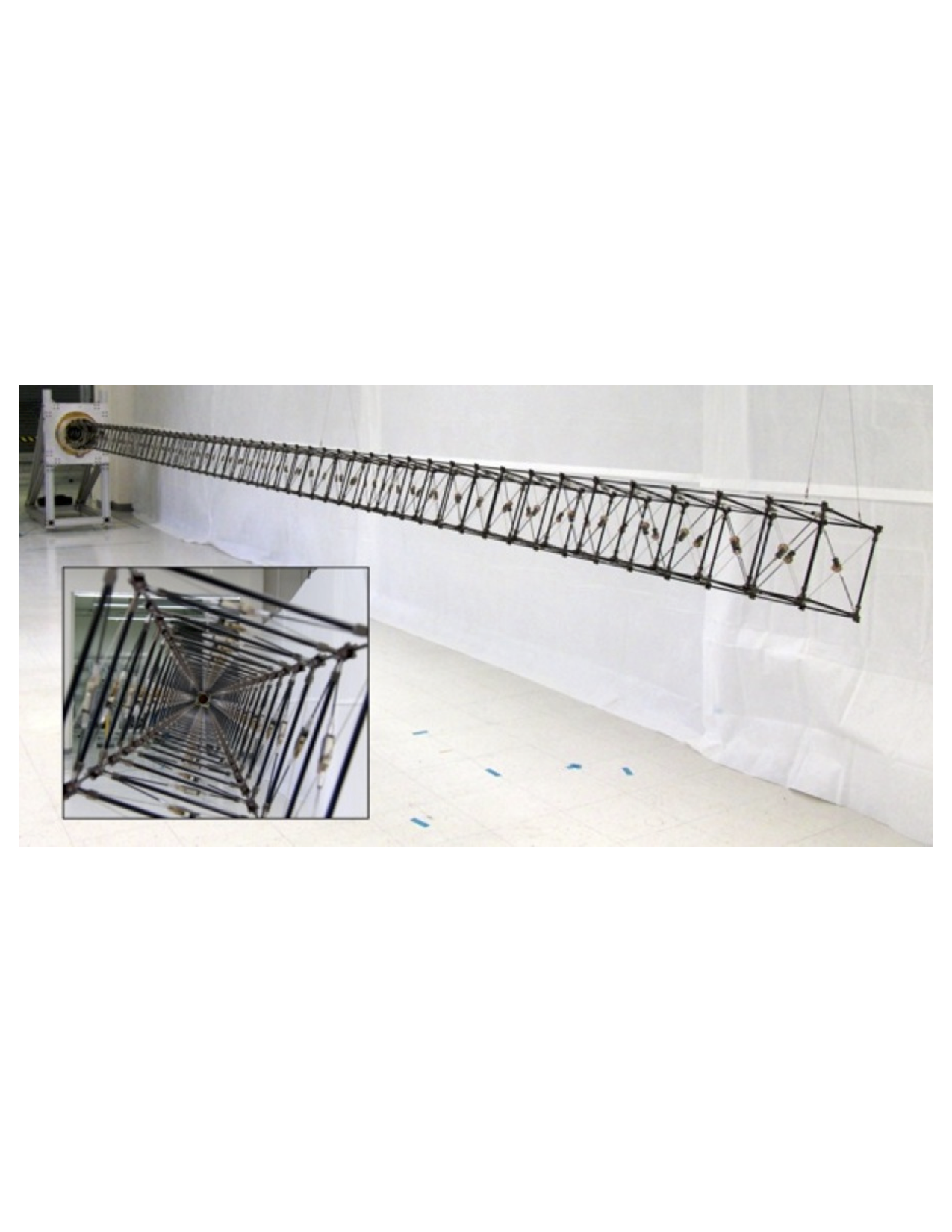}
\vspace*{-4cm}
\caption{Essential to the \ps and \ns designs is a
deployable mast which extends to 10.14-m after launch. 
Using a deployable structure allowed \ns and would allow \ps to
launch on a Pegasus XL rocket.  This extendable mast was built by
ATK Goleta, specializing in space-based deployable structures.
These images are from a full deployment test of the \ns flight
mast at ATK Goleta in August 2009.
\label{fig:mast}}
\end{figure}

{\bf Extendable mast and structures:}  \polstar\ uses a build-to-print
copy of the canister, deployment mechanism and mast used by \nustar\
to provide the required 10.14-m focal length (Figure~\ref{fig:mast}).
These would be fabricated at ATK-Goleta using the same team, facility
and processes as used for \nustar. The flight-validated design
provides the required on-orbit stiffness with a near-zero coefficient
of thermal expansion (CTE).

\polstar\ uses identical benches and structures to \nustar, as well
as a mast adjustment mechanism (MAM) at the optics end of the
extendable mast that allows tip-tilt adjustment to the optics unit.
The MAM provides for on-orbit refinement to the optical axis
\citep{2013ApJ...770..103H}.  The \nustar\ benches were designed for two
telescopes (\ie, two optics modules, focusing light on two focal
plane modules).  In order to maximize heritage with \nustar\,
to allow for a build-to-print design, and to reduce costs, \polstar\ uses the same benches,
but leaves one of the optical arms empty.

{\bf Polarimeter:}  The polarimeter (Figure~{\ref{fig:polarimeter}) consists of
a LiH scattering element, CZT detectors, readout electronics, and an active CsI shield. 
The cylindrical LiH scattering element has a diameter and length of 1~cm and 6~cm, respectively.
LiH combines a low mean atomic number (implying a high probability for scatterings 
relative to photoelectric effect interactions) with a relatively high density of 0.82 g cm$^{-2}$ 
(implying a high interaction probability for a 6~cm long cylinder).
LiH has flight heritage from being used as a shield on the Department of Energy's 
{\it Systems Nuclear Auxiliary Power} mission.
The diameter is driven by the competing desires to have a thin stick to minimize
internal absorption, but to have it large compared to the PSF and
pointing errors. The current design provides a balance matched
to the size of the CZT detector assembly.

LiH reacts with water and oxygen. The stick would therefore be packaged in a thin 
Be shield, 0.5~mm in thickness along the sides and the rear end, with a 10 $\mu$m 
entrance window at the front (mirror) end. This is sufficient to prevent moisture (on the ground) 
and atomic oxygen (in orbit) from diffusing into the stick. 
Launch loads are estimated not to be an issue for the small cylinder.  
The Be housing, included in the Monte Carlo instrument simulations, has minimal 
impact on throughput. We are currently evaluating the merits of making the rear end 
of the scattering slab of Be rather than LiH. Mostly higher energy ($>$10 keV) photons 
reach the rear end. The higher density of Be (1.85 g cm$^{-3}$) 
compared to that of LiH (0.82 g cm$^{-3}$) leads to an increased fraction of 
high-energy photons Compton scattering in the rear end of the stick.

\polstar\ uses 17 flight-proven $32 \times 32$~pixel CZT hybrid
detectors. Each pixel is attached to a readout circuit on a custom-designed 
low-noise application-specific integrated circuit (ASIC).  
\nustar\ has two focal plane modules, each constructed
of a $2 \times 2$ array of detectors on wedge-shaped ceramic boards
(Figure~\ref{fig:fpm}).  \polstar\  packages the detectors
slightly differently, with 16 detectors arranged in four $1 \times
4$ array modules forming the box that surrounds the LiH stick
(Figure~\ref{fig:detectionprinciple}).  The final detector (``Detector
17'') is located behind the LiH stick, perpendicular to the incident
photon path as a tail catcher, enabling the imaging of the observed 
source with photons not interacting in the scattering element.  
\polstar\ uses these images to verify pointing throughout an observation.

Each pixel in the CZT detector has an independent discriminator,
and individual X-ray photons trigger the readout process. On-board
processors, one for each detector module, identify the row and
column with the largest pulse height and read out pulse height
information from this pixel as well as its eight neighbors, as on
\nustar.  Unlike CCDs, CZT detectors are non-integrating and
self-triggering: for each each incident photon the charge deposited in the detectors
is collected within $\le$1/2~$\mu$s  and is subsequently read out with an electronic 
processing time of 2.5~ms per event. The event time is recorded to an
accuracy of 2~$\mu$s relative to the on-board clock 
and with an absolute timing accuracy of $\le 2$~ms 
limited by the stability of the spacecraft clock. 
The design replicates the timing capabilities of \nustar\  which surpass those of
X-ray telescopes with CCD detectors and have led to numerous discoveries 
\citep[\eg][]{2013ApJ...770L..23M,2014Natur.514..202B,2015ApJ...800..109B,2009SPIE.7435E..03R}. 
Additional details about the \nustar\ detectors and ASICs were described by 
Hailey et al.~\citeyearpar{2010SPIE.7732E..0SH}, 
Kitaguchi et al.~\citeyearpar{2010SPIE.7732E..0SH}, and 
Harrison et al.~\citeyearpar{2010SPIE.7732E..0SH}.
%
\begin{figure}[t]
\begin{center}
\includegraphics[width=3.5in]{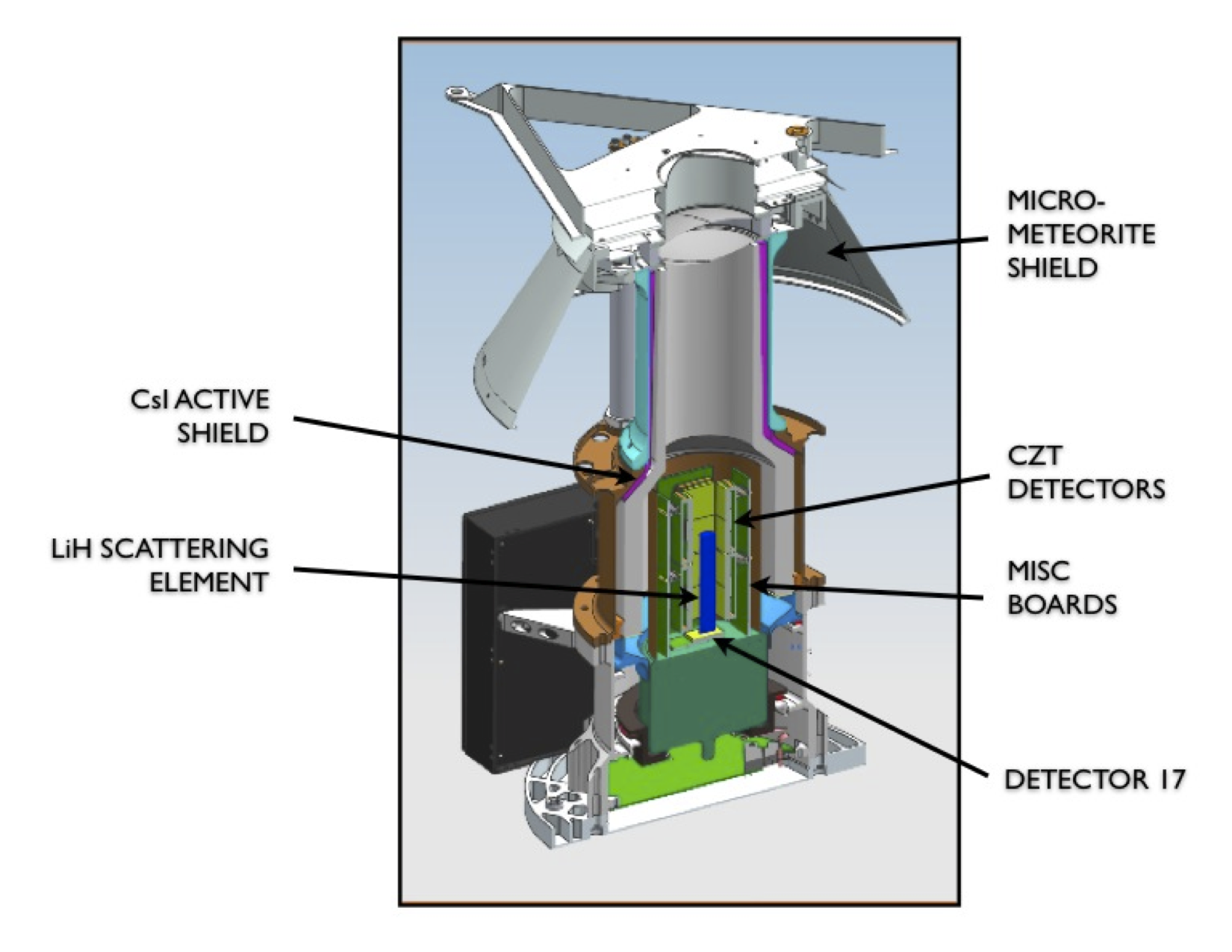}
\end{center}
\caption{Design of the \ps polarimeter showing the LiH scattering element 
surrounded by a CZT detector assembly inside a fully active CsI shield.
\label{fig:polarimeter}}
\end{figure}

{\bf Metrology system:}  \polstar\ uses the \nustar\ build-to-print
metrology system, consisting of two infrared lasers mounted on the
optics bench that focus beams on two corresponding detectors on the
focal plane bench. The lasers spots are measured to an accuracy of
10~$\mu$m (0.1\arcsec) and, combined with the instrument star camera
data, track the thermal mast motions and enable accurate knowledge
of the X-ray focal point position.  The metrology system can be used to track
the movement of the focal spot during the observations (Figure~\ref{fig:motion}).
\citet{2012OptEn..51d3605L} gives a detailed
description of the system.

{\bf Shield module:}  \polstar's equatorial orbit, identical to
that of \nustar, provides a low cosmic-ray flux and minimizes South
Atlantic Anomaly (SAA) passage, thereby enabling re-use of the
\nsg low mass, cost-effective shield configuration.  The polarimeter
is contained inside a CsI active anti-coincidence shield with a
photomultiplier tube, essentially identical to the one used on
\nustar, but with an elongated geometry to accommodate the polarimeter.
The rear portion of the CsI shield is 1.5~cm thick,  and the side walls are 1.2~cm thick 
adjacent to the CZT detectors. The front (collimator) portion of the active shield 
has a wall thickness of 0.9~cm close to the CZT detectors tapering to 0.4~cm at the front side.

{\bf Calibration source:}  \polstar\ uses the same radioactive
$^{155}$Eu calibration source and deployable mounting as \nustar.
The 10~$\mu$Ci source is mounted on the side of the shield and can
be moved into the field-of-view to monitor the gain and functionality
of the detectors.  When not deployed, the detectors are shielded
from the source. As was done with \nustar, \polstar\ would use the
calibration unit extensively during integration and testing, but
only rarely on orbit.

\subsection{Mission and Mission Operations}
\begin{figure}[t]
\includegraphics[width=3.5in]{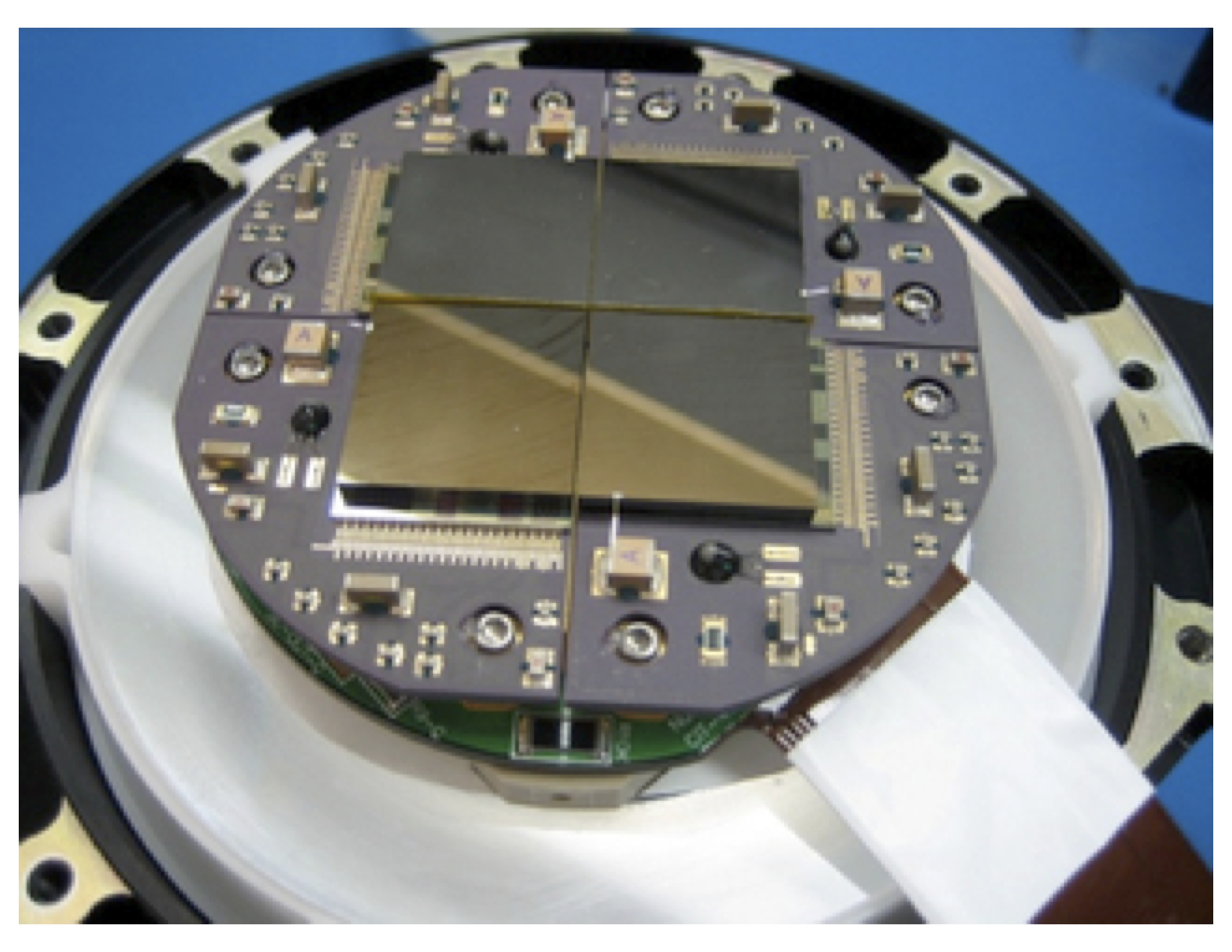}
\caption{\ns focal plane module, consisting of a $2 \times 2$ array of
CZT detectors.  \ps uses identical detectors.
\label{fig:fpm}}
\end{figure}
\ps would be in a similar low-Earth ($\sim$530 km), 6\deg inclination, near-circular orbit to \nsc. 
On orbit, \ps has two operational modes. During solar-eclipse portions of orbits, 
\ps would point at science targets and slowly rotate along 
the optical axis as a systematic error mitigation strategy. During the Sun-illuminated portions of orbits, 
the spacecraft would maintain the same pointing with respect to the science target, but would stop rotation. 
The solar panel array can have a fixed attitude with respect to the Sun and charge the batteries. 
The typical eclipse rotation rate would be once per 10 min, or three full rotations per eclipse; for observing efficiency, 
the rotation rate would be tuned on a per-target basis to maintain an integer number of rotations per science observation. 
The settle time as the observatory goes in and out of eclipse is $<$15 sec. 
Note that this is a conservative approach, providing significant power margins for charging the batteries. 
\ns nominally observes during both solar-illuminated and eclipse portions of orbits.  
We plan for a baseline mission with 18 months of science observations. 
\begin{figure}
\includegraphics[width=3.5in]{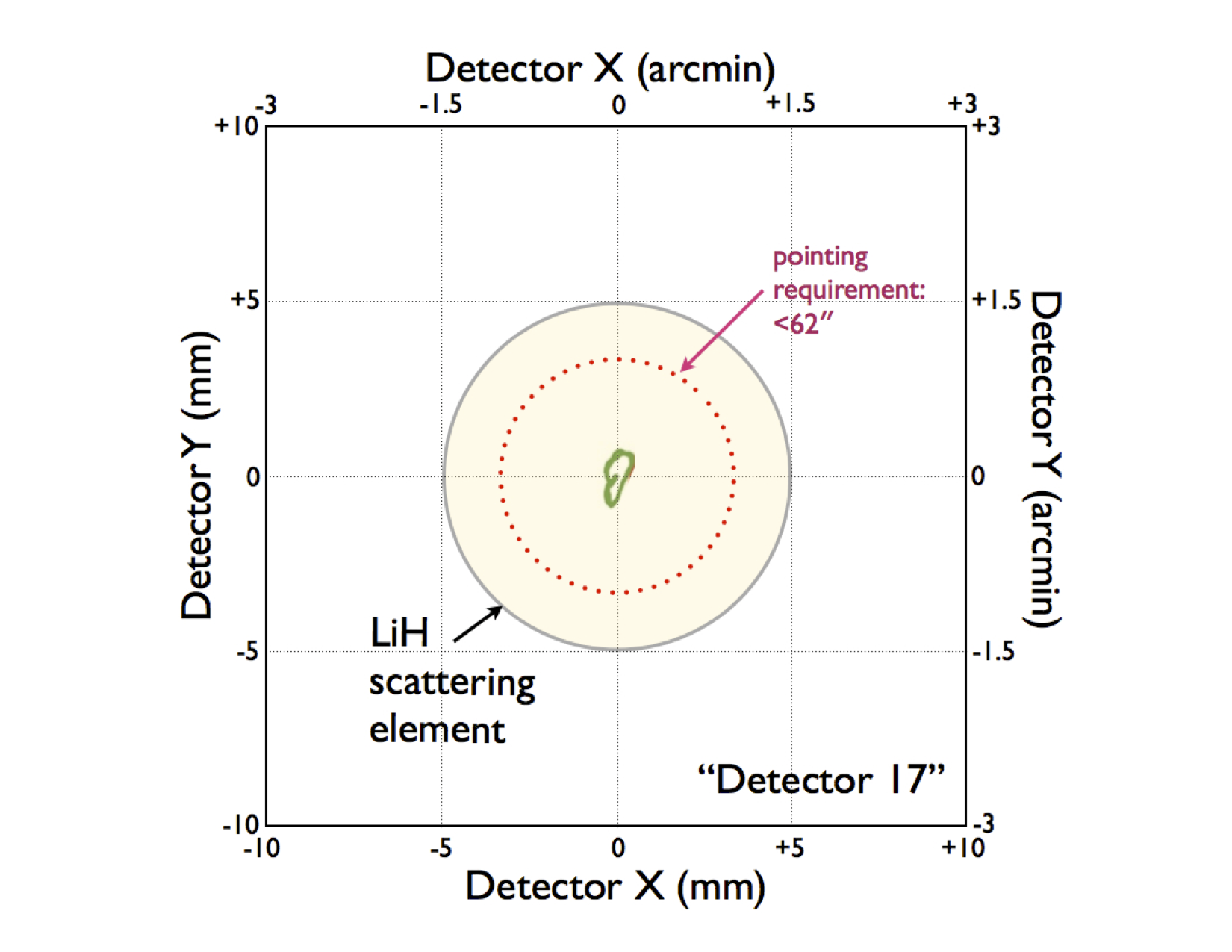}
\caption{The \ns focal spot moves as the spacecraft moves into and out of Earth's shadow. 
The green line compares the \ns focal spot for six orbits of the spacecraft to the 
size of the LiH scattering element and the 17$^{\rm th}$ CZT detector. 
\label{fig:motion}}
\end{figure}
\subsection{Returned Data}
 As per \nsc, for each event \ps records the pulse height measured in the triggered CZT detector pixel 
 and eight adjacent pixels, a time tag, and a bit indicating if the active CsI shield was triggered (54 bytes in total). 
 \ps can use \ns algorithms to measure the energy of the photon striking the CZT detector based on the signal 
 in the triggered and adjacent pixels. The processed \ps data products include a list of events, and for each event 
 the location of the energy deposition in the CZT detector, the sum $E_{\rm i}$ of all the energies recorded 
 in the CZT detectors owing to one or multiple interactions of the photon with the detector material, 
 the time of the event trigger, the offsets of the optical axis from the center 
 of the scatterer in detector coordinates, and the roll angle of the spacecraft. The data products include 
 the instrument response matrices, i.e., lookup tables giving the distribution of the observables as function of energy, 
 polarization fraction and angle, the offsets and roll angle. The science products consist of the constraints on 
 the model parameters, i.e., the parameters describing the flux, polarization fraction and polarization 
 angle as function of energy. The constraints are derived from  
 forward folding the model with the instrument response matrices and comparing 
 the resulting distributions with the measured ones (see next section).
\section{Analysis Methods, Instrument Simulations, and Projected Performance}
\label{S:perf}
\subsection{Analysis Methods}
The polarization properties of a quasi-monochromatic electromagnetic wave can be described entirely by the four Stokes parameters (all having the units of intensity): the beam intensity $I$, the parameters $Q = p_0\,I\cos(2\psi_0)$ and $U = p_0\,I\sin(2\psi_0)$, describing linear polarization, where $p_0$ and $\psi_0$ are the polarization fraction and angle, and the circular polarization $V$~\citep{stokes_gg_1852}. \ps measures $I$, $Q$ and $U$ but is not sensitive to $V$. 
We use an analysis based on assigning Stokes parameters to each individual event \citep{kislat_etal_2015a}.
The main advantage of the analysis described in the following over alternative methods (i.e. fitting the 
azimuthal scattering angle distribution with a suitable template) is (besides the ease of the involved calculations) 
that the Stokes parameters are normally distributed with a mean centered on 0 if the signal is unpolarized. 
The significance of a polarization detection and the confidence intervals on derived parameters (i.e.\ the
polarization fraction and angle) thus follow from the well understood properties of Gaussian distributions. 
Assuming that the $i^{\rm th}$ detected photon scattered along the optical axis, we use the position $x_i$, $y_i$ 
of the photon detection in the plane perpendicular to the optical axis to calculate the azimuthal scattering angle $\alpha_i$ 
measured relative to the celestial North direction.
We then define
\begin{equation}\begin{split}
  q_i &= \cos\left(2\left(\alpha_i - \tfrac{\pi}{2}\right)\right), \\
  u_i &= \sin\left(2\left(\alpha_i - \tfrac{\pi}{2}\right)\right),
\end{split}\end{equation}
where the terms $\tfrac{\pi}{2}$ account for the fact that photons scatter preferentially perpendicular to the polarization
direction. As the Stokes parameters are additive, the Stokes parameters of the 
signal are simply the sum of the $q_i$ and $u_i$ of all
$N$  observed events which pass the analysis cuts (i.e. which were not flagged as background events which triggered the CsI shield).
We define the reduced Stokes parameters:
\begin{equation}\begin{split}
  \QQ_r &= \frac{2}{\mu N} \sum_{i=1}^Nq_i, \\
  \UU_r &= \frac{2}{\mu N} \sum_{i=1}^Nu_i
\end{split}\end{equation}
with $\mu$ being the modulation factor, i.e. the fractional amplitude of the azimuthal scattering angle distribution
for a 100\% linearly polarized X-ray beam.  The measured polarization fraction and angle are then given by 
\begin{align}
  \label{eq:p0}p &= \sqrt{\QQ_r^2 + \UU_r^2}, \\
  \label{eq:psi}\psi &= \frac{1}{2}\arctan\frac{\UU_r}{\QQ_r}.
\end{align}

The Stokes parameters follow a Gaussian distribution, and the one sigma measurement error is given by
\citep{kislat_etal_2015a}:
\begin{equation}\label{eq:sigma_stokes}\begin{split}
  \sigma(\QQ_r) &= \sqrt{\frac{1}{N-1}\left(\frac{2}{\mu^2}-\QQ_r^2\right)}, \,{\rm and}\\
  \sigma(\UU_r) &= \sqrt{\frac{1}{N-1}\left(\frac{2}{\mu^2}-\UU_r^2\right)}.
\end{split}\end{equation}
The polarization fraction is restricted to values $p \geq 0$, with the probability distribution of the measurement $p$ given the true polarization fraction $p_0$~\citep{vinokur_m_1965,weisskopf_mc_2006,krawczynski_h_2011a,kislat_etal_2015a}:
\begin{equation}\label{eq:p0_prob}
  P(p|p_0) = \frac{Np\mu^2}{2}e^{-\frac{N\mu^2}{4}(p^2 + p_0^2)} \,\\ I_0\left(\frac{N\mu^2 \, p \, p_0}{2}\right),
\end{equation}
where $I_0$ is the modified Bessel function of order zero.
We use Equation~\ref{eq:p0_prob} to estimate the measurement errors on the polarization fractions measured with \ps.
For this purpose,we numerically integrate Eq.~\eqref{eq:p0_prob} and find the range $[p_1, p_2]$ 
within which $67\%$ of measurements are expected, such that $P(p_1|p_0) = P(p_2|p_0)$.

The polarization angle is described by a normal distribution with
\begin{equation}
  \sigma(\psi) \approx \frac{1}{p\mu\sqrt{2(N-1)}}.
\end{equation}
The minimum detectable polarization (MDP) is defined as the $99\%$ confidence level upper limit found for an unpolarized source~\citep{weisskopf_mc_2006,krawczynski_h_2011a,kislat_etal_2015a},
\begin{equation}
  \text{MDP} \approx \frac{4.29}{\mu\sqrt{N}}.
\end{equation}

\subsection{Measurements in the presence of backgrounds}
\ps intersperses science observations with slightly offset ($< 1^\circ$) observations to
measure the local background (rate of events not initiated by the source, e.g. by cosmic rays and photons from
the cosmic X-ray background), with offset pointing durations tailored on a per-target basis. 
The background regions would be chosen to avoid bright X-ray sources in the field of view.

For all but the brightest sources, $\QQ_r$ and $\UU_r$ are then measured independently 
for on-source and off-source observations.The Stokes parameters of the source are then
\begin{equation}\begin{split}
  \QQ_{\text{source}} &= \QQ_{\text{on}} - w_\text{off}\,\QQ_{\text{off}}, \\
  \UU_{\text{source}} &= \UU_{\text{on}} - w_\text{off}\,\UU_{\text{off}}, 
\end{split}\end{equation}
where the weight $w_\text{off}$ is the ratio of the on-source observation time divided by the off-source observation time:
\begin{equation}
  w_\text{off} = \frac{t_\text{on}}{t_\text{off}}.
\end{equation}
The optimal choice of $w_\text{off}$ depends on the expected signal rate $R_\text{S}$ and background rate $R_\text{BG}$ 
for a given source.
By minimizing the expected uncertainties on $\QQ_\text{source}$ and $\UU_\text{source}$ from Equation~\eqref{eq:sigma_stokes}, one finds the optimum value~\citep{kislat_etal_2015a}
\begin{equation}
  w_\text{off} = \sqrt{1 + R_\text{S/B}},
\end{equation}
with $R_\text{S/B} = R_S/R_{BG}$.
Accounting for the statistical errors on the Stokes parameters of the signal and the background, 
the MDP becomes
\begin{equation}
  \text{MDP} = \frac{4.29 \sqrt{R_S + R_{BG}}}{\mu \sqrt{T} (R_S + R_{BG} - \sqrt{R_{BG}(R_{BG}+R_S)})}
\end{equation}
with $T = t_\text{on} + t_\text{off}$ being the total on-source and background observation time.
Based on the worst-case estimate of the \ps background, \ps would spend $\sim 15\%$ of the observation time
of the baseline mission on background observations. 

An offset of the focal point from the center of the scattering element leads to an asymmetry in the azimuthal scattering distribution 
owing to the direction dependent absorption in the scattering element and a geometrical bias from folding the mirror 
PSF with the cross section of the scattering element \citep{beilicke_etal_2014}. 
The full \ps data analysis corrects for pointing offsets with the help of a forward-folding analysis. The latter uses a template
library of simulated events generated according to a~$E^{-1}$ power-law spectrum for a matrix of focal spot offsets 
with Stokes parameters $Q,U = \pm 1$. A particular observation is modeled by drawing events from the library 
to mimic the pointing history during the observation. Events with certain $Q$ and $U$ values are drawn to simulate 
a beam with certain net Stokes $Q$ and $U$ values, and are weighted according to the assumed energy spectrum.
A chi-square minimization is then used to find the best-fit model parameters and to derive model uncertainties.
The details of this analysis are the subject of a future paper.
\subsection{\ps Instrument Simulations and Performance}
\label{S:perfX}
We estimate the \ps performance by combining \ns pre-flight and in-orbit results with Monte Carlo 
simulations of the polarimeter response. The simulations use the {\em GEANT-4} simulation package 
with the Livermore Low-Energy Electromagnetic Models physics list \citep{2003NIMPA.506..250A}.
The simulations use the results from a ray-tracing code developed for \nustar\ and include 
the \nustar-measured energy-dependent quantum efficiency of the CZT detectors.
The latter include the effects of photon absorption by the cathode and the inactive
transition layer between the cathode and the CZT. The simulations account for the possibility 
of multiple interactions of the high-energy photons within the scatterer and the detectors.
%
\begin{figure}
\includegraphics[width=3.5in]{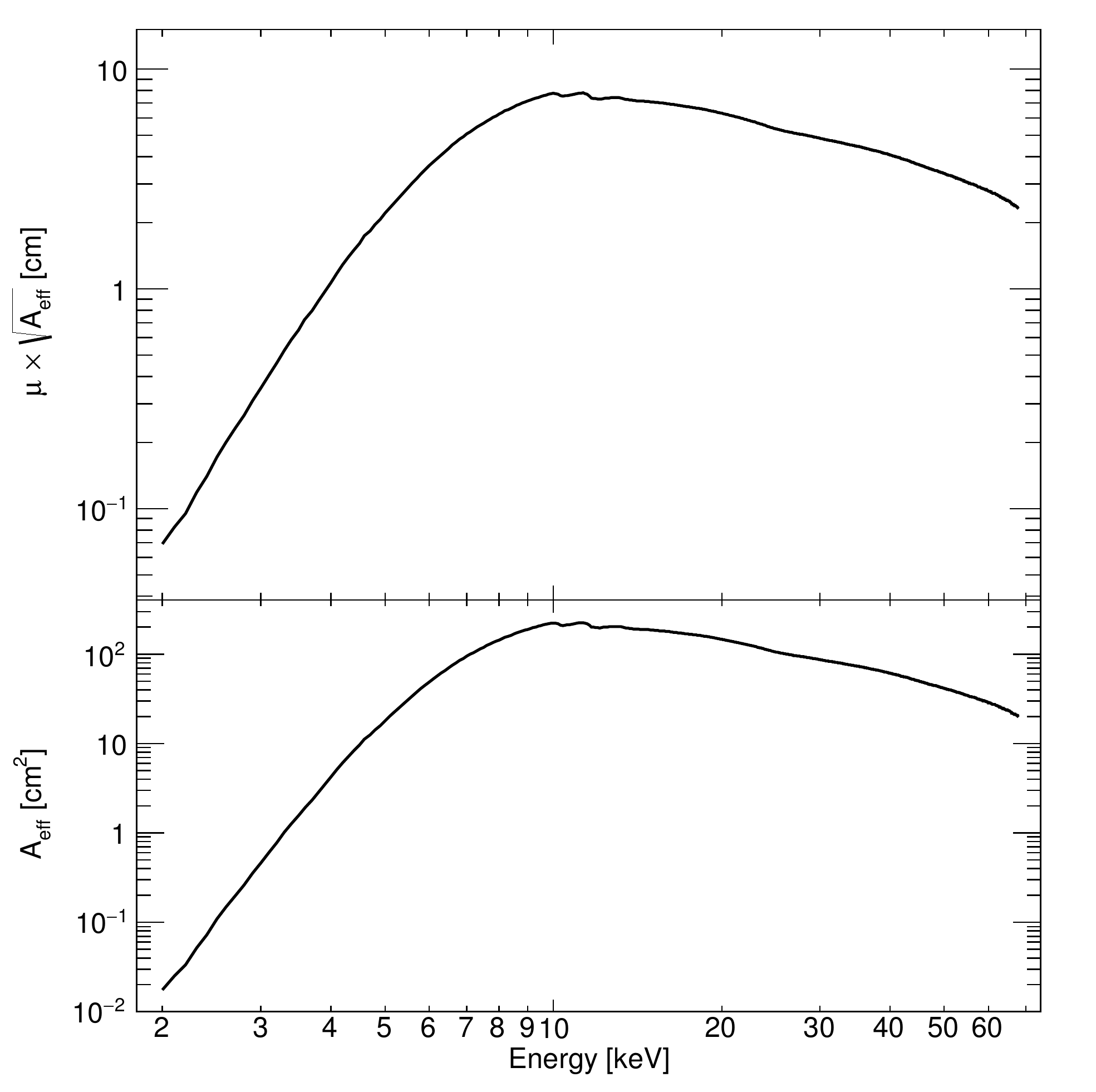}
\caption{The upper panel shows \psg modulation factor $\mu$ times the 
square root of the effective detection area $A_{\rm eff}$ as function of energy, 
and the lower panel shows $A_{\rm eff}$ as function of energy.
For comparison, {\it OSO-8} had a peak continuum radiation effective area of 
0.6~cm$^2$ at 9~keV \citep{1976ApJ...210..805K}.
A Crab-like source leads to a flux density rate $> 1$ count sec$^{-1}$
keV$^{-1}$  over the entire range 3.5-25~keV.
\label{fig:effectivearea}}
\end{figure}

\begin{figure*}[tbh]
\begin{center}
\includegraphics[width=5in]{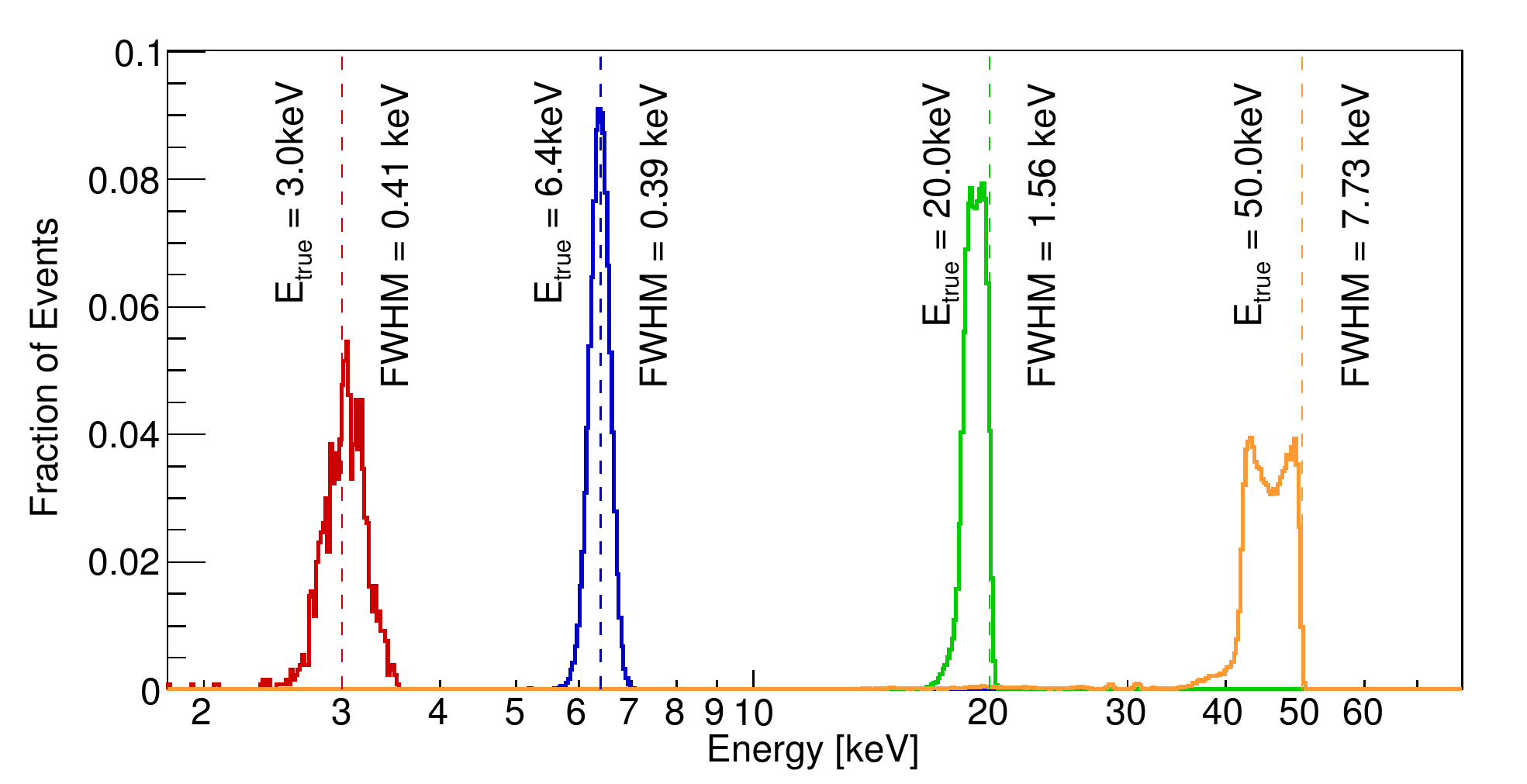}
\end{center}
\caption{Energies deposited in the CZT detectors for different incident photon energies. 
\label{fig:energyRes}}
\end{figure*}

Figure~\ref{fig:effectivearea} presents the total effective area $A_{\rm eff}$ of \polstar\ 
(including the optics, scatterer, and detector efficiencies) as a function of energy 
as well the product of the modulation factor $\mu$ times the square root of the effective area.
The latter product can be used as figure of merit characterizing the differential sensitivity of a polarimeter 
as the minimum detectable polarization fraction follows the scaling law MDP~$\propto\, (\mu\sqrt{A_{\rm eff}})^{-1}$.
For a source with a Crab like flux and energy spectrum  the 3-15 keV detection rate is 
$R_{\rm src}=108$ Hz (CBE, 97~Hz requirement). 
The modulation factor $\mu$ gives the amplitude of the sinusoidal modulation of the azimuthal scattering angle distribution 
for a 100\% polarized signal. For \psc, the simulations give $\mu\approx0.52$ largely independent of energy.

\psg energy resolution is limited by the CZT detector/readout resolution to about  0.4 keV FWHM at low 
($<$10 keV) energies. At higher energies ($>$10 keV) an increasing fraction of the primary photon's energy 
is given to the Compton electron.  
Figure~\ref{fig:energyRes} shows the detector response for a few exemplary incident photon energies. 
After re-normalizing the energies deposited in the CZT detectors to the incident
photon energy, we infer energy resolutions of 0.4 keV FWHM at $<$10 keV, 
1.65 keV FWHM at 20 keV and 8.5 keV FWHM at 50 keV. 
The full (forward folding) analysis makes use of the fact that each ring of pixels surrounding 
the scattering element sees photons preferentially from the front-end of the scattering element 
with an energy dependent exponentially suppressed distribution of scattering locations 
deeper into the element. Taking into account where the photons strike the CZT detector assembly, 
the energy resolutions can be improved for a subset of the events. For example, the energy of 50 keV events 
detected at the front end of the second detector ring (counted from the front end) can be 
reconstructed with an effective energy resolution of 4 keV FWHM.
\begin{figure}[t]
\includegraphics[width=3.5in]{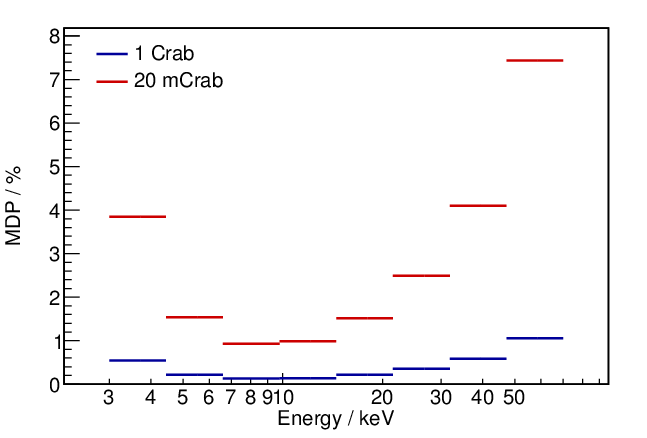}
\caption{\psg polarization sensitivity for dim (upper red bars) 
and bright (lower blue bars) sources. The lines show the 
Minimum Detectable Polarization fraction (MDP, 99\% CL) as a function of source flux 
in eight statistically independent energy bins assuming 860 ksec (10 days) 
of on-source exposure time for a source with a Crab-like spectrum.
\label{F:sens}}
\end{figure}
%
\begin{figure}[tbh]
\includegraphics[width=3.5in]{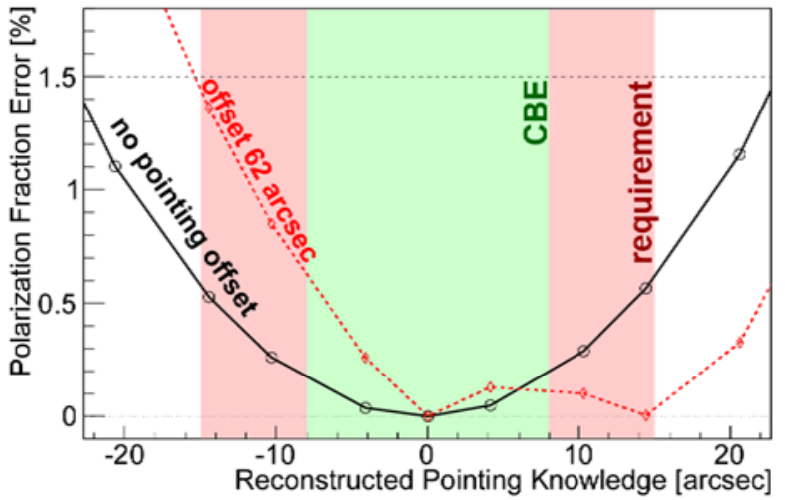}
\caption{
Monte Carlo simulation of the systematic 3-15~keV polarization fraction error as a function of the pointing knowledge error. 
For a nominal position centered on the LiH stick (black circles), a 15\arcsec (8\arcsec ) positional uncertainty translates into a systematic polarization fraction uncertainty of 0.6\% (0.2\%). 
For a nominal position centered 62\arcsec\ from the LiH stick center (red triangles), a 15\arcsec\ positional uncertainty 
can produce polarization fraction errors as high as 1.4\%. All systematic uncertainties quoted at the 99.7\% CL.
\label{fig:montecarlo}}
\end{figure}

For dim sources, \psg sensitivity depends on the level of background counts.
We conservatively assume the background per detector measured in-orbit with the \ns CZT detectors (Wik et al. 2014).
For \nustar\, stray cosmic X-ray background light leaking through the aperture stop
dominates the background below $\sim 20$~keV. Internal radiation activated by the
orbital environment dominates above 20 keV. 
The \ps background should be substantially lower than the \ns background as the detectors 
do not see the sky directly but are oriented towards the scattering slab, and the latter absorbs 
a considerable fraction of the events at $\le$~10~keV energies.
The \nsc-based estimate predicts a worst-case 5-20 keV background rate of $R_{\rm bkg}=$0.94 Hz.  
Simulations are underway to improve this estimate.
%
\begin{table*}[tbh]
\centering
{\small
\begin{tabular}{|p{6cm}|p{3cm}|p{3cm}|p{3cm}|}
\hline
Effect & \multicolumn{2}{|p{6cm}|}{Pol. fraction error for non-rotating instrument{[}\%{]}}           & Error suppressed by rotation? \\  \cline{2-3}
Energy range (keV):  & 3-15            & 15-50           &  \\ \hline \hline
Detector and background inhomogeneities & 0.25 & 0.25 & Y \\ \hline
LiH mechanical tolerances (0.2 mm)   & 0.04           & 0.13            & Y \\ \hline
PSF unc. (pre-/post-launch \nustar\ comparison)     & 0.28           & 0.24            & Y \\ \hline
0.1 keV energy calibration error, limiting pointing offset corrections & 0.1           &  0.001           & Y \\ \hline
Pointing knowledge error &  1.4/0.25~$^{\rm a}$ & 1.3/0.1~$^{\rm a}$           & N \\ \hline 
\multicolumn{4}{p{15cm}}{\footnotesize ~~$^{\rm a}$~Two values are given: the first one is for the required performance (62\arcsec offset, 15\arcsec knowledge), and the second for the CBE performance (17\arcsec offset, 8\arcsec knowledge)}.\\
\end{tabular}}
\caption{\ps worst-case systematic errors (99.7\% CL) for a source consistently offset by 3~mm (62\arcsec) from the center of the
 field of view.}
\label{table:rotation}
\end{table*}
Bright sources within $5^\circ$ of a target cause additional stray light issues for \nustar\, which is
only a concern for five targets in the \ps baseline observation program (\S~\ref{S:Science}). This can be mitigated
through a combination of modeling, off-target measurements, and
data censoring.  We are also evaluating the merits of incorporating 
a flight-ready stray light baffle that was built too late for \nustar\ to be
incorporated.

The \ps sensitivity is best between 5-15 keV (see Figure~\ref{F:sens}). The sensitivity 
decreases at lower energies owing to the limited scattering efficiency in the scattering element. 
At higher energies, the assumed steep energy spectrum of the astrophysical source 
and the declining mirror effective area limit the sensitivity.  

We used simulated data sets to estimate systematic errors. The relevant figure of merit is the spurious polarization 
measured for an unpolarized X-ray beam. Table~\ref{table:rotation} lists the main sources of systematic errors 
before accounting for their suppression through the spacecraft roll. 
The largest error stems from the practical limitations of flat-fielding the detectors.
Note that the spacecraft roll suppresses most systematic errors by averaging over detector non-uniformities
and spacecraft asymmetries. The main contribution to the residual systematic error is expected 
to come from the pointing knowledge error which varies as the spacecraft rolls. 
Figure~\ref{fig:montecarlo} presents the resulting spurious polarization for a target both centered on-axis and
offset by 3 mm (62\arcsec). Based on these simulations, we find that in order to meet the requirement 
that systematic polarization fraction errors be $\leq 1.5\%$, we are required to maintain the
target within 62\arcsec\ of the LiH stick center and have reconstructed
pointing knowledge $\leq 15$\arcsec, where all numbers are at the 99.7\% CL.  
%
\begin{table*}[tbh]
\centering
{\small
\begin{tabular}{|l|c|c|c|}
\hline
                          & \begin{tabular}[c]{@{}l@{}}\polstar\ Req't\\ (99.7\% CL)\end{tabular} & \begin{tabular}[c]{@{}l@{}}\polstar\ CBE\\ (99.7\% CL)\end{tabular} & \begin{tabular}[c]{@{}l@{}}\nustar\\ Actuals\end{tabular} \\ \hline
Control                   & $\leq 40$\arcsec                                                           & 11.3\arcsec                                                        & 8\arcsec                                                  \\ \hline
Stability - science orbit & $\leq 40$\arcsec                                                           & 12.9\arcsec                                                        & 11\arcsec                                                 \\ \hline
Combined  & $\leq 62 $\arcsec                                                           & 17.1\arcsec                                                        & 13.6\arcsec                                                 \\ \hline \hline

\end{tabular}}
\caption{\polstar\ pointing requirements.}
\label{table:budget}
\end{table*}
Table~\ref{table:budget}  presents the budgeted requirements on the
pointing system derived from the science constraints on the systematic errors 
and compares them to the \ps CBE and the \ns actuals.  
By requiring pointing control $\le$40\arcsec (from the spacecraft) and pointing stability $\le$40\arcsec 
(from the combined motions of the spacecraft and mast), we can maintain the target within 62\arcsec 
of the desired location aligned with the LiH stick center (Root of Sum of Squares (RSS)=57\arcsec; CBE is 17\arcsec). 
Using the metrology and star tracker data on the ground, \ns achieves reconstructed pointing 
knowledge to an accuracy of 8\arcsec\ (99.7\% CL). The rotation of \polstar\ is sufficiently slow that this
performance is not degraded.
\begin{table*}[tbh]
{\small
\begin{tabular}{ |  p{2.6cm} |  p{2.6cm} |p{1.7cm} |p{1.7cm}| p{1.7cm} |p{1.7cm} |}
\hline
Source & 
Source Type & 
Flux 2-12 keV $[\rm mCrab]$& 
On-source time $[\rm days] $& 
MDP $[\rm\%]$ 
3-15 keV, 99\%CL & 
Pol. Fraction Error $[\rm\%]$ 3-15keV\\ \hline \hline
Cyg X-1 & Stellar BH & 414 & 3.4 & 0.2 & 0.09\\
GRS 1915+105 & Stellar BH & 717 & 15.2 & 0.1 & 0.03\\
LMC X-3 & Stellar BH & 20.1 & 13.3 & 0.7 & 0.35\\
Cyg X-3 & Stellar BH & 168 & 0.9 & 0.7 & 0.30\\
Flaring source & Stellar BH & 82.8  & 2.1 & 0.7 & 0.31 \\ \hline
NGC 4151 & Supermass.\ BH & 5.9  & 5.5 & 2.6 & 1.51\\
MCG-5-23-16 & Supermass.\ BH & 4.4 & 6.3 & 3.0 & 1.84\\
MCG-6-30-15 & Supermass.\ BH  & 3.9 & 7.8 & 3.0 & 1.85\\ \hline
Mrk 421  & HSP Blazar & 14 & 5.7  & 1.3  & 0.71\\
Mrk 501 & HSP Blazar & 5.1 & 10.0 & 2.1 & 1.28\\
3C 273 & FSRQ & 4.9 & 1.9 & 5.1 & 3.05\\
PKS 1510-08 & FSRQ  & 2.57 & 5.9 & 5.0 & 3.13\\
Flaring Blazar & Blazar & 10 & 0.5 & 7.2 & 4.30\\ \hline
1E 2259+586  & AXP & 10.2 & 9.3 & 1.3 & 0.72\\
4U 0142+61 & AXP & 4.8 & 10.1 & 2.3 & 1.35\\
SGR 1806-20 & SGR & 4.5 & 11.2 & 2.2 & 1.35\\ \hline
Vela X-1 & Acc.\ Pulsar & 48.3 & 0.7 & 1.8 & 0.81\\
GX 301-2 & Acc.\ Pulsar & 22 & 1.9 & 1.7 & 0.86\\
Her X-1 & Acc.\ Pulsar & 13.5 & 3.8 & 1.6 & 0.89\\
Sco X-1 & Acc.\ NS & 1173 & 0.1 & 1.0 & 0.35 \\
4U 1700-377 & Acc.\ NS & 50.7 & 0.6 & 1.8 & 0.81\\
Cyg X-2 & Acc.\ NS & 50 & 0.6 & 1.8 & 0.81\\
Crab & Rot.\ Pulsar & 1000 & 0.1 & 1.0 & 0.35\\
Vela & Rot.\ Pulsar & 7.9 & 2.2 & 3.2 & 1.85\\ \hline
\end{tabular}}
\caption{\label{T:sources} The 24 targets of the baseline \ps mission. 
Tabulated minimum detectable polarization fraction (MDP) and polarization fraction 
errors represent the statistical uncertainties for mission requirements.}
\end{table*}
\section{Science Investigations}
\label{S:Science}
We designed \ps for a mission duration of 18 months. In this time, \ps can observe the 24 sources listed in~Table~\ref{T:sources}. The sources include the brightest and best-studied sources of several source classes as well as two targets of opportunity. The \ps observations can be used for physics-type experiments, validating or falsifying the leading paradigms
of where and how the X-ray emission originates in these sources.  The science objectives can be summarized as follows:\\[1ex]
\hspace*{0.5cm} {\bf Objective 1:} Reveal black hole accretion flows: \ps combines spectroscopic, timing, and polarimetric information to map the innermost accretion flows around stellar mass and supermassive black holes, where gravitational potential energy is converted into radiation and mechanical energy. \psg 3-D information about the structure of the accretion flow tests theories of black hole accretion \citep{2004MNRAS.355.1005D,2008MNRAS.391...32D,2009ApJ...691..847L,2009ApJ...701.1175S,2010ApJ...712..908S,2011ApJ...731...75D}.
\psg energy band is ideally suited to decomposing the spectra of accreting objects into components from accretion disks (3-8 keV; only in the case of stellar mass black holes), hot coronal regions (3-50 keV), coronal emission reflected and re-processed off the accretion disk (6-50 keV, including the $>$10 keV Compton hump), and, in some cases jets (which in some scenarios may contribute significantly above $\sim$40 keV). Einstein's theory of General Relativity (GR) makes as-yet untested predictions about the behavior of matter and radiation in the extremely curved and twisted spacetime around black holes. \ps can search for the predicted signatures of the general relativistic Lense-Thirring precession and Bardeen-Peterson warp of the inner accretion flow in the strong gravity regime.\\[1ex]
\hspace*{0.5cm} {\bf Objective 2:} Reveal the magnetic backbone of blazar jets: \ps can test leading theories of how actively accreting supermassive black holes form, accelerate, and collimate powerful outflows (jets) by accurately measuring the time evolution of the polarization angle of the X-ray emission from blazars (supermassive black holes with jets pointing at us). The theories invoke a helical magnetic field moving through the X-ray emission region, and predict smooth swings of the polarization angle of the synchrotron continuum emission over time \citep{2008Natur.452..966M}. Polarization angle swings are predicted to be more pronounced in the X-ray band than in the radio or optical bands because the X-ray emitting regions are smaller and more uniform, as evidenced by their fast, large amplitude flares \citep{2013arXiv1303.7158K}.\\[1ex]
\hspace*{0.5cm} {\bf Objective 3:} Explore the new physics of strongly
magnetized neutron stars: \ps observations of anomalous X-ray pulsars (AXPs)
and soft gamma-ray repeaters (SGRs) test the magnetar hypothesis, which
posits that the high-energy emission from these objects is driven by extremely
strong ($10^{14}-10^{15}$ G) neutron star magnetic fields. The magnetar model
predicts extremely high ($\sim$20-100\%) polarization fractions in the \ps
energy band \cite[e.g.][]{1999MNRAS.306..333S,2011ApJ...730..131F,2014MNRAS.438.1686T}.
Phase-resolved polarization measurements (i.e., folding the polarization data
with the rotation period of the neutron star) can constrain the magnetic field
configuration in the magnetosphere, distinguishing a pure dipole field from a
twisted magnetic field \citep{2011ApJ...730..131F}.  \ps observations of
magnetars and strongly magnetized accreting neutron stars can afford, for the
first time, the capability to detect vacuum birefringence, a prediction of
quantum electrodynamics (QED) in ultra-strong magnetic fields that cannot be
tested in terrestrial laboratories
\citep{1987PASJ...39..781K,1988ApJ...324.1056M,2014MNRAS.438.1686T}. Again, \psg broad energy band
pass is key for this study as the effects manifest themselves in the 5-50 keV
energy band.  In accreting pulsars with magnetic field strengths of $\sim
10^{12}$~G, this band covers the vacuum resonance (where plasma and vacuum
birefringence compete) as well as the cyclotron line energy
\citep{1987PASJ...39..781K,1992hrfm.book.....M,2011APh....34..550K,
2013arXiv1301.5514G}.

\begin{figure*}[tb]
\begin{center}
\includegraphics[width=3.5in]{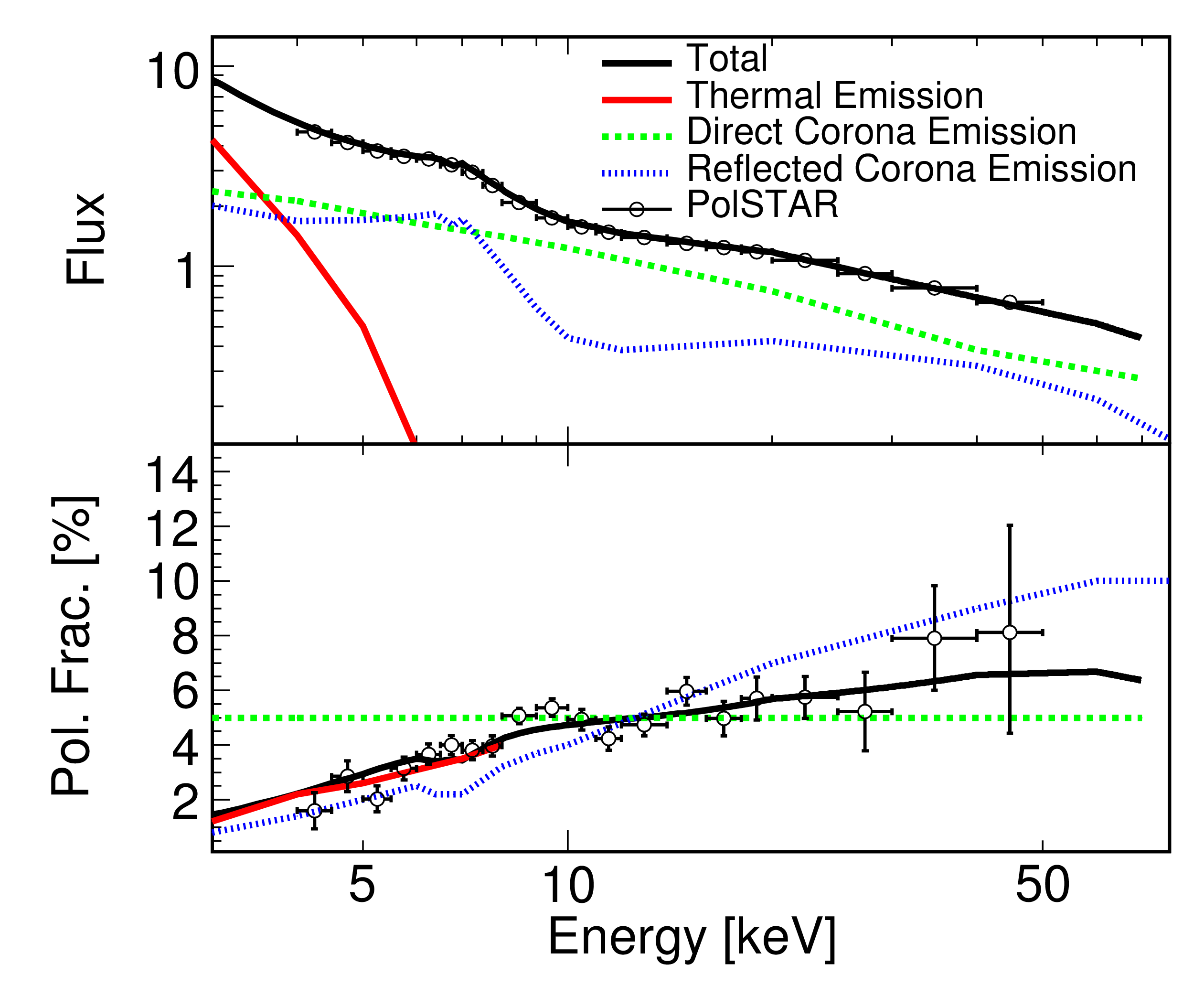}
\includegraphics[width=3.5in]{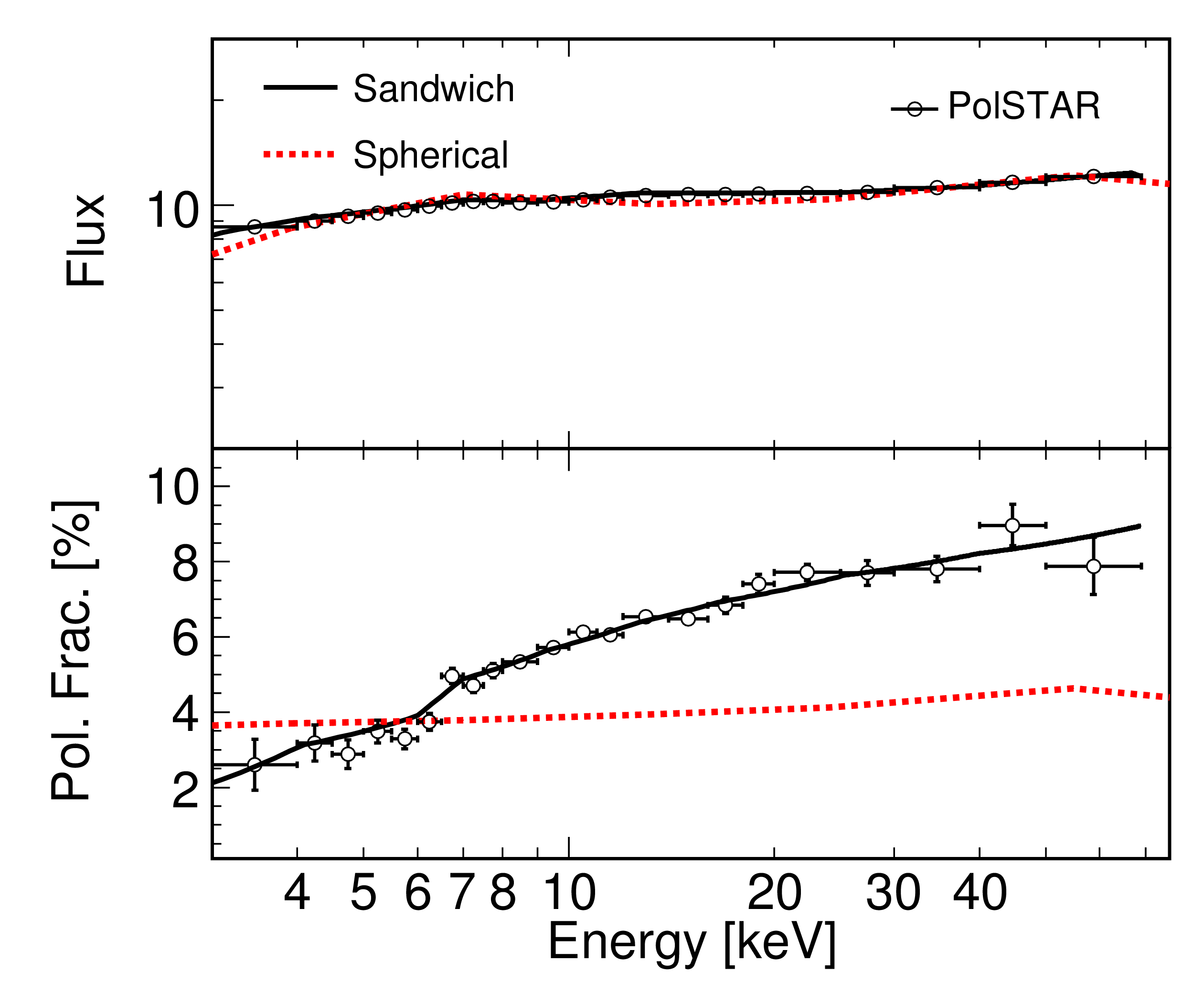}
\caption{\label{F:stellarMassBH} The broad energy range of \ps covers all primary X-ray 
emission components of accreting compact objects, and provides a wide lever arm for distinguishing models. 
{\bf Left:} Simulated 3-day observation of the black hole Cyg X-1 in the soft state, illustrating the distinct spectral and spectropolarimetric properties of the three primary emission components of accreting stellar mass black holes: 
thermal disk emission, direct coronal emission, and reflected coronal emission.
The simulated data set (data points) assumed the model predictions for the total emission (solid line).    
(energy spectrum from \citealp{2014ApJ...780...78T}; polarization fractions and angles roughly 
consistent with the simulation results of \citealp{2009ApJ...701.1175S,2010ApJ...712..908S} for
a 10 solar mass black hole with a spin of $a\,=$~0.9 accreting at 10\% of the Eddington luminosity 
as seen at an inclination of $i\,=$~75$^{\circ}$).
{\bf Right:} Simulated three-day observation of the black hole GRS~1915+105 in the power law state, 
showing that \ps can distinguish between the modeled sandwich and spherical corona models.
The simulated data set (data points) assumed the model predictions for the sandwich corona (solid line).    
The polarization fractions and angles are from the simulations of a 10 solar mass black hole 
with a spin of $a\,=$~0.9, accreting at 10\% of the Eddington luminosity as seen at an 
inclination of $i\,=$~75$^{\circ}$ \citep[][Fig.\ 3]{2010ApJ...712..908S}.
In both figures, the upper flux panels refer to energy flux per logarithmic energy interval in 
units of $10^{-9}$~erg~cm$^{-2}$~s$^{-1}$.}
\end{center}
\end{figure*}
In the following, we discuss the observations for each of the three objectives.
\subsection{Dissect the Black Hole Accretion Flows onto Stellar Mass Black Holes}
\label{SS:stellarBHs}
\subsubsection*{\ps Can Measures the Polarization Properties of Multiple Emission Components}
The observation plan includes five bright stellar mass black holes in Galactic X-ray binaries with particularly deep 
observations of the bright systems Cyg X-1 and GRS 1915+105. 
\ps can measure polarization fractions with statistical accuracies of 0.5\% (1$\sigma$) in as many as 40 (Cyg X-1) to 320 
(GRS 1915+105) independent temporal and energy bins. The high signal-to-noise data sets sample the polarization properties as functions of time, flux, and emission state. 
Figure \ref{F:stellarMassBH} (left) shows a simulated three-day observation of Cyg X-1 in the soft state highlighting the quality of the data with detailed information about the polarization properties of the thermal disk emission and the direct and reflected coronal emission. Observing the source in different states can disentangle the polarization of the individual emission components.

\subsubsection*{Studies of Black Hole Coronas}
Spectroscopic observations of black holes require the presence of a hot plasma to explain the 
power law spectral component dominating the emission at higher energies as Comptonized  
accretion disk photons \citep[see e.g.][]{1973Sunyaev,1975ApJ...195L.101T,1976ApJ...204..187S,1976ApJ...206..910K,1980A&A....86..121S}.  
Although coronas have been the subject of intense studies, their shapes, origin, and 
the roles they play in accretion systems are still a matter of debate 
\citep[e.g.][]{2013FrPhy...8..630Z,2014SSRv..183..121G}. 
X-ray polarimetry offers additional information that can be used to constrain the corona properties.
Figure~\ref{F:stellarMassBH} (right) shows the simulated results of a three-day observation of GRS 1915+105 in the power law state. The polarization properties of the two corona models differ markedly. \psg results enable using the corona as a diagnostic tool to understand the processes driving black hole accretion and jet formation. 

\begin{figure*}[tb]
\begin{center}
\includegraphics[width=3.0in]{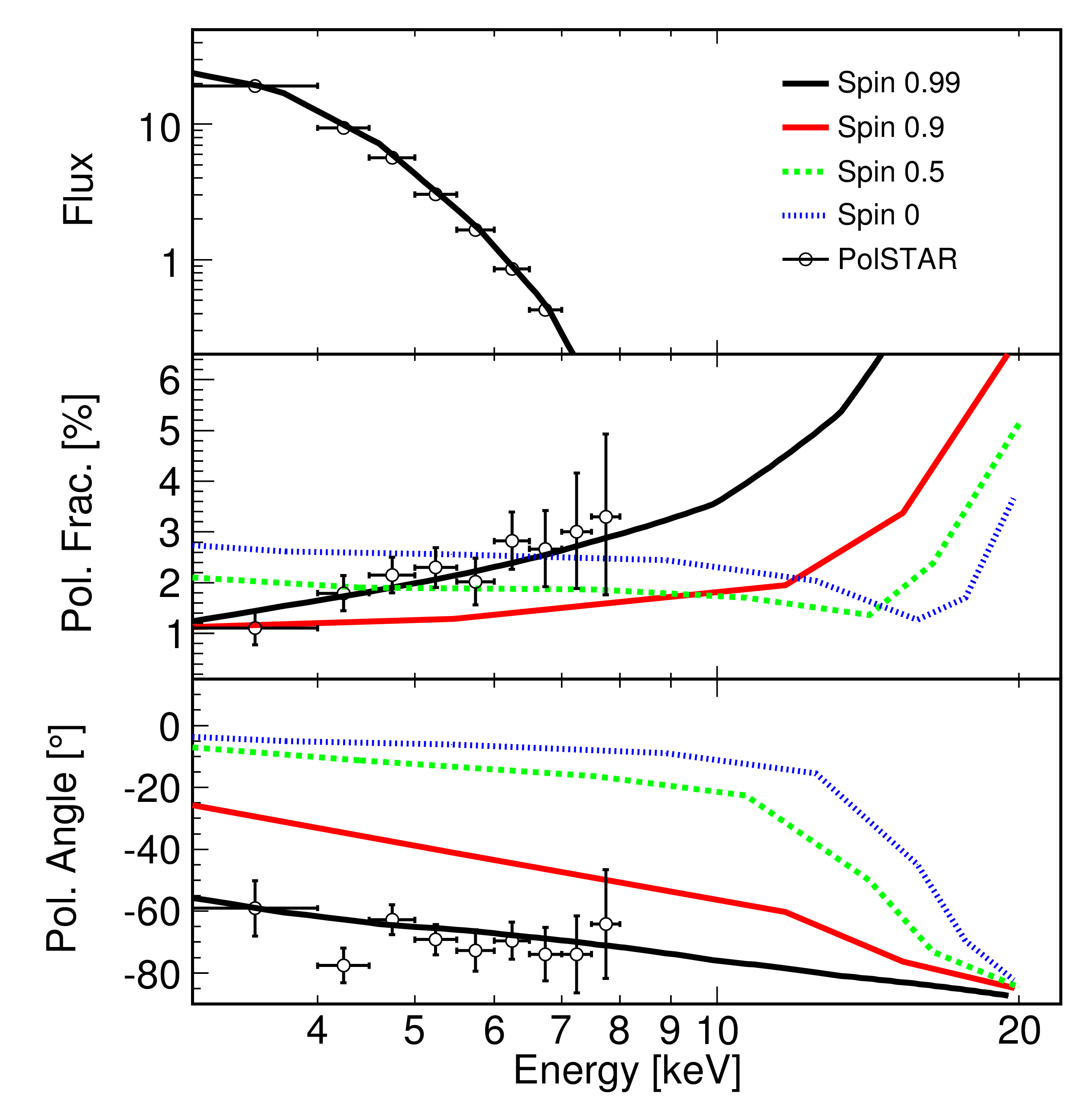}
\includegraphics[width=4.0in]{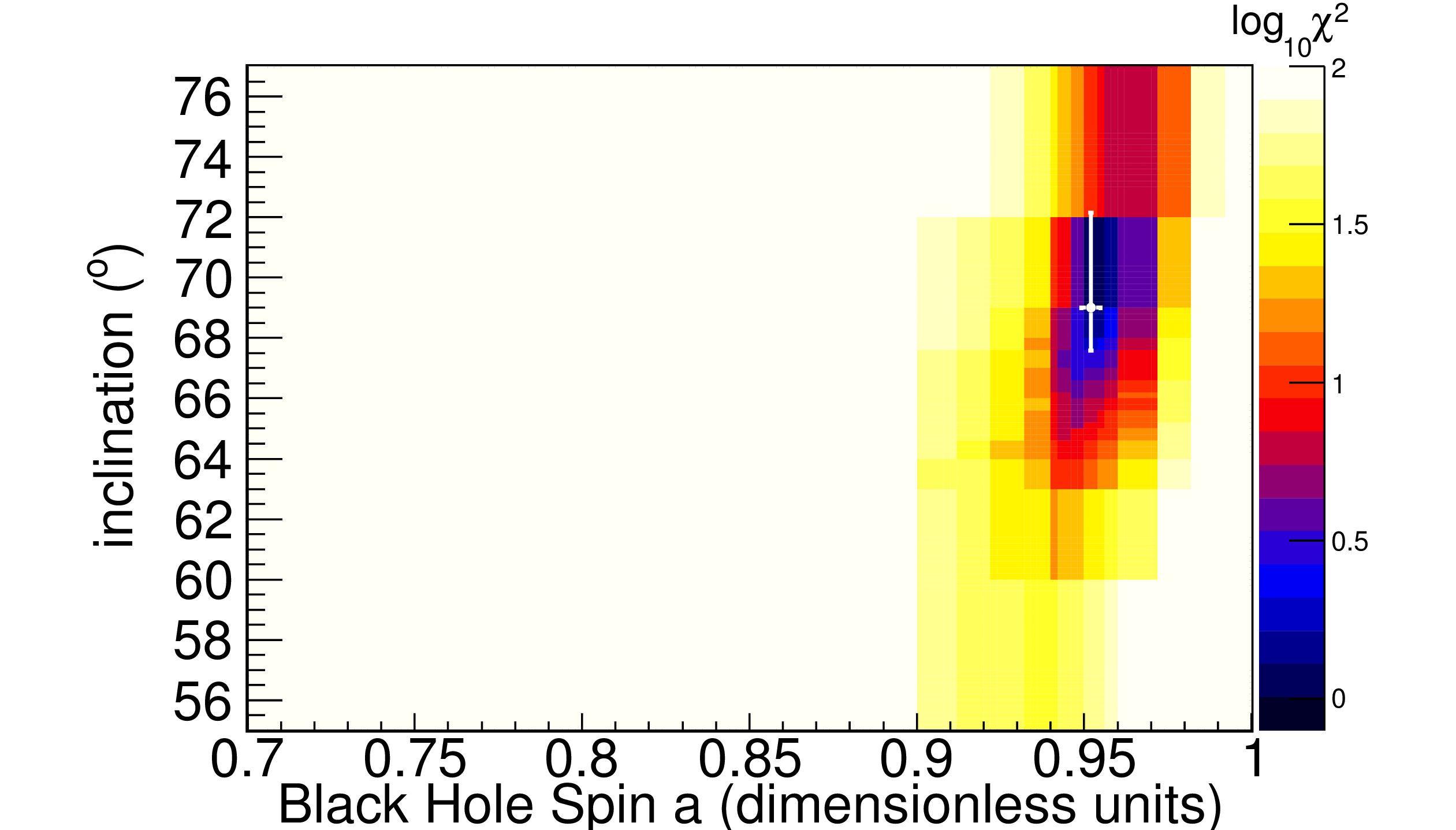}
\caption{\label{F:spin} \ps can provide an independent, geometric measure of black hole spin.
{\bf Left:} Simulated seven-day observation of GRS~1915+105 in the thermal state.
We used the measured energy spectrum of \citep{2010ApJ...713..257U}. 
The lines show the polarization fractions and angles 
from the simulations of a 10 solar mass black hole accreting at 10\% of the Eddington luminosity 
for different values of the black hole spin as seen at an inclination of $i\,=\,75^{\circ}$ 
\citep[from][Fig.\ 7]{2009ApJ...701.1175S}. 
The simulated data points assume the polarization properties for the model with spin $a=0.99$ 
shown by the solid line. The flux is given in the same units as in Figure~\ref{F:stellarMassBH}. 
{\bf Right:} results of a quantitative analysis of simulated \ps data, using a least squares fit to determine the black hole 
spin, inclination, and the orientation of the spin axis in the sky (not shown). The white dot shows the best-fit value of the 
$\chi^2$ fit at $a=0.952$, inclination=$69^{\circ}$ (input values: $a=0.95$ and inclination=66$^{\circ}$). 
The white cross shows the combined 1$\sigma$ statistical and systematic errors, 
which are primarily systematic for this bright source.}
\vspace*{-0.7cm}
\end{center}
\end{figure*}
\subsubsection*{Measurement of Black Hole Spin and the Orientation of the Inner Accretion Disk}
Figure~\ref{F:spin} shows that a seven-day exposure of GRS 1915+105 in the thermal state promises to 
measure its black hole spin and inner accretion disk orientation. 
We performed a quantitative analysis of how well \ps can measure the spin by 
generating a library of fitting templates derived from modeling the thermal emission from
an optically thick, geometrically thin accretion disk for an array of black hole spin values.
The modeling assumes the standard general relativistic Novikov-Thorne emissivity profile, and 
traces photons emanating from the accretion disk forward in time \citep[see][for a detailed description of the 
general relativistic ray tracing code]{2012ApJ...744...30K}. The initial polarization and the polarization change 
associated with photon scattering off the accretion disk are modeled with the classical Chandrasekhar equations
giving the Stokes parameters for the emission and reflection by a optically thick atmosphere \citep{1960ratr.book.....C}.     
The fitting templates are generated by folding the simulated Stokes parameter energy spectra with
the detector response. After simulating one observed \ps data set, a least squares fit is performed 
to determine the best-fit black hole parameters and the associated errors.
The least squares analysis (see the right panel of Fig.\ \ref{F:spin}) recovers the input dimension-less spin parameter 
and inclination and position angle of the inner accretion disk with 1$\sigma$ 
(combined statistical and systematic) accuracies of 0.02 (spin) and $\sim$2-3$^{\circ}$ (inclination and position angle). 
The measurement errors for such a bright source are dominated by the systematic uncertainties (if we neglect 
astrophysical uncertainties);  achieving measurements at this accuracy drive the systematic error 
requirements for \psc. Spin measurements based on modeling the thermal X-ray continuum assume that 
the inner accretion disk is aligned with the binary orbit and that the disk emission from within the
innermost stable circular orbit is negligible \citep[e.g.\ see][and references therein]{2011ApJ...742...85G,2014ApJ...790...29G}.
Measurements based 
on modeling the iron  line and reflection component rely on a number of 
assumptions regarding the geometry of the corona, the illumination profile, 
and the thermo-dynamic state of the reflecting disk material \citep[e.g.\ see][and references therein]{2014ApJ...780...78T}. 
\ps can provide independent checks of these assumptions. 

The orientation of the inner accretion disk can be compared to the orientation of the binary 
orbit and jet, if present (Figure~\ref{F:jet}). 
A misalignment between the inner accretion disk and the orbital plane would provide strong evidence 
for the general relativistic Bardeen-Peterson effect, while a non-zero angle between the spin axis of the 
inner disk and the jet would suggest mechanisms to bend the jet away from its original direction.
  
\subsubsection*{Testing General Relativity's Prediction of Lense-Thirring Precession}

Low frequency quasi-periodic oscillations (hereafter QPOs) are regularly observed in the X-ray 
light curves of accreting black holes \citep[e.g.][and references therein]{2006ARA&A..44...49R}. 
The QPO and power spectral break frequencies are correlated in black hole X-ray binaries \citep{1999ApJ...514..939W}, 
and the QPO amplitude depends on system inclination \citep{2015MNRAS.447.2059M,2015MNRAS.448.3348H}. 
Both of these properties suggest that the QPOs are geometric in origin independent of any specific model. Broadband X-ray polarization probes geometry, and so can test the nature of black hole QPOs.

Perhaps the most successful model to explain QPOs considers precession of the inner accretion flow 
due to the relativistic effect of frame dragging. A spinning black hole twists up the surrounding space-time, 
inducing Lense-Thirring precession in test mass orbits. The model of \citet{2009MNRAS.397L.101I}
assumes that the entire inner accretion flow ($r<~20r_{\rm g}$) precesses as a solid body, motivated 
by the simulations of \citet{2007ApJ...668..417F}. If this is indeed the true QPO mechanism, the X-ray polarization 
signature from black holes should also contain a QPO. 
\psg broadband sensitivity is ideally suited for this kind of study. The precession leads to quasi-periodic large-amplitude changes of the polarization of the Comptonized 10-20 keV emission. Since the thin accretion disk is not expected to precess and dominates the soft X-ray emission, low-energy ($<$10 keV) observations are poorly poised to address this question, while \psg broadband sensitivity can establish the physical basis of the QPO phenomenon 
as predicted by the calculations of \citet{2015arXiv150500015I}.

\begin{figure}[tbh]
\begin{center}
\includegraphics[width=3in]{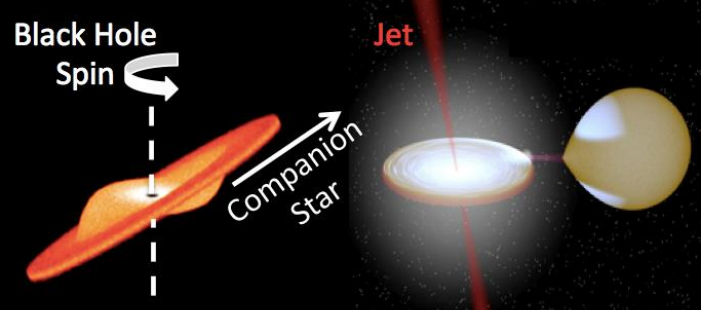}
\caption{\label{F:jet} \ps can use polarization to measure the orientation of the inner accretion disk 
of stellar mass black holes with an accuracy of a few degrees. {\bf Left:} \ps can measure the amount of misalignment between the inner disk and the binary plane owing to the general relativistic Bardeen-Peterson effect \citep[warped accretion disk image from][]{2010MNRAS.405.1212L}. {\bf Right:} \ps is also 
sensitive to misalignments between the disk spin axis and the jet axis. 
(image from Rob Hynes, www.phys.lsu.edu$/\!\sim$rih$/$).
}
\vspace*{-0.5cm}
\end{center}
\end{figure}

GRS 1915+105 displays very strong QPOs with periods ranging from $\sim$10-0.1 s \citep[e.g.][and references therein]{2015AJ....149...82Z}. The brightness of the source, and the amplitude of the QPOs peak when the QPO period is $t_{\rm qpo}\sim1$~s. 
However, accurate time resolved polarization measurements require fairly large time bins, and therefore this favors longer QPO periods. We choose an example of $t_{\rm qpo}\,=\,2$~s as these considerations balance well for this period. 
\ps measures a count rate of $\sim$70 c$s^{-1}$ from GRS 1915+105. 
We assume the modulations in flux, polarization degree and polarization angle calculated 
by \citet{2015arXiv150500015I} for an inclination angle of $i\,=\,70^{\circ}$, which is appropriate for GRS 1915+105.

\begin{figure}[tb]
\begin{center}
\vspace*{-4.5cm}
\includegraphics[width=3.5in]{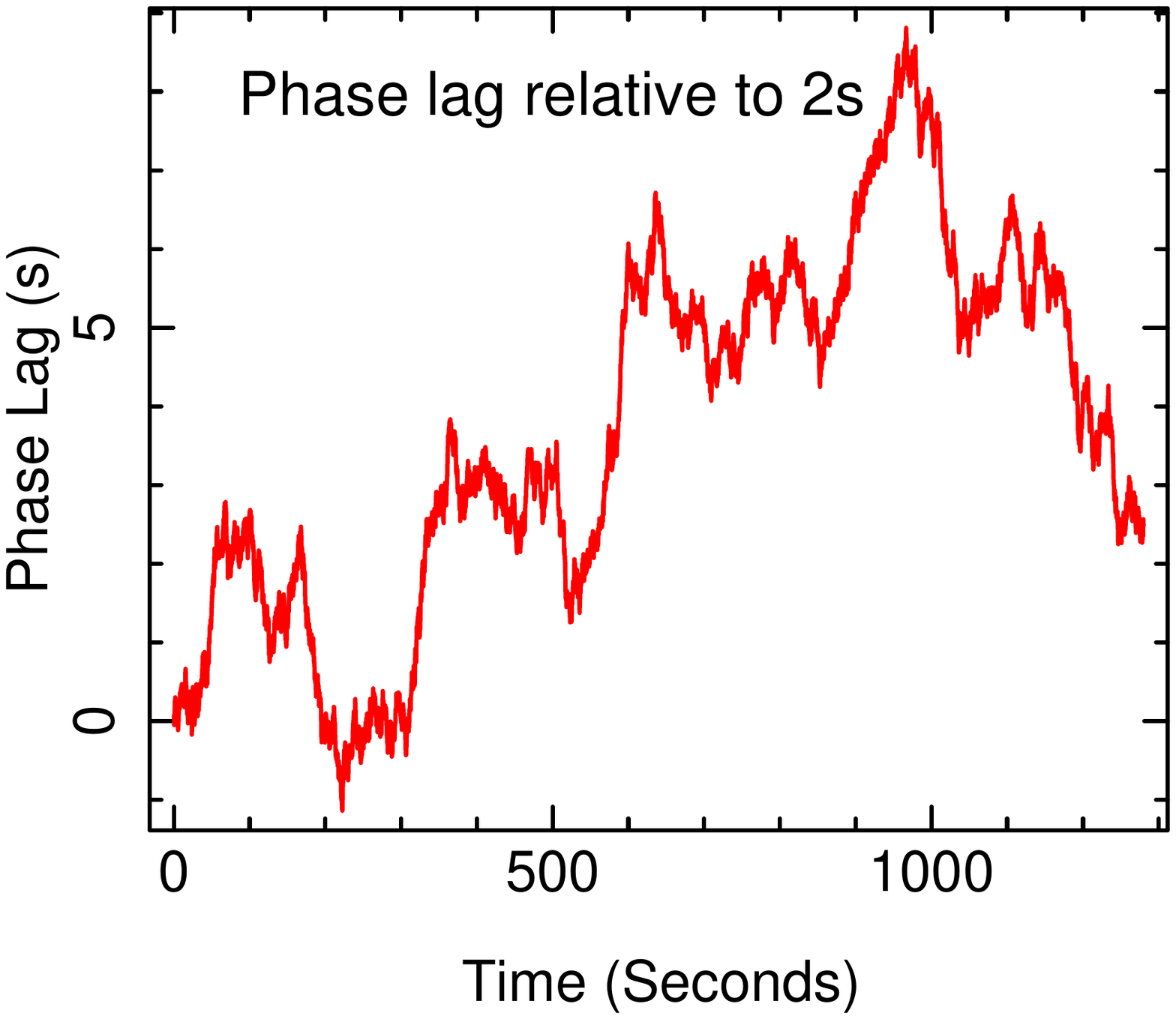}
\includegraphics[width=3.0in]{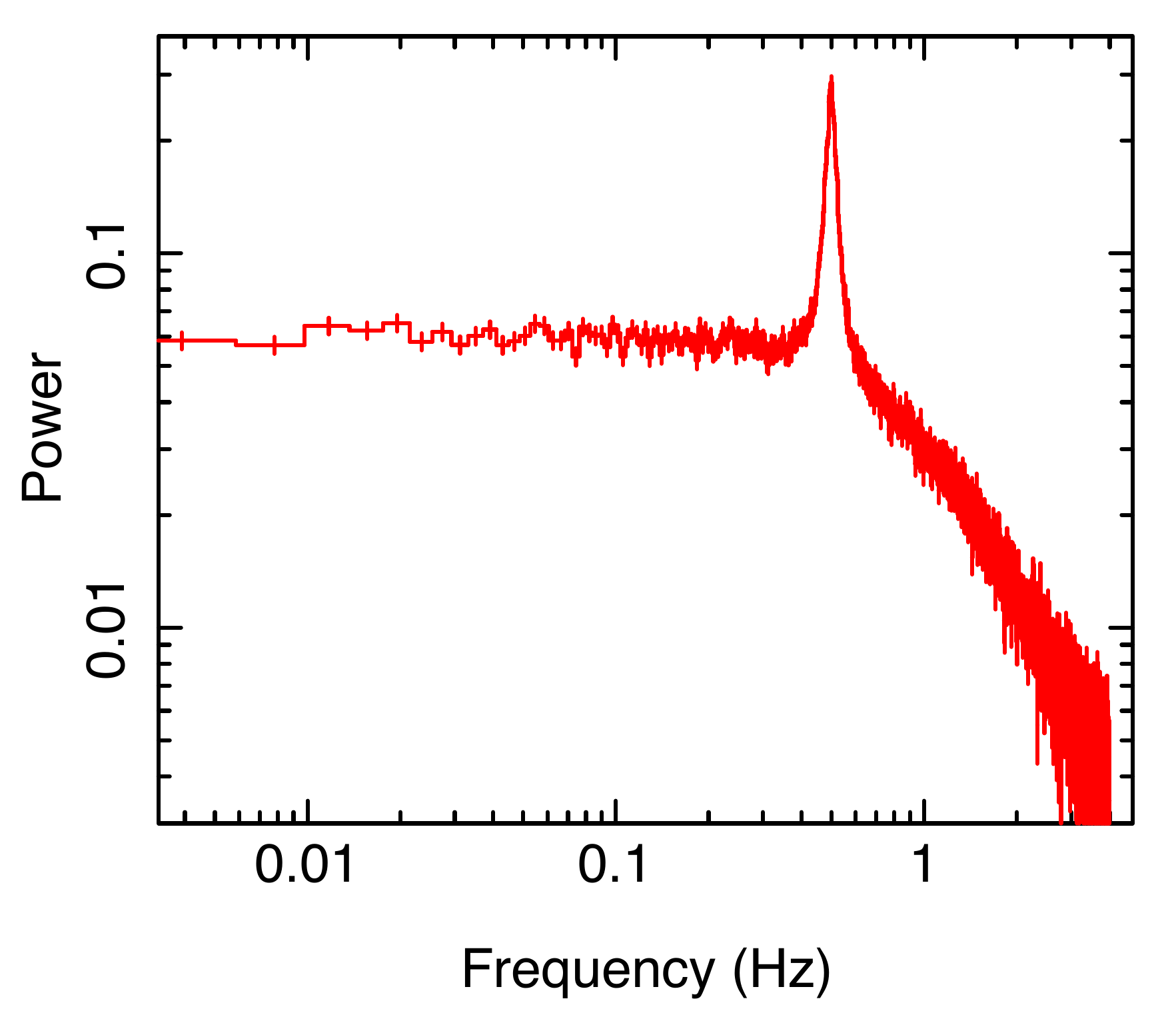}
\caption{\label{F:qpo1}
Top: Phase of a simulated QPO relative to a perfectly periodic function. It is this phase drift that makes QPOs quasi-periodic.
Bottom: Power spectrum of the simulated light curve. The QPO is seen as a peak at 0.5 Hz. When fit with a Lorentzian function, the QPO is measured to have a quality factor of Q=12.5, consistent with what is observed for GRS 1915+105.}
\vspace*{-0.8cm}
\end{center}
\end{figure}

\begin{figure}[tb]
\begin{center}
\vspace*{-1cm}
\includegraphics[width=3.5in]{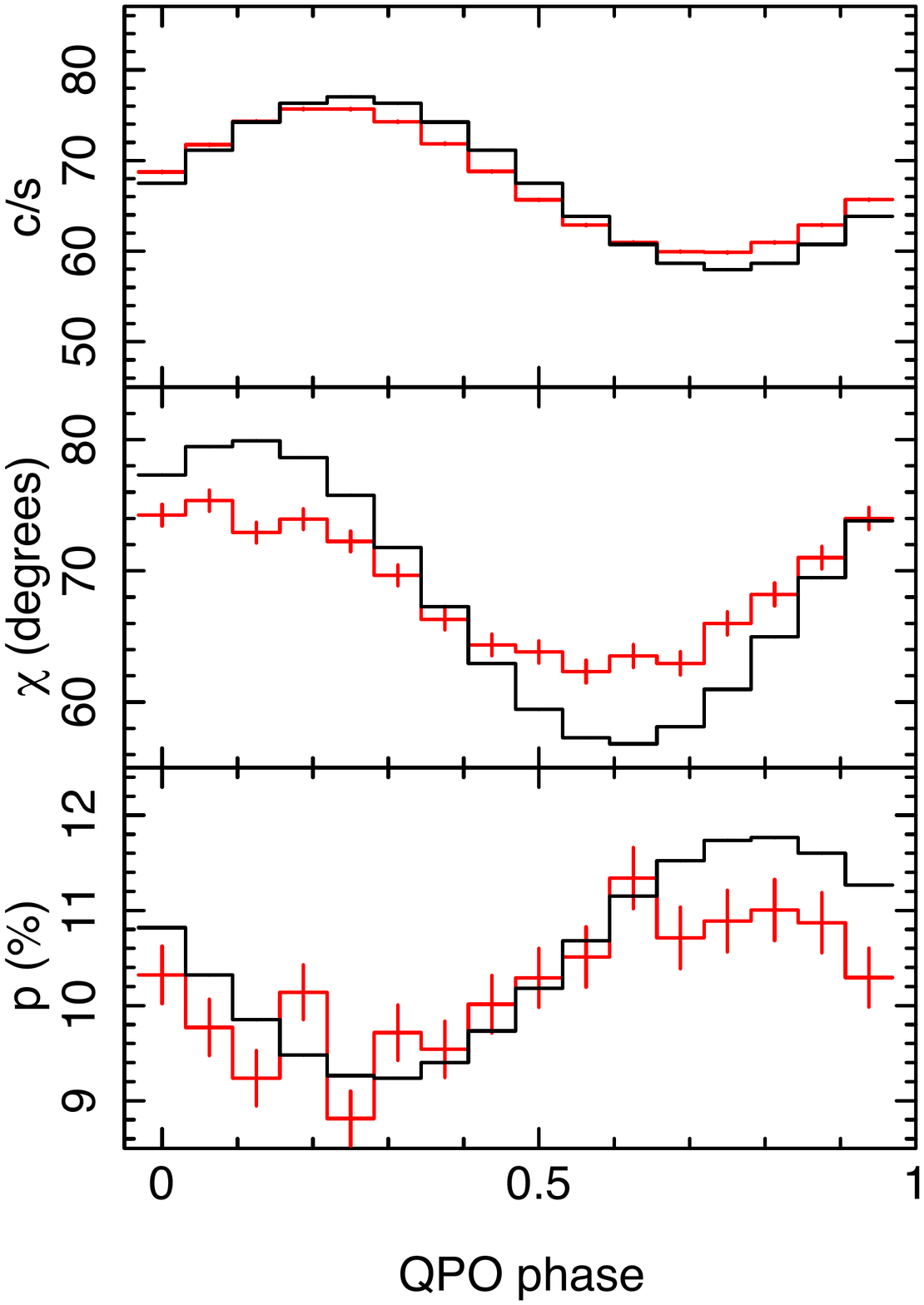}
\vspace*{-1cm}
\caption{\label{F:qpo3}
Count rate, polarization angle and polarization degree plotted against QPO phase. Red points show the values based
on reconstructing the simulated \ps data, and the black lines show the input values. 
Constant polarization degree and angle can be rejected with a high statistical confidence.}
\vspace*{-0.8cm}
\end{center}
\end{figure}

We simulate the QPOs taking into account that they are quasi-periodic as opposed to periodic and are observed coincident with broadband noise, which is intrinsic to the source rather than instrumental. 
We simulate the phase of the QPO to drift on a random walk away from that of a strictly periodic sine wave (Figure~\ref{F:qpo1}, top panel), as is observed for QPOs in GRS 1915+105 \citep{1997ApJ...482..993M}. 
We also simulate noise, which has a broad Lorentzian power spectrum but exhibits the statistical correlations observed in the light curves of accreting objects \citep[the so-called linear rms-flux relation:][]{2001MNRAS.323L..26U}. 
To do this we use the {\it exponentiation method} of \citet{2005MNRAS.359..345U}. 
We multiply the QPO and broad band noise light curves together 
\citep[again to mimic statistical properties observed in the data:][]{2013MNRAS.434.1476I} to obtain the expected number of photon counts per time bin. Finally, we simulate \ps detecting an integer number of photons per time bin by choosing a Poisson random variable from the calculated expectation value for every time bin. For each photon in each time bin, we then simulate where that photon lands on the \ps detector based on the true polarization vector at that time and the modulation factor of \psc, $\mu\,=\,0.5$.

\begin{figure}[tb]
\begin{center}
\includegraphics[width=3.5in]{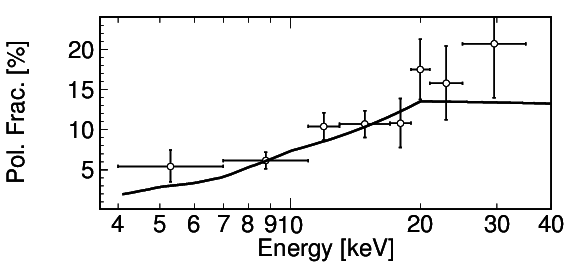}
\caption{\label{F:agn} \ps is sensitive at high energies (5-40 keV), where the reflected coronal emission is highly polarized. This plot shows that advantage for a simulated 544 ksec \ps observation of MCG-5-23-16 \citep[model from:][assuming a lamppost corona]{2011ApJ...731...75D}.
}
\vspace*{-0.8cm}
\end{center}
\end{figure}
The lower panel of Fig.~\ref{F:qpo1} shows the power spectrum of the 205~ks simulated light curve. We see a QPO and broadband noise, consistent with real observations. Measurement of Stokes parameters in 0.125~s time bins are too noisy to discover a QPO in the polarization signature by simply taking a power spectrum of a Stokes parameters time series. 
We use the phase-folding method of \citet{2001ApJ...548..401T}: we filter timescales much longer and shorter than the QPO period out of the flux time series and identify peaks in the filtered time series as QPO peaks. We then assign a QPO phase value to every time bin based on the time since the last peak relative to the time until the next peak. We then stack into 16 QPO phase bins. Figure~\ref{F:qpo3} shows the mean count rate (top), polarization angle (middle) and polarization degree (bottom) in each phase bin. Red points are recovered from the data and black lines are the input models. We see that the modulations in both polarization degree and angle are recovered and constant polarization properties can be ruled out with a high statistical significance.
\ps can thus detect the precession of the inner accretion flow. The \ps detection would constrain accretion 
disk and corona models and would confirm a strong-field prediction of general relativity \citep[e.g.][]{2005AN....326..782A}.

\subsubsection*{Numerical Modeling of the Data}

The studies of black hole accretion disks, corona, and jets would benefit and be in dialog with the rapidly progressing field
of numerical simulations of accreting black hole systems. General relativistic magnetohydrodynamic
simulations can provide a self-consistent physical model for accretion flows and 
jets \citep[e.g.,][]{2003ApJ...589..458D,2003ApJ...589..444G} and can be used to derive testable predictions. 
Simulations predict that ordered magnetic fields with large fluxes (area time magnetic field strength) lead to 
more powerful jets \citep{2011MNRAS.418L..79T}, and that black holes with misaligned spin and accretion disk axes
may lead to twisted jets \citep{2013Sci...339...49M}. Both of these predictions can be tested with \ps data.

Recent improvements in radiation physics \citep{2015arXiv150300654S,2015arXiv150804980S,2014MNRAS.441.3177M,2014ApJ...796..106J} and thermodynamics \citep{2013ApJ...769..156S,2015MNRAS.451.1661Z} at high luminosities, and radiative cooling \citep{2015ApJ...807...31R} and plasma physics \citep{2015arXiv150800878C} at low luminosities can soon be combined with radiative transfer codes \citep[e.g.,][]{2013ApJ...777...11S,2012ApJ...755..133S}. 
Such general relativistic radiation magnetohydrodynamic codes make it possible to 
derive energy spectra, light curves, and polarization for the vast majority of the observed accreting black holes, 
including the \ps targets. Detailed comparisons of simulated and observational (\ps and other) data test the 
numerical models, and can be used to derive more  robust constrains on parameters such as the mass accretion rate, 
the magnetic field strength and configuration, the black hole spin, and the alignment between 
the accreting material and the black hole spin axes. 
\subsection{Dissect the Black Hole Accretion Flows onto Supermassive Mass Black Holes}
\label{SS:supermassiveBHs}
Accretion is key in understanding how black holes grow and influence the galaxies in which they reside.
The energy released from accretion on to a supermassive black hole is 100 times greater than the gravitational 
binding energy of its host galaxy, and yet the mass of the black hole is 1000 times less than that of the galaxy's bulge.
Most of the energy released in accretion is concentrated within a few tens of gravitational radii from the central black hole.  
This corresponds to microparsec scales, which are far smaller than the angular resolution power of current or future telescopes.  Therefore, we must rely on other properties of the emission in order to understand the geometry and kinematics of the complex regions around supermassive black holes.

Traditional spectral analysis has revealed two clear components that help us characterize the inner accretion flow: the broad $\sim$5-8 keV \ika emission line and the strong Compton hump component above 10~keV.   These two spectral features are produced through the irradiation of the inner accretion disc by the X-ray corona, and are broadened due to relativistic effects in the strong potential well of the central black hole.  \nsc, with its large effective area and broad energy coverage, has been ideal for measuring these two important spectral components, which has allowed for the most precise measurements of black hole spin to date \citep[e.g.][]{2013Natur.494..449R,walton14,marinucci14,parker14}.  Furthermore, \ns\ measurements of the reverberation time delays associated with the broad \ika line and Compton hump indicate a small light travel distance between the continuum emitting corona and the inner accretion disc \citep{zoghbi14,kara15}.  

While reverberation mapping has given us a new way to constrain the accretion flow, there is still a fundamental degeneracy between the height of the corona above the accretion disc and the mass of the black hole \citep{cackett14}.  \ps could break
that degeneracy. The polarization fraction and angle associated with the broad \ika line and Compton hump are highly dependent on the height of the corona and are mass-invariant.  The Compton hump is particularly vital in helping break this degeneracy because the polarization fraction increases from $<$1\% at 3 keV to 15\% at $>$20 keV (Figure~\ref{F:agn}, \citealp{2011ApJ...731...75D}).  By combining spectral, timing, and polarization, \ps can constrain coronal parameters such as spatial extent, relative location, optical thickness and clumpiness of the corona \citep[e.g.][]{2010ApJ...712..908S}.

The observation plan includes three of the brightest bare Seyfert galaxies, NGC 4151, MCG 5-23-16, and MCG 6-30-15.  They have been observed with \ns and show strong Compton hump components \citep{keck15,balokovic15,marinucci14b}. 
All three sources are also highly variable making them ideal targets for \psg simultaneous spectral, reverberation and polarization capabilities.  {\em XMM-Newton} observations of NGC~4151 and MCG-5-23-16 already reveal strong \ika reverberation lags, and \ns observations of MCG-5-23-16 reveal the associated lag of the Compton hump \citep{zoghbi12,zoghbi14}.  MCG-6-30-15 is of particular interest. While it is highly variable, the continuum emission does not appear to vary with the broad \ika emission, and therefore reverberation is not detected \citep{vaughan03, papadakis05, kara14}.  \citet{miniutti03} suggests that this behaviour is due to strong relativistic light bending effects from a corona within a few gravitational radii of the black hole. \ps can test this conjecture since the scenario predicts large polarization degrees for the reflected emission.  The observed data can also be compared to simulated data from 
general relativistic radiation magnetohydrodynamic (GRRMHD) simulations \citep{2013MNRAS.429.3533S,2014MNRAS.441.3177M}.
\subsection{Reveal the Magnetic Back-bone of Blazar Jets}
\label{SS:blazars}
\begin{figure}[tb]
\begin{center}
\includegraphics[width=3.5in]{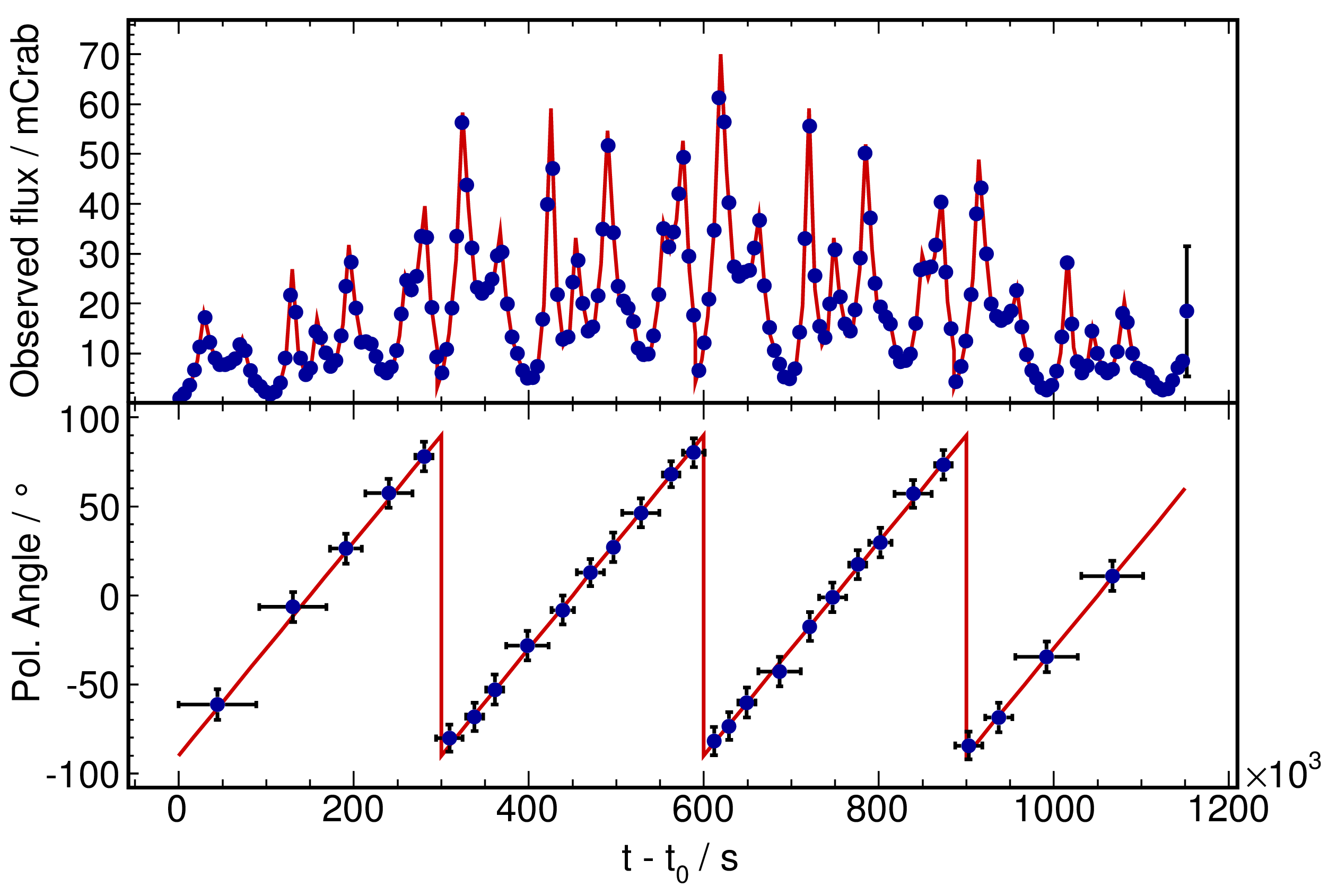}
\caption{\label{F:mrk421} \ps observations of blazars can reveal the helical magnetic field by detecting 
swings of the polarization angle. This simulation of \ps observations of a Mrk 421 flaring epoch assumes 
a polarization fraction of 6\% \citep[similar to what is measured at optical wavelengths, e.g.,][]{1998A&A...339...41T,2013A&A...559A..75B}.
}
\vspace*{-0.8cm}
\end{center}
\end{figure}
Active galactic nuclei (AGN) can launch extremely powerful and highly relativistic out-flows (jets). These jets appear to play an important 
role in galaxies and galaxy clusters as they can heat the interstellar and intracluster medium 
and thus impact the rate 
at which gas cools to form stars and feed the central black hole \citep[e.g.][and references therein]{2012ARA&A..50..455F}. 
Although jets have been studied intensively, their 
governing physics is still largely unknown \citep{2012rjag.book.....B}. 
Over the last decade, the magnetic model of jet formation has emerged as 
the standard paradigm to explain the observed jet characteristics and relativistic velocities \citep[][and references 
therein]{2011AIPC.1381..227S}. 
The model invokes a helical magnetic field threading the jet. Magnetic stresses along 
the field lines accelerate the jet material to velocities close to the speed of light and help collimate the jets.
General relativistic magnetohydrodynamic simulations seem to confirm 
that magnetically-dominated jets can form from the accretion process and propagate to large distances 
\citep{2005ApJ...620..878D,2006MNRAS.368.1561M,2009MNRAS.394L.126M}.  
\ps observations of blazars can test this model. In high synchrotron-peaked blazars (HSPs), the X-ray emission is 
produced by synchrotron emission and is polarized perpendicular to the projection of the magnetic field lines onto 
the plane of the sky. \psg observation plan includes the two bright HSPs, Mrk 421 and Mrk 501. 

Optical blazar polarimetry occasionally reveals swings of the polarization angle correlated with multiwavelength 
flaring activity \citep[e.g.][]{2008Natur.452..966M,2010Natur.463..919A}. 
Models predict that X-ray polarimetric observations show such swings more consistently than 
optical observations since the X-ray emitting regions are more compact (yet still optically thin) because high-energy 
electrons lose their energy on shorter time scales. Figure \ref{F:mrk421} illustrates simulated \ps observations 
of a Mrk 421 flaring epoch, assuming a polarization fraction of 6\% \citep[similar to what is measured at optical 
wavelengths;][]{1998A&A...339...41T,2013A&A...559A..75B}. It clearly shows that \ps has sufficient sensitivity to 
detect such polarization angle swings (Figure \ref{F:mrk421}), which would provide clear evidence for a helical 
magnetic field topology \citep[][]{2014ApJ...789...66Z,2015ApJ...804...58Z}. 

Even if \ps does not detect ubiquitous polarization swings, it can still constrain magnetic fields inside 
jets \citep{2013arXiv1303.7158K}. If flares are associated with the shock-compression of magnetic fields, a 
correlation of X-ray flux and polarization fraction is predicted. Relatedly, correlations between the X-ray 
spectral index and polarization fraction constrain the magnetic field homogeneity in the emitting region. 
Finally, multiwavelength studies, comparing simultaneous optical and X-ray polarization measurements, provide 
information on the co-spatiality of the optical and X-ray emitting regions. This is an essential ingredient 
for modeling blazar physics.

The \ps baseline program includes two flat-spectrum radio-loud quasars (FSRQs), 3C~273 and PKS~1510-08. 
These observations can help solve a second long-standing question in the blazar community by distinguishing 
between the two main flavors of radiation models to explain the X-ray to gamma-ray emission: leptonic models 
(including both synchrotron self-Compton and external Compton models) vs. hadronic models. As hadronic models 
predict higher polarization fractions (up to $\sim 70$\% in the case of a perfectly ordered magnetic field in
the high-energy emission region) than leptonic models (less than $\sim 30$\%), a measurement of 
very high X-ray polarization fractions would vindicate the hadronic model 
\citep{2012ApJ...744...30K,2013ApJ...774...18Z}. This in turn would imply that blazars can accelerate 
ions to $>10^{19}$~eV, comparable to the energies of ultra-high energy cosmic rays.
\subsection{Study of Extremely Magnetized Neutron Stars}
\label{SS:magnetars}
The observation plan includes two AXPs and one SGR. Both types of sources are explained by
the \emph{magnetar} hypothesis as neutron stars with extremely strong ($10^{14}-10^{15}$~G) 
magnetic fields \citep{1992ApJ...392L...9D}. The bright flares from this class of objects are thought to be 
magnetically powered events in which the field stresses deform or break
the stellar crust, releasing a large amount of energy which leads to
a reconfiguration of the magnetosphere \citep{thompson1995}.
The magnetic fields are so strong that they lead to a unique
phenomenology with telltale observational signatures, both for flare emission, 
and also for the steady, persistent signal that \ps can focus on through pointed observations. 

\begin{figure}[tb]
\begin{center}
\includegraphics*[width=3.5in]{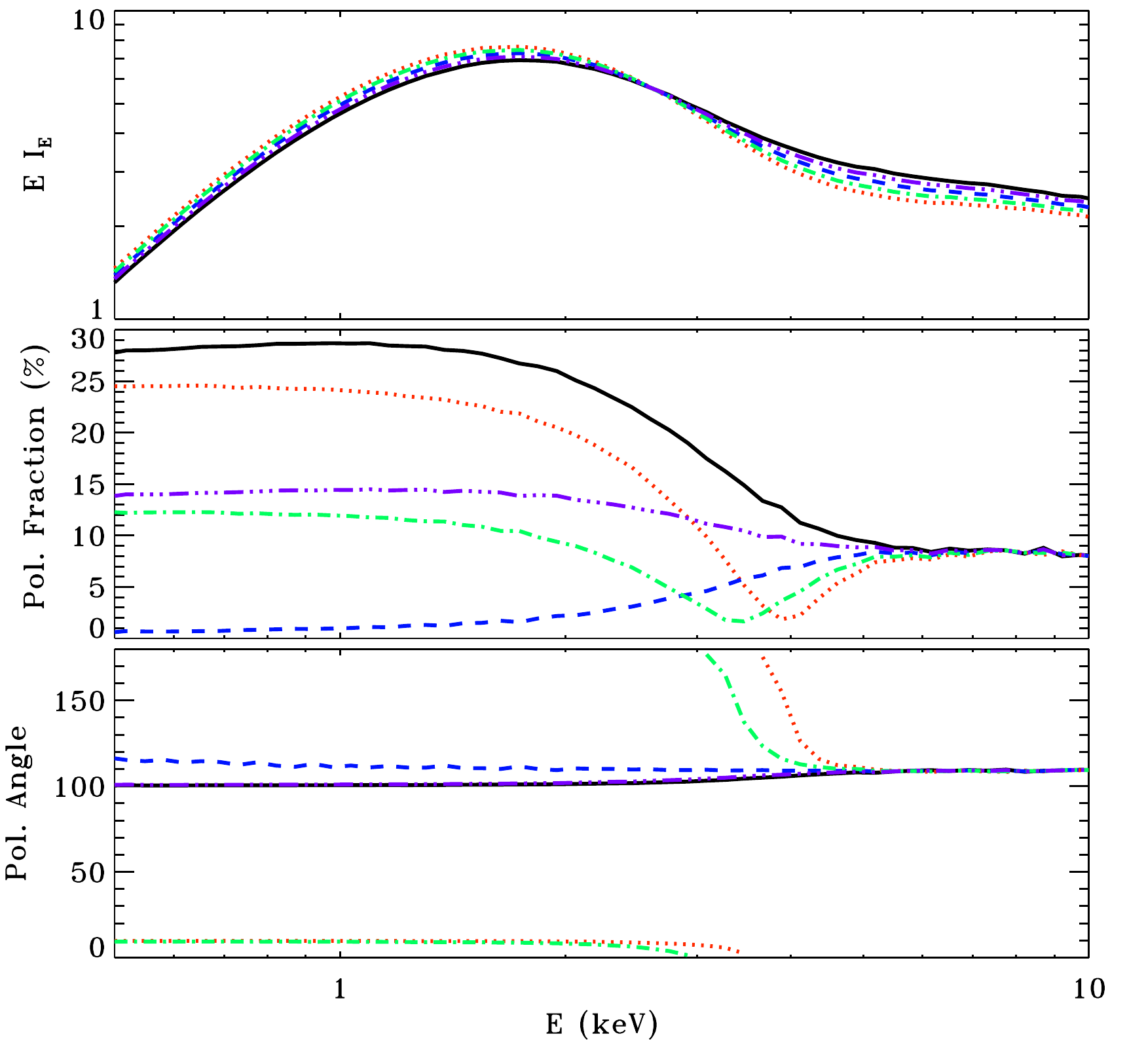}
\caption{Intensity (top), linear polarization fraction (middle), and polarization
angle (bottom) as a function of energy for phase-averaged observations. 
The magnetar model predicts that the soft X-ray emission is a combination of highly polarized
thermal photons at energies $<4$~keV, with resonant Compton upscattering providing
a non-thermal tail at energies $>4$~keV. The different lines shows predictions for different 
polarization properties of the thermal seed photons from the 
magnetar surface \citep[from][\textcopyright AAS, reproduced with permission]{2011ApJ...730..131F}.}
\label{F:magnetar_phase_averaged}
\end{center}
\end{figure}
The quiescent thermal ($<$4 keV) X-ray emission from magnetars is predicted to 
be nearly 100\% polarized since the extremely magnetized plasma near
the neutron star surface is birefringent, with the lowest opacity
mode carrying most of the radiation (e.g., \citealt{2001ApJ...563..276O,lai2003}). 
The non-thermal, low-energy X-rays ($\sim$4-10 keV) are thought to be produced
by resonant cyclotron/Compton upscattering of thermal photons in the 
inner magnetosphere, a process enabled by extremely strong magnetic fields. The scattering process
is expected to impart a moderate polarization amplitude ($\sim10-30\%$) to
the outgoing photons, provided the observer samples viewing perspectives at 
significant angles to the local field direction. Phase-averaged observations by \ps can test this
basic emission mechanism since they will provide a polarization signal averaged 
over a variety of instantaneous magnetic viewing angles (Figure~\ref{F:magnetar_phase_averaged}).

\begin{figure}[tb]
\begin{center}
\includegraphics[width=3.5in]{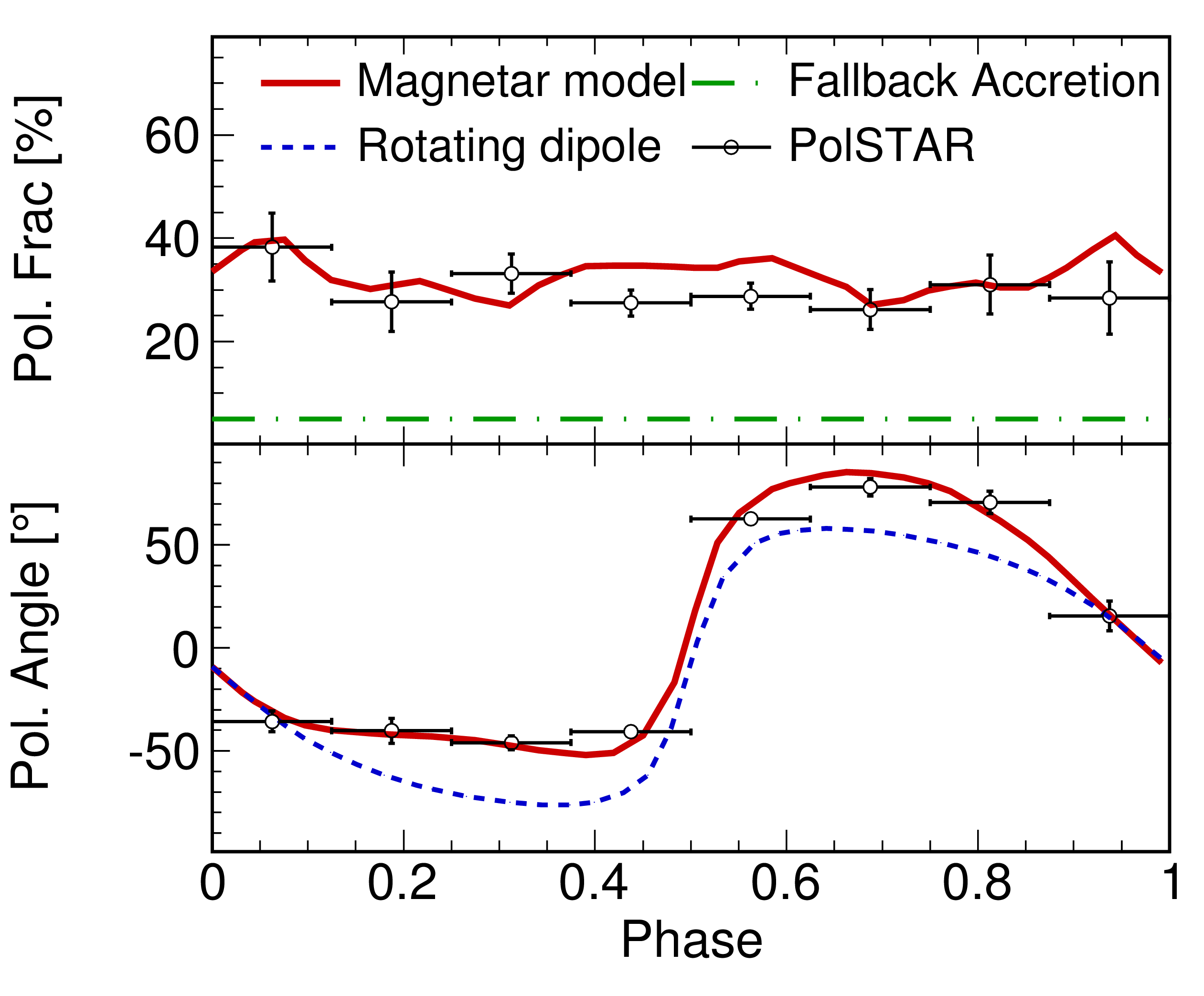}
\caption{\label{F:magnetar} \ps observations of the AXP 1E 2259+586 can
distinguish the magnetar model (top panel, red line) from the fallback
accretion model (top panel, green line). Both panels use the 3-15 keV data.
The lower panel compares the modeled polarization direction (red line) and \ps
measurements (black data points) to the time-reversed model (blue dashed line).
The detection of such a time reversal asymmetry implies the presence of a
toroidal magnetic field component in the magnetosphere. The magnetar model predictions
are from Fern\'andez and Davis (2011).  No detailed predictions for the fallback model are
available, but modest ($\lesssim 10\%$) values are expected if polarization arises from
Compton scattering as is assumed in the fallback model \citep{2010A&A...518A..46T}.}
\end{center}
\end{figure}
\psg detection of the extremely polarized ($\sim$30-80\%)
phase-resolved emission predicted by the magnetar model would distinguish between
alternatives such as the \emph{fallback accretion} model, which attributes the non-thermal emission
to a combination of thermal and bulk Compton scattering in the accretion flow \citep{2010A&A...518A..46T,2013ApJ...764...49T},
leading to lower ($<10\%$) polarization fractions  (Figure~\ref{F:magnetar}).
This low expectation is an estimate based on non-magnetic Comptonization scenarios 
in laminar geometries, for which computed polarization degrees typically in this range were 
obtained by \citet{1985A&A...143..374S} in the context of black hole accretion disks.

The phase-resolved 3-15 keV polarization angle measurements 
can distinguish between a pure dipole field geometry and the twisted magnetic 
field at the heart of the magnetar model \citep{2011ApJ...730..131F}. 
This is because each phase corresponds to a particular viewing orientation relative to the mean 
magnetic field direction sampled in the emission region, for either surface/atmospheric signals or radiation 
generated in the inner magnetosphere. \ps can contrast
polarization signatures (fraction and direction) between quiescent and
post-outburst states, and search for evidence of a magnetic reconfiguration
initiated by the outburst activity. At even higher ($>$10 keV) energies, \ps
can detect the flat hard X-ray tails of magnetars \citep{2004ApJ...613.1173K,2006A&A...449L..31G,2008A&A...489..245D}
believed to be polarized to $>$50\% owing to resonant inverse Compton/cyclotron
scattering \citep{2007Ap&SS.308..109B}. In such upscattering models, the polarization 
degree increases at higher photon energies due to an increased contribution from 
photons backscattered in an electron's rest frame.

The \ps observations of magnetars and accreting pulsars can probe the dielectric properties 
of the strongly magnetized quantum vacuum to test predictions of the theory of 
Quantum Electrodynamics (QED) in extremely strong magnetic fields not accessible in terrestrial laboratories.
At field strengths above the quantum
critical field, 4.41$\times$10$^{13}$~G, the vacuum itself is polarized by QED
coupling to virtual $e^+e^-$ pairs, thereby rendering the magnetosphere
birefringent, i.e., the refractive index differs for the two photon
polarizations \citep[e.g.\ see][]{2006RPPh...69.2631H,2011ApJ...730..131F}.
X-ray photons of different polarizations therefore travel at slightly different speeds, 
and their electric field vectors can re-orient during propagation through the 
magnetosphere, rotating about the local magnetic field direction.
This effect is more pronounced in magnetars than for pulsars of lower magnetization.

\begin{figure}[tb]
\begin{center}
\includegraphics[width=3.5in]{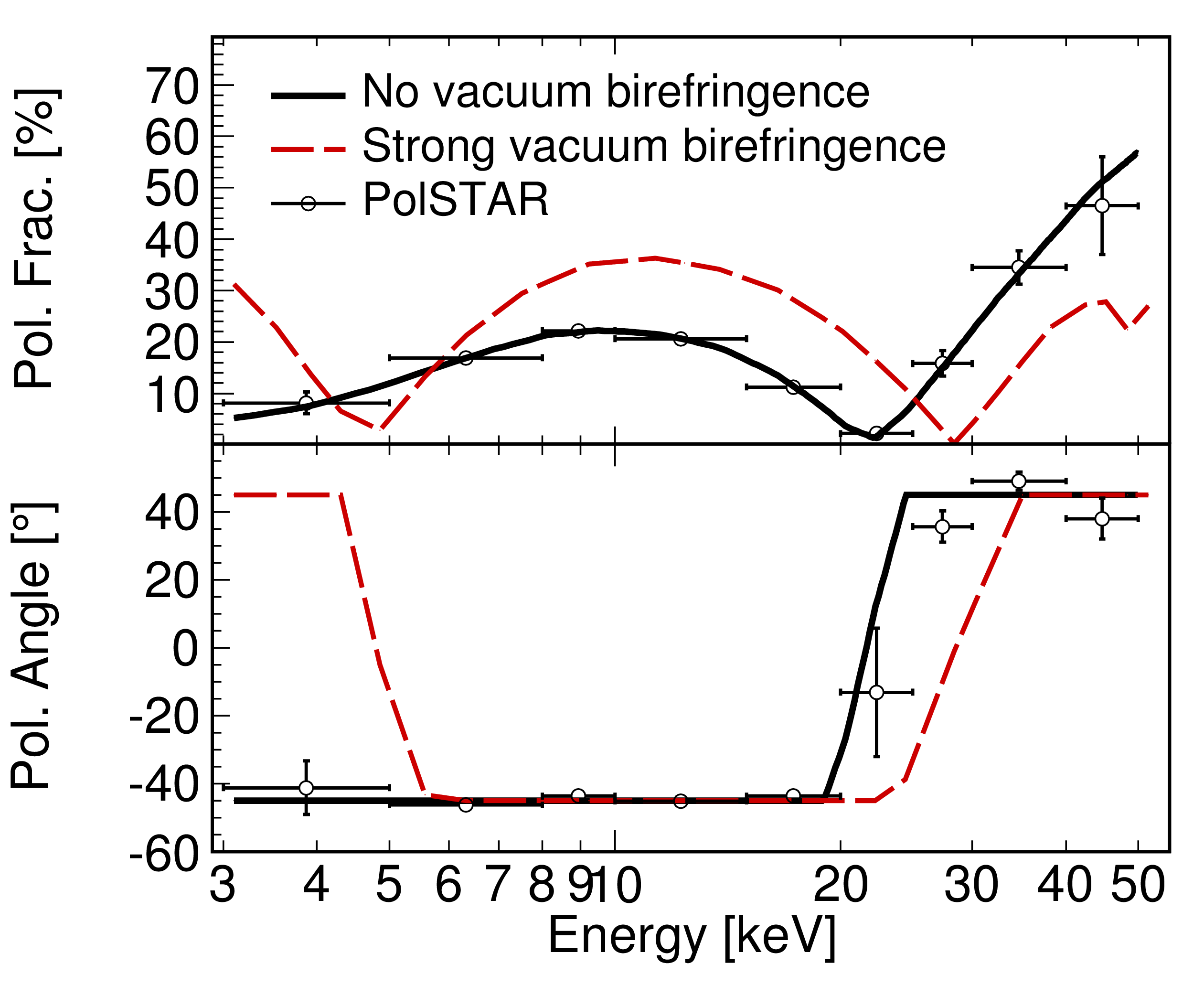}
\caption{\label{F:herx1} \ps is sensitive to the distinct polarization signatures of QED vacuum birefringence. This simulated 22 ksec \ps observation of Her X-1 in its high-state distinguishes between the model with (dashed red line) and without (solid black line) vacuum birefringence. The models predictions are from Kii (1987).
}
\end{center}
\end{figure}

While magnetars serve as the best type of neutron stars with which to
study magnetospheric propagation influences of vacuum birefringence,
accreting pulsars with much lower magnetic field strengths of around
$10^{12}$~G present another opportunity to test fundamental QED predictions
of dispersion of the magnetized vacuum.  This is in reference to the
so-called {\it vacuum resonance} feature, discussed in detail in the
review by \citet{2006RPPh...69.2631H}. This corresponds to the critical photon
frequency $\omega$ where the vacuum modifications to the refractive index
$n$ (which depend only on the magnetic field $B$) approximately equal the 
dispersion correction imposed by a plasma, which scales as $\omega_p^2/\omega^2$, 
where $\omega_p$ is the familiar plasma frequency.  The vacuum resonance frequency
therefore depends on the plasma density and the magnetic field strength.
Accretion columns impacting neutron stars provide density and $B$ values
for which the feature naturally appears above a few keV.  
In contrast, magnetars will elicit such a feature in their atmospheres at energies
below 1~keV, a band that generally leads to obscuration of its
signatures. 

As the photon frequency rises through the resonance from
plasma to vacuum dispersion domains, the character of the photon
propagation eigenstates changes profoundly, leading to distinctive
swings in polarization degree and angle \citep{1987PASJ...39..781K,1992hrfm.book.....M}
when integrated over emission volumes. \ps is well-suited to conduct
a search for QED signatures in the 5-50 keV polarization spectra of
accreting pulsars at the energies of the vacuum resonance and cyclotron
lines where these effects are strongest (see Figure~\ref{F:herx1} for examples).
Observational demonstration of the existence of the vacuum resonance
feature through X-ray polarimetry would be a huge advance for the
fundamental theory of QED in strong field domains.
\subsection{Observations of Accretion Powered Neutron Stars and Pulsars}
The nature of the accretion flow onto neutron stars with lower surface
magnetic fields (e.g. $B \leq 10^{9}$~Gauss) remains uncertain,
despite decades of X-ray timing and spectroscopy.  Persistently
accreting neutron stars trace well-known ``Z'' and ``atoll'' tracks in
X-ray color-color diagrams over time \citep{zatoll}, indicating that
accretion is somehow regulated.  Both strong quasi-periodic
oscillations tied to the inner accretion disk and relativistic radio
jets change in characteristic ways along these tracks \citep{migliari}.
It is likely the case that the Z/atoll tracks, QPOs, and jet
production are all affected or even driven by the interaction of the
neutron star magnetic field with accreting matter, but strong evidence
of this has remained elusive.  By sensitively searching for changes in
polarization fraction and angle along color-color tracks, \ps 
brings an entirely new diagnostic tool to bear on neutron star accretion.
Observations of bright sources such as Scorpius X-1 and Cygnus X-2,
among others, can achieve this science within reasonable observing times.

\ps observations of accretion-powered pulsars can demonstrate the
cyclotron nature of the hard X-ray absorption lines seen in neutron stars based
on energy-dependent polarization measurements across the lines. 
\psg energy resolution should be sufficiently good as the cyclotron lines
exhibit typical widths of 5-10 keV. The results refine neutron star magnetic field 
strength measurements and probe the geometry of the accretion shock (and the particle acceleration region) by
distinguishing between pencil beam and fan beam approximations to the accretion
column geometry \citep{1992hrfm.book.....M}.

\subsection{Observations of the Crab and Vela Pulsars and Pulsar Wind Nebulas}
\label{Crab}
The rotation powered Crab pulsar and nebula remains a prime target of  high-energy astrophysical research. 
The system is a paragon of a high-energy astrophysical source, e.g. 
AGN and GRB models are based on models developed for explaining 
Crab observations. The recent discovery of flares with the {\it AGILE} and {\it Fermi} gamma-ray telescopes
\citep{2011Sci...331..736T,2011Sci...331..739A} may require a re-evaluation of the basic assumptions 
about the physical processes accelerating high-energy emitting particles in these sources  
\citep{2012MNRAS.426.1374C,2012ApJ...746..148C,2012MNRAS.427.1497L}.
As one of the brightest sources in the X-ray sky \citep[flux of $3\times 10^{-8}$ erg cm$^{-2}$ sec$^{-1}$ in the 1-10keV range,][]{1971IAUS...46...42K}, it is the only one for which X-ray polarization has been measured with a high degree of confidence \citep{1978ApJ...220L.117W}. 

\ps would observe the Crab and Vela pulsars and pulsar wind nebulae.  Although \ps does not have imaging capabilities, 
phase-resolved analyses can be used to constrain the polarization properties of the magnetospheric emission.
The phase- and energy-resolved polarization fraction and angle can discern synchrotron and curvature emission 
(because of their contrasting position angle sweeps with pulse phase) and provide excellent diagnostics 
on the locale of the magnetospheric emission 
\citep[see][and references therein]{2004ApJ...606.1125D}.
\ps is not able to image the polarization properties of the nebular emission.   A next-generation imaging X-ray 
polarimeter with one to two orders of magnitude higher throughput than \psc, would be able to access the 
polarization properties of the innermost particle acceleration regions. 
Recent measurements of 60\% optical polarization of the inner knot - \citep{2013MNRAS.433.2564M} significantly 
exceeded theoretical expectations  \citep{2015arXiv150404613R}.
This suggests filamentary field structure as opposed to a more chaotic morphology.  
Discerning differences between optical and X-ray polarization in the nebula 
would yield fascinating insights into the scale-dependence of the field geometry.
\section{Summary}
\label{S:Discussion}
\begin{figure}[tb]
\begin{center}
\hspace*{-1cm}
\includegraphics[width=4.5in]{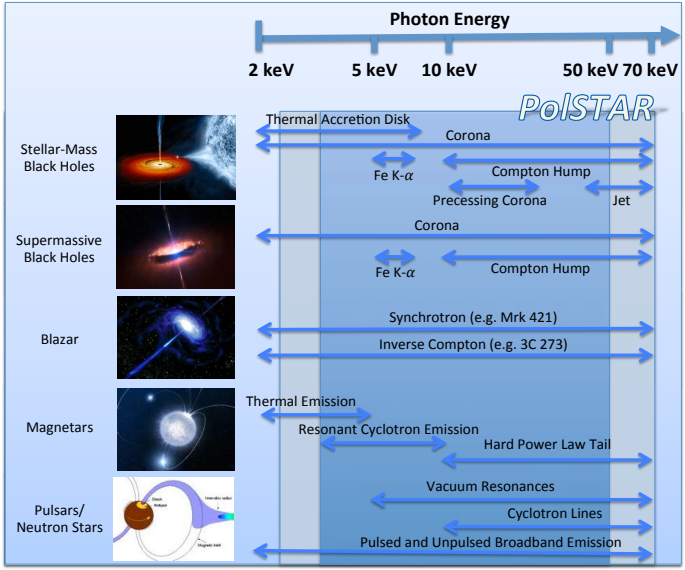}
\caption{\label{F:energy} \ps covers the most important emission components of accreting black holes, blazars, 
and accretion, rotation, and magnetically powered neutron stars. The measurement of the polarization properties
of several emission components allows for powerful tests of the leading models.}
\end{center}
\end{figure}
The \ps X-ray polarimeter offers the capability to measure the linear polarization of X-rays 
over the broad 2.5-70 keV energy range. The approach combines a Si/W multilayer coated X-ray mirror assembly
with a scattering element and CZT solid state detectors, all covering the 2.5-70 keV energy range.
The combination leads to a high O(100\%) efficiency at energies exceeding 10 keV. The modest scattering
efficiency at $<$10 keV energies is offset by the large and approximately constant modulation factor 
of $\mu\approx 0.5$.  For the brightest objects of the baseline observation program (accreting stellar mass black holes, neutron stars, blazars), \ps delivers data sets with very high signal to noise ratios allowing dynamic studies of
such phenomena as QPOs and blazar flares. Even for the dimmest of the objects of the baseline observation program 
(AGNs and magnetars), the polarization measurements can answer important astrophysical questions. 

A scattering polarimeter like \ps offers a broad energy range covering the most important emission components of
black holes, jets, magnetars and neutron stars (Figure~\ref{F:energy}). Simultaneous observations of 
multiple emission components are not only crucial for disentangling the polarization properties of the individual components, 
but also for measuring the relative polarization angles and thus the relative orientation of different emission regions. 
The measurements of the polarization of key accreting black hole components can be used to tighten existing constraints 
on the black hole parameters (including the black hole spin) and to study the corona whose geometry is poorly 
understood and plays an important role in X-ray spectroscopy and timing black hole studies. 
Furthermore, the hard X-ray coverage has the potential to reveal the origin of low-frequency 
QPOs and possibly to detect Lense-Thirring precession in the extreme spacetime of rapidly spinning black holes. 
The observations of blazar jets have the potential to validate the paradigm that helical magnetic fields accelerate and
collimate AGN jets. The observations of magnetars enable the confirmation of the magnetar paradigm and the
detection of polarization from resonant cyclotron scattering. Furthermore, a mission like \ps can scrutinize the 
emission for imprints of the birefringent QED vacuum.  The science described here can best be addressed 
with a broad energy bandpass polarimeter like \psc. As mentioned in the introduction, we are considering
proposing a somewhat larger version of \ps to the next MIDEX announcement of opportunity.
   
At the time of writing this article we are preparing a first one-day long science flight of the 
balloon-borne {\it X-Calibur} experiment planned for Fall 2016.
As mentioned above, {\it X-Calibur} is (like \psc) a scattering polarimeter.  The one-day flight should allow us to observe the Crab, the stellar mass black holes GRS 1915+105 and Cyg X-1, and the accreting neutron star Sco X-1. 
Depending on the flux state of the sources during the observations, {\it X-Calibur} will be able to measure the 
polarization of the 30-60 keV X-ray emission down to 5-10\% polarization fractions. 
The most significant results of the one-day flight will be the polarization properties of the 
power law (presumably corona) emission from GRS 1915+105 and Cyg X-1, and the energy resolved 
polarization properties of the Crab emission. 
The one-day balloon flight will be followed up by a $\sim$28-day long LDB (long duration balloon) 
flight from McMurdo (Ross Island) in December 2018-January 2019.
The longer flight will be used to measure the X-ray polarization of a sample of accreting X-ray pulsars, 
flaring X-ray binaries, and the extragalactic radio galaxy Cen A.  

X-ray polarimetry missions can build on the successes of the past and current X-ray missions 
and are complementary to future X-ray missions. For example, traditional X-ray spectroscopy observations 
have revealed the broad \ika line at $\sim$5-8 keV and the Compton hump above 10~keV in the energy spectra 
of stellar mass and supermassive black holes \citep[e.g.][and references therein]{2014SSRv..183..277R}.  
These features can be well described through a relativistic reflection model in which an irradiated inner accretion disc produces fluorescence and Compton scattering that appears to the observer to be blurred due to relativistic effects close to the central black hole \citep{fabian89,1993MNRAS.261...74R}. The AGN \ika observations constrain black hole spins to 
$\sim$1\% {\em statistical} uncertainty \citep[e.g.][]{2013Natur.494..449R, marinucci14}. Recently, AGN X-ray reverberation mapping 
has indeed confirmed the basic premises of this model \citep{kara13,zoghbi14,2014A&ARv..22...72U}. 
X-ray polarimetry would allow us to further refine our knowledge about the geometry of the inner flow, 
and thus improve on the {\em systematic} uncertainties associated with the spin measurements 
\citep{2004MNRAS.355.1005D,2008MNRAS.391...32D,2009ApJ...691..847L,2009ApJ...701.1175S,2010ApJ...712..908S,2011ApJ...731...75D}.  These measurements are of particular interest in light of ESA's planned {\em Athena} mission 
(planned to be launched in 2028) \citep[e.g.][]{2014SPIE.9144E..5XW}.  
{\em Athena}, with unprecedented effective area (2 m$^{2}$ at 1~keV and 0.25 m$^{2}$ at 6 keV), 
has the ability to put strong constraints on the broad iron line in a large sample of supermassive black holes out to a 
redshifts of 2 or 3.  Measuring the distribution of spin in high-redshift quasars constrains the growth of supermassive 
black holes in the universe (e.g. \citealt{berti08}).  The X-ray polarimetric observations of AGNs 
in our local Universe would directly benefit those studies.
\section*{Acknowledgements}
HK thanks the McDonnell Center for the Space Sciences at Washington University for sponsoring 
the design and fabrication of the X-Calibur prototype polarimeter, and NASA grant \#NNX14AD19G 
for the support of the X-Calibur balloon project. 
MGB acknowledges NASA grants NNX13AQ82G and NNX13AP08G for support.
CD acknowledges STFC support from grant ST/L00075X/1.
JCM acknowledges NASA/NSF/TCAN grant NNX14AB46G, NSF/XSEDE/TACC (TGPHY120005),
and NASA/Pleiades (SMD-14-5451).
AI acknowledges support from the Netherlands Organization for Scientific Research (NWO) Veni Fellowship.
RF acknowledges support from the University of California Office of the President, and from NSF grant AST-1206097.
\section*{References}
\bibliographystyle{elsarticle-harv-abbrv} 
\bibliography{references}

\begin{thebibliography}{181}
\expandafter\ifx\csname natexlab\endcsname\relax\def\natexlab#1{#1}\fi
\expandafter\ifx\csname url\endcsname\relax
  \def\url#1{\texttt{#1}}\fi
\expandafter\ifx\csname urlprefix\endcsname\relax\def\urlprefix{URL }\fi

\bibitem[{{Abdo} et~al.(2011){Abdo}, {Ackermann}, {Ajello}, {Allafort},
  {Baldini}, {Ballet}, {Barbiellini}, {Bastieri}, {Bechtol}, {Bellazzini},
  {Berenji}, {Blandford}, {Bloom}, {Bonamente}, {Borgland}, {Bouvier},
  {Brandt}, {Bregeon}, {Brez}, {Brigida}, {Bruel}, {Buehler}, {Buson},
  {Caliandro}, {Cameron}, {Cannon}, {Caraveo}, {Casandjian}, {{\c C}elik},
  {Charles}, {Chekhtman}, {Cheung}, {Chiang}, {Ciprini}, {Claus},
  {Cohen-Tanugi}, {Costamante}, {Cutini}, {D'Ammando}, {Dermer}, {de Angelis},
  {de Luca}, {de Palma}, {Digel}, {do Couto e Silva}, {Drell}, {Drlica-Wagner},
  {Dubois}, {Dumora}, {Favuzzi}, {Fegan}, {Ferrara}, {Focke}, {Fortin},
  {Frailis}, {Fukazawa}, {Funk}, {Fusco}, {Gargano}, {Gasparrini}, {Gehrels},
  {Germani}, {Giglietto}, {Giordano}, {Giroletti}, {Glanzman}, {Godfrey},
  {Grenier}, {Grondin}, {Grove}, {Guiriec}, {Hadasch}, {Hanabata}, {Harding},
  {Hayashi}, {Hayashida}, {Hays}, {Horan}, {Itoh}, {J{\'o}hannesson},
  {Johnson}, {Johnson}, {Khangulyan}, {Kamae}, {Katagiri}, {Kataoka}, {Kerr},
  {Kn{\"o}dlseder}, {Kuss}, {Lande}, {Latronico}, {Lee}, {Lemoine-Goumard},
  {Longo}, {Loparco}, {Lubrano}, {Madejski}, {Makeev}, {Marelli}, {Mazziotta},
  {McEnery}, {Michelson}, {Mitthumsiri}, {Mizuno}, {Moiseev}, {Monte},
  {Monzani}, {Morselli}, {Moskalenko}, {Murgia}, {Nakamori}, {Naumann-Godo},
  {Nolan}, {Norris}, {Nuss}, {Ohsugi}, {Okumura}, {Omodei}, {Ormes}, {Ozaki},
  {Paneque}, {Parent}, {Pelassa}, {Pepe}, {Pesce-Rollins}, {Pierbattista},
  {Piron}, {Porter}, {Rain{\`o}}, {Rando}, {Ray}, {Razzano}, {Reimer},
  {Reimer}, {Reposeur}, {Ritz}, {Romani}, {Sadrozinski}, {Sanchez},
  {Parkinson}, {Scargle}, {Schalk}, {Sgr{\`o}}, {Siskind}, {Smith}, {Spandre},
  {Spinelli}, {Strickman}, {Suson}, {Takahashi}, {Takahashi}, {Tanaka},
  {Thayer}, {Thompson}, {Tibaldo}, {Torres}, {Tosti}, {Tramacere}, {Troja},
  {Uchiyama}, {Vandenbroucke}, {Vasileiou}, {Vianello}, {Vitale}, {Wang},
  {Wood}, {Yang}, and {Ziegler}}]{2011Sci...331..739A}
{Abdo}, A.~A., {Ackermann}, M., {Ajello}, M., et~al., Feb. 2011. {Gamma-Ray
  Flares from the Crab Nebula}. Science 331, 739--.

\bibitem[{{Abdo} et~al.(2010){Abdo}, {Ackermann}, {Ajello}, {Axelsson},
  {Baldini}, {Ballet}, {Barbiellini}, {Bastieri}, {Baughman}, {Bechtol}, and
  et~al.}]{2010Natur.463..919A}
{Abdo}, A.~A., {Ackermann}, M., {Ajello}, M., et~al., Feb. 2010. {A change in
  the optical polarization associated with a {$\gamma$}-ray flare in the blazar
  3C279}. \nat 463, 919--923.

\bibitem[{{Abramowicz}(2005)}]{2005AN....326..782A}
{Abramowicz}, M.~A., Nov. 2005. {QPO as the Rosetta Stone for understanding
  black hole accretion}. Astronomische Nachrichten 326, 782--786.

\bibitem[{{Agostinelli} et~al.(2003){Agostinelli}, {Allison}, {Amako},
  {Apostolakis}, {Araujo}, {Arce}, {Asai}, {Axen}, {Banerjee}, {Barrand},
  {Behner}, {Bellagamba}, {Boudreau}, {Broglia}, {Brunengo}, {Burkhardt},
  {Chauvie}, {Chuma}, {Chytracek}, {Cooperman}, {Cosmo}, {Degtyarenko},
  {Dell'Acqua}, {Depaola}, {Dietrich}, {Enami}, {Feliciello}, {Ferguson},
  {Fesefeldt}, {Folger}, {Foppiano}, {Forti}, {Garelli}, {Giani},
  {Giannitrapani}, {Gibin}, {G{\'o}mez Cadenas}, {Gonz{\'a}lez}, {Gracia
  Abril}, {Greeniaus}, {Greiner}, {Grichine}, {Grossheim}, {Guatelli},
  {Gumplinger}, {Hamatsu}, {Hashimoto}, {Hasui}, {Heikkinen}, {Howard},
  {Ivanchenko}, {Johnson}, {Jones}, {Kallenbach}, {Kanaya}, {Kawabata},
  {Kawabata}, {Kawaguti}, {Kelner}, {Kent}, {Kimura}, {Kodama}, {Kokoulin},
  {Kossov}, {Kurashige}, {Lamanna}, {Lamp{\'e}n}, {Lara}, {Lefebure}, {Lei},
  {Liendl}, {Lockman}, {Longo}, {Magni}, {Maire}, {Medernach}, {Minamimoto},
  {Mora de Freitas}, {Morita}, {Murakami}, {Nagamatu}, {Nartallo}, {Nieminen},
  {Nishimura}, {Ohtsubo}, {Okamura}, {O'Neale}, {Oohata}, {Paech}, {Perl},
  {Pfeiffer}, {Pia}, {Ranjard}, {Rybin}, {Sadilov}, {Di Salvo}, {Santin},
  {Sasaki}, {Savvas}, {Sawada}, {Scherer}, {Sei}, {Sirotenko}, {Smith},
  {Starkov}, {Stoecker}, {Sulkimo}, {Takahata}, {Tanaka}, {Tcherniaev}, {Safai
  Tehrani}, {Tropeano}, {Truscott}, {Uno}, {Urban}, {Urban}, {Verderi},
  {Walkden}, {Wander}, {Weber}, {Wellisch}, {Wenaus}, {Williams}, {Wright},
  {Yamada}, {Yoshida}, {Zschiesche}, and {G EANT4
  Collaboration}}]{2003NIMPA.506..250A}
{Agostinelli}, S., {Allison}, J., {Amako}, K., et~al., Jul. 2003. {GEANT4 -- a
  simulation toolkit}. Nuclear Instruments and Methods in Physics Research A
  506, 250--303.

\bibitem[{{Angel}(1969)}]{1969ApJ...158..219A}
{Angel}, J.~R.~P., Oct. 1969. {Polarization of Thermal X-Ray Sources}. \apj
  158, 219.

\bibitem[{{Bachetti} et~al.(2015){Bachetti}, {Harrison}, {Cook}, {Tomsick},
  {Schmid}, {Grefenstette}, {Barret}, {Boggs}, {Christensen}, {Craig},
  {Fabian}, {F{\"u}rst}, {Gandhi}, {Hailey}, {Kara}, {Maccarone}, {Miller},
  {Pottschmidt}, {Stern}, {Uttley}, {Walton}, {Wilms}, and
  {Zhang}}]{2015ApJ...800..109B}
{Bachetti}, M., {Harrison}, F.~A., {Cook}, R., et~al., Feb. 2015. {No Time for
  Dead Time: Timing Analysis of Bright Black Hole Binaries with NuSTAR}. \apj
  800, 109.

\bibitem[{{Bachetti} et~al.(2014){Bachetti}, {Harrison}, {Walton},
  {Grefenstette}, {Chakrabarty}, {F{\"u}rst}, {Barret}, {Beloborodov}, {Boggs},
  {Christensen}, {Craig}, {Fabian}, {Hailey}, {Hornschemeier}, {Kaspi},
  {Kulkarni}, {Maccarone}, {Miller}, {Rana}, {Stern}, {Tendulkar}, {Tomsick},
  {Webb}, and {Zhang}}]{2014Natur.514..202B}
{Bachetti}, M., {Harrison}, F.~A., {Walton}, D.~J., et~al., Oct. 2014. {An
  ultraluminous X-ray source powered by an accreting neutron star}. \nat 514,
  202--204.

\bibitem[{{Balokovi{\'c}} et~al.(2015){Balokovi{\'c}}, {Matt}, {Harrison},
  {Zoghbi}, {Ballantyne}, {Boggs}, {Christensen}, {Craig}, {Esmerian},
  {Fabian}, {F{\"u}rst}, {Hailey}, {Marinucci}, {Parker}, {Reynolds}, {Stern},
  {Walton}, and {Zhang}}]{balokovic15}
{Balokovi{\'c}}, M., {Matt}, G., {Harrison}, F.~A., et~al., Feb. 2015. {Coronal
  Properties of the Seyfert 1.9 Galaxy MCG-05-23-016 Determined from Hard X-Ray
  Spectroscopy with NuSTAR}. \apj 800, 62.

\bibitem[{{Baring} and {Harding}(2007)}]{2007Ap&SS.308..109B}
{Baring}, M.~G., {Harding}, A.~K., Apr. 2007. {Resonant Compton upscattering in
  anomalous X-ray pulsars}. Astrophysics and Space Science 308, 109--118.

\bibitem[{{Beilicke} et~al.(2014{\natexlab{a}}){Beilicke}, {Kislat}, {Zajczyk},
  {Guo}, {Endsley}, {Stork}, {Cowsik}, {Dowkontt}, {Barthelmy}, {Hams},
  {Okajima}, {Sasaki}, {Zeiger}, {de Geronimo}, {Baring}, and
  {Krawczynski}}]{2014JAI.....340008B}
{Beilicke}, M., {Kislat}, F., {Zajczyk}, A., et~al., 2014{\natexlab{a}}.
  {Design and Performance of the X-ray Polarimeter X-Calibur}. Journal of
  Astronomical Instrumentation 3, 40008.

\bibitem[{{Beilicke} et~al.(2014{\natexlab{b}}){Beilicke}, {Kislat}, {Zajczyk},
  {Guo}, {Endsley}, {Stork}, {Cowsik}, {Dowkontt}, {Barthelmy}, {Hams},
  {Okajima}, {Sasaki}, {Zeiger}, {de Geronimo}, {Baring}, and
  {Krawczynski}}]{beilicke_etal_2014}
{Beilicke}, M., {Kislat}, F., {Zajczyk}, A., et~al., 2014{\natexlab{b}}.
  {Design and Performance of the X-ray Polarimeter X-Calibur}. Journal of
  Astronomical Instrumentation 3, 40008.

\bibitem[{{Bellazzini} et~al.(2010){Bellazzini}, {Costa}, {Matt}, and
  {Tagliaferri}}]{2010xrp..book.....B}
{Bellazzini}, R., {Costa}, E., {Matt}, G., {Tagliaferri}, G., Jul. 2010. {X-ray
  Polarimetry}. Cambridge University Press.

\bibitem[{{Berti} and {Volonteri}(2008)}]{berti08}
{Berti}, E., {Volonteri}, M., Sep. 2008. {Cosmological Black Hole Spin
  Evolution by Mergers and Accretion}. \apj 684, 822--828.

\bibitem[{{Blasi} et~al.(2013){Blasi}, {Lico}, {Giroletti}, {Orienti},
  {Giovannini}, {Cotton}, {Edwards}, {Fuhrmann}, {Krichbaum}, {Kovalev},
  {Jorstad}, {Marscher}, {Kino}, {Paneque}, {Perez-Torres}, {Piner}, and
  {Sokolovsky}}]{2013A&A...559A..75B}
{Blasi}, M.~G., {Lico}, R., {Giroletti}, M., et~al., Nov. 2013. {The TeV blazar
  Markarian 421 at the highest spatial resolution}. \aap 559, A75.

\bibitem[{{Boettcher} et~al.(2012){Boettcher}, {Harris}, and
  {Krawczynski}}]{2012rjag.book.....B}
{Boettcher}, M., {Harris}, D.~E., {Krawczynski}, H., Jan. 2012. {Relativistic
  Jets from Active Galactic Nuclei}. Wiley-VCH.

\bibitem[{{Cackett} et~al.(2014){Cackett}, {Zoghbi}, {Reynolds}, {Fabian},
  {Kara}, {Uttley}, and {Wilkins}}]{cackett14}
{Cackett}, E.~M., {Zoghbi}, A., {Reynolds}, C., et~al., Mar. 2014. {Modelling
  the broad Fe K{$\alpha$} reverberation in the AGN NGC 4151}. \mnras 438,
  2980--2994.

\bibitem[{{Cerutti} et~al.(2012){Cerutti}, {Uzdensky}, and
  {Begelman}}]{2012ApJ...746..148C}
{Cerutti}, B., {Uzdensky}, D.~A., {Begelman}, M.~C., Feb. 2012. {Extreme
  Particle Acceleration in Magnetic Reconnection Layers: Application to the
  Gamma-Ray Flares in the Crab Nebula}. \apj 746, 148.

\bibitem[{{Chandra} et~al.(2015){Chandra}, {Gammie}, {Foucart}, and
  {Quataert}}]{2015arXiv150800878C}
{Chandra}, M., {Gammie}, C.~F., {Foucart}, F., {Quataert}, E., Aug. 2015. {An
  Extended Magnetohydrodynamics Model for Relativistic Weakly Collisional
  Plasmas}. ArXiv e-prints.

\bibitem[{{Chandrasekhar}(1960)}]{1960ratr.book.....C}
{Chandrasekhar}, S., 1960. {Radiative transfer}.

\bibitem[{{Christensen} et~al.(2011){Christensen}, {Jakobsen}, {Brejnholt},
  {Madsen}, {Hornstrup}, {Westergaard}, {Momberg}, {Koglin}, {Fabricant},
  {Stern}, {Craig}, {Pivovaroff}, and {Windt}}]{2011SPIE.8147E..0UC}
{Christensen}, F.~E., {Jakobsen}, A.~C., {Brejnholt}, N.~F., et~al., Sep. 2011.
  {Coatings for the NuSTAR mission}. In: Society of Photo-Optical
  Instrumentation Engineers (SPIE) Conference Series. Vol. 8147 of Society of
  Photo-Optical Instrumentation Engineers (SPIE) Conference Series. p.~0.

\bibitem[{{Clausen-Brown} and {Lyutikov}(2012)}]{2012MNRAS.426.1374C}
{Clausen-Brown}, E., {Lyutikov}, M., Oct. 2012. {Crab nebula gamma-ray flares
  as relativistic reconnection minijets}. \mnras 426, 1374--1384.

\bibitem[{{Costa} et~al.(2006){Costa}, {Bellazzini}, {Soffitta}, {di Persio},
  {Feroci}, {Morelli}, {Muleri}, {Pacciani}, {Rubini}, {Baldini}, {Bitti},
  {Brez}, {Cavalca}, {Latronico}, {Massai}, {Omodei}, {Sgro'}, {Spandre},
  {Matt}, {Perola}, {Santangelo}, {Celotti}, {Barret}, {Vilhu}, {Piro},
  {Fraser}, {Courvoisier}, and {Barcons}}]{2006astro.ph..3399C}
{Costa}, E., {Bellazzini}, R., {Soffitta}, P., et~al., Mar. 2006. {Opening a
  New Window to Fundamental Physics and Astrophysics: X-ray Polarimetry}. ArXiv
  Astrophysics e-prints.

\bibitem[{{Craig} et~al.(2011){Craig}, {An}, {Blaedel}, {Christensen},
  {Decker}, {Fabricant}, {Gum}, {Hailey}, {Hale}, {Jensen}, {Koglin}, {Mori},
  {Nynka}, {Pivovaroff}, {Sharpe}, {Stern}, {Tajiri}, and
  {Zhang}}]{2011SPIE.8147E..0HC}
{Craig}, W.~W., {An}, H., {Blaedel}, K.~L., et~al., Sep. 2011. {Fabrication of
  the NuSTAR flight optics}. In: Society of Photo-Optical Instrumentation
  Engineers (SPIE) Conference Series. Vol. 8147 of Society of Photo-Optical
  Instrumentation Engineers (SPIE) Conference Series. p.~0.

\bibitem[{{De Villiers} and {Hawley}(2003)}]{2003ApJ...589..458D}
{De Villiers}, J.-P., {Hawley}, J.~F., May 2003. {A Numerical Method for
  General Relativistic Magnetohydrodynamics}. \apj 589, 458--480.

\bibitem[{{De Villiers} et~al.(2005){De Villiers}, {Hawley}, {Krolik}, and
  {Hirose}}]{2005ApJ...620..878D}
{De Villiers}, J.-P., {Hawley}, J.~F., {Krolik}, J.~H., {Hirose}, S., Feb.
  2005. {Magnetically Driven Accretion in the Kerr Metric. III. Unbound
  Outflows}. \apj 620, 878--888.

\bibitem[{{Dean} et~al.(2008){Dean}, {Clark}, {Stephen}, {McBride}, {Bassani},
  {Bazzano}, {Bird}, {Hill}, {Shaw}, and {Ubertini}}]{2008Sci...321.1183D}
{Dean}, A.~J., {Clark}, D.~J., {Stephen}, J.~B., et~al., Aug. 2008. {Polarized
  Gamma-Ray Emission from the Crab}. Science 321, 1183--.

\bibitem[{{den Hartog} et~al.(2008){den Hartog}, {Kuiper}, {Hermsen}, {Kaspi},
  {Dib}, {Kn{\"o}dlseder}, and {Gavriil}}]{2008A&A...489..245D}
{den Hartog}, P.~R., {Kuiper}, L., {Hermsen}, W., et~al., Oct. 2008. {Detailed
  high-energy characteristics of AXP 4U 0142+61. Multi-year observations with
  INTEGRAL, RXTE, XMM-Newton, and ASCA}. \aap 489, 245--261.

\bibitem[{{Dov{\v c}iak} et~al.(2004){Dov{\v c}iak}, {Karas}, and
  {Matt}}]{2004MNRAS.355.1005D}
{Dov{\v c}iak}, M., {Karas}, V., {Matt}, G., Dec. 2004. {Polarization
  signatures of strong gravity in active galactic nuclei accretion discs}.
  \mnras 355, 1005--1009.

\bibitem[{{Dov{\v c}iak} et~al.(2008){Dov{\v c}iak}, {Muleri}, {Goosmann},
  {Karas}, and {Matt}}]{2008MNRAS.391...32D}
{Dov{\v c}iak}, M., {Muleri}, F., {Goosmann}, R.~W., {Karas}, V., {Matt}, G.,
  Nov. 2008. {Thermal disc emission from a rotating black hole: X-ray
  polarization signatures}. \mnras 391, 32--38.

\bibitem[{{Dov{\v c}iak} et~al.(2011){Dov{\v c}iak}, {Muleri}, {Goosmann},
  {Karas}, and {Matt}}]{2011ApJ...731...75D}
{Dov{\v c}iak}, M., {Muleri}, F., {Goosmann}, R.~W., {Karas}, V., {Matt}, G.,
  Apr. 2011. {Light-bending Scenario for Accreting Black Holes in X-ray
  Polarimetry}. \apj 731, 75.

\bibitem[{{Duncan} and {Thompson}(1992)}]{1992ApJ...392L...9D}
{Duncan}, R.~C., {Thompson}, C., Jun. 1992. {Formation of very strongly
  magnetized neutron stars - Implications for gamma-ray bursts}. \apjl 392,
  L9--L13.

\bibitem[{{Dyks} et~al.(2004){Dyks}, {Harding}, and
  {Rudak}}]{2004ApJ...606.1125D}
{Dyks}, J., {Harding}, A.~K., {Rudak}, B., May 2004. {Relativistic Effects and
  Polarization in Three High-Energy Pulsar Models}. \apj 606, 1125--1142.

\bibitem[{{Enoto} et~al.(2014){Enoto}, {Black}, {Kitaguchi}, {Hayato}, {Hill},
  {Jahoda}, {Tamagawa}, {Kaneko}, {Takeuchi}, {Yoshikawa}, {Marlowe},
  {Griffiths}, {Kaaret}, {Kenward}, and {Khalid}}]{2014SPIE.9144E..4ME}
{Enoto}, T., {Black}, J.~K., {Kitaguchi}, T., et~al., Jul. 2014. {Performance
  verification of the Gravity and Extreme Magnetism Small explorer (GEMS) x-ray
  polarimeter}. In: Society of Photo-Optical Instrumentation Engineers (SPIE)
  Conference Series. Vol. 9144 of Society of Photo-Optical Instrumentation
  Engineers (SPIE) Conference Series. p.~4.

\bibitem[{{Fabian}(2012)}]{2012ARA&A..50..455F}
{Fabian}, A.~C., Sep. 2012. {Observational Evidence of Active Galactic Nuclei
  Feedback}. \araa 50, 455--489.

\bibitem[{{Fabian} et~al.(1989){Fabian}, {Rees}, {Stella}, and
  {White}}]{fabian89}
{Fabian}, A.~C., {Rees}, M.~J., {Stella}, L., {White}, N.~E., May 1989. {X-ray
  fluorescence from the inner disc in Cygnus X-1}. \mnras 238, 729--736.

\bibitem[{{Fern{\'a}ndez} and {Davis}(2011)}]{2011ApJ...730..131F}
{Fern{\'a}ndez}, R., {Davis}, S.~W., Apr. 2011. {The X-ray Polarization
  Signature of Quiescent Magnetars: Effect of Magnetospheric Scattering and
  Vacuum Polarization}. \apj 730, 131.

\bibitem[{{Forot} et~al.(2008){Forot}, {Laurent}, {Grenier}, {Gouiff{\`e}s},
  and {Lebrun}}]{2008ApJ...688L..29F}
{Forot}, M., {Laurent}, P., {Grenier}, I.~A., {Gouiff{\`e}s}, C., {Lebrun}, F.,
  Nov. 2008. {Polarization of the Crab Pulsar and Nebula as Observed by the
  INTEGRAL/IBIS Telescope}. \apjl 688, L29--L32.

\bibitem[{{Fragile} et~al.(2007){Fragile}, {Blaes}, {Anninos}, and
  {Salmonson}}]{2007ApJ...668..417F}
{Fragile}, P.~C., {Blaes}, O.~M., {Anninos}, P., {Salmonson}, J.~D., Oct. 2007.
  {Global General Relativistic Magnetohydrodynamic Simulation of a Tilted Black
  Hole Accretion Disk}. \apj 668, 417--429.

\bibitem[{{Fukazawa} et~al.(2014){Fukazawa}, {Tajima}, {Watanabe}, {Blandford},
  {Hayashi}, {Harayama}, {Kataoka}, {Kawaharada}, {Kokubun}, {Laurent},
  {Lebrun}, {Limousin}, {Madejski}, {Makishima}, {Mizuno}, {Mori}, {Nakamori},
  {Nakazawa}, {Noda}, {Odaka}, {Ohno}, {Ohta}, {Saito}, {Sato}, {Sato},
  {Takeda}, {Takahashi}, {Takahashi}, {Tanaka}, {Terada}, {Uchiyama},
  {Uchiyama}, {Yamaoka}, {Yatsu}, {Yonetoku}, and
  {Yuasa}}]{2014SPIE.9144E..2CF}
{Fukazawa}, Y., {Tajima}, H., {Watanabe}, S., et~al., Jul. 2014. {Soft
  gamma-ray detector (SGD) onboard the ASTRO-H mission}. In: Society of
  Photo-Optical Instrumentation Engineers (SPIE) Conference Series. Vol. 9144
  of Society of Photo-Optical Instrumentation Engineers (SPIE) Conference
  Series. p.~2.

\bibitem[{{F{\"u}rst} et~al.(2013){F{\"u}rst}, {Grefenstette}, {Staubert},
  {Tomsick}, {Bachetti}, {Barret}, {Bellm}, {Boggs}, {Chenevez}, {Christensen},
  {Craig}, {Hailey}, {Harrison}, {Klochkov}, {Madsen}, {Pottschmidt}, {Stern},
  {Walton}, {Wilms}, and {Zhang}}]{2013ApJ...779...69F}
{F{\"u}rst}, F., {Grefenstette}, B.~W., {Staubert}, R., et~al., Dec. 2013. {The
  Smooth Cyclotron Line in Her X-1 as Seen with Nuclear Spectroscopic Telescope
  Array}. \apj 779, 69.

\bibitem[{{F{\"u}rst} et~al.(2015){F{\"u}rst}, {Pottschmidt}, {Miyasaka},
  {Bhalerao}, {Bachetti}, {Boggs}, {Christensen}, {Craig}, {Grinberg},
  {Hailey}, {Harrison}, {Kennea}, {Rahoui}, {Stern}, {Tendulkar}, {Tomsick},
  {Walton}, {Wilms}, and {Zhang}}]{2015ApJ...806L..24F}
{F{\"u}rst}, F., {Pottschmidt}, K., {Miyasaka}, H., et~al., Jun. 2015.
  {Distorted Cyclotron Line Profile in Cep X-4 as Observed by NuSTAR}. \apjl
  806, L24.

\bibitem[{{F{\"u}rst} et~al.(2014{\natexlab{a}}){F{\"u}rst}, {Pottschmidt},
  {Wilms}, {Kennea}, {Bachetti}, {Bellm}, {Boggs}, {Chakrabarty},
  {Christensen}, {Craig}, {Hailey}, {Harrison}, {Stern}, {Tomsick}, {Walton},
  and {Zhang}}]{2014ApJ...784L..40F}
{F{\"u}rst}, F., {Pottschmidt}, K., {Wilms}, J., et~al., Apr.
  2014{\natexlab{a}}. {NuSTAR Discovery of a Cyclotron Line in KS 1947+300}.
  \apjl 784, L40.

\bibitem[{{F{\"u}rst} et~al.(2014{\natexlab{b}}){F{\"u}rst}, {Pottschmidt},
  {Wilms}, {Tomsick}, {Bachetti}, {Boggs}, {Christensen}, {Craig},
  {Grefenstette}, {Hailey}, {Harrison}, {Madsen}, {Miller}, {Stern}, {Walton},
  and {Zhang}}]{2014ApJ...780..133F}
{F{\"u}rst}, F., {Pottschmidt}, K., {Wilms}, J., et~al., Jan.
  2014{\natexlab{b}}. {NuSTAR Discovery of a Luminosity Dependent Cyclotron
  Line Energy in Vela X-1}. \apj 780, 133.

\bibitem[{{Gammie} et~al.(2003){Gammie}, {McKinney}, and
  {T{\'o}th}}]{2003ApJ...589..444G}
{Gammie}, C.~F., {McKinney}, J.~C., {T{\'o}th}, G., May 2003. {HARM: A
  Numerical Scheme for General Relativistic Magnetohydrodynamics}. \apj 589,
  444--457.

\bibitem[{{Ghosh} et~al.(2013){Ghosh}, {Angelini}, {Baring}, {Baumgartner},
  {Black}, {Dotson}, {Harding}, {Hill}, {Jahoda}, {Kaaret}, {Kallman},
  {Krawczynski}, {Krolik}, {Lai}, {Markwardt}, {Marshall}, {Martoff}, {Morris},
  {Okajima}, {Petre}, {Poutanen}, {Reynolds}, {Scargle}, {Schnittman},
  {Serlemitsos}, {Soong}, {Strohmayer}, {Swank}, {Tawara}, and
  {Tamagawa}}]{2013arXiv1301.5514G}
{Ghosh}, P., {Angelini}, L., {Baring}, M., et~al., Jan. 2013. {White Paper on
  GEMS Study of Polarized X-rays from Neutron Stars}. ArXiv e-prints.

\bibitem[{{Gilfanov} and {Merloni}(2014)}]{2014SSRv..183..121G}
{Gilfanov}, M., {Merloni}, A., Sep. 2014. {Observational Appearance of Black
  Holes in X-Ray Binaries and AGN}. \ssr 183, 121--148.

\bibitem[{{G{\"o}tz} et~al.(2006){G{\"o}tz}, {Mereghetti}, {Tiengo}, and
  {Esposito}}]{2006A&A...449L..31G}
{G{\"o}tz}, D., {Mereghetti}, S., {Tiengo}, A., {Esposito}, P., Apr. 2006.
  {Magnetars as persistent hard X-ray sources: INTEGRAL discovery of a hard
  tail in SGR 1900+14}. \aap 449, L31--L34.

\bibitem[{{Gou} et~al.(2011){Gou}, {McClintock}, {Reid}, {Orosz}, {Steiner},
  {Narayan}, {Xiang}, {Remillard}, {Arnaud}, and {Davis}}]{2011ApJ...742...85G}
{Gou}, L., {McClintock}, J.~E., {Reid}, M.~J., et~al., Dec. 2011. {The Extreme
  Spin of the Black Hole in Cygnus X-1}. \apj 742, 85.

\bibitem[{{Gou} et~al.(2014){Gou}, {McClintock}, {Remillard}, {Steiner},
  {Reid}, {Orosz}, {Narayan}, {Hanke}, and
  {Garc{\'{\i}}a}}]{2014ApJ...790...29G}
{Gou}, L., {McClintock}, J.~E., {Remillard}, R.~A., et~al., Jul. 2014.
  {Confirmation via the Continuum-fitting Method that the Spin of the Black
  Hole in Cygnus X-1 Is Extreme}. \apj 790, 29.

\bibitem[{{Guo} et~al.(2013){Guo}, {Beilicke}, {Garson}, {Kislat}, {Fleming},
  and {Krawczynski}}]{2013APh....41...63G}
{Guo}, Q., {Beilicke}, M., {Garson}, A., et~al., Jan. 2013. {Optimization of
  the design of the hard X-ray polarimeter X-Calibur}. Astroparticle Physics
  41, 63--72.

\bibitem[{{Hailey} et~al.(2010{\natexlab{a}}){Hailey}, {An}, {Blaedel},
  {Brejnholt}, {Christensen}, {Craig}, {Decker}, {Doll}, {Gum}, {Koglin},
  {Jensen}, {Hale}, {Mori}, {Pivovaroff}, {Sharpe}, {Stern}, {Tajiri}, and
  {Zhang}}]{2010SPIE.7732E..0TH}
{Hailey}, C.~J., {An}, H., {Blaedel}, K.~L., et~al., Jul. 2010{\natexlab{a}}.
  {The Nuclear Spectroscopic Telescope Array (NuSTAR): optics overview and
  current status}. In: Society of Photo-Optical Instrumentation Engineers
  (SPIE) Conference Series. Vol. 7732 of Society of Photo-Optical
  Instrumentation Engineers (SPIE) Conference Series. p.~0.

\bibitem[{{Hailey} et~al.(2010{\natexlab{b}}){Hailey}, {An}, {Blaedel},
  {Brejnholt}, {Christensen}, {Craig}, {Decker}, {Doll}, {Gum}, {Koglin},
  {Jensen}, {Hale}, {Mori}, {Pivovaroff}, {Sharpe}, {Stern}, {Tajiri}, and
  {Zhang}}]{2010SPIE.7732E..0SH}
{Hailey}, C.~J., {An}, H., {Blaedel}, K.~L., et~al., Jul. 2010{\natexlab{b}}.
  {The Nuclear Spectroscopic Telescope Array (NuSTAR): optics overview and
  current status}. In: Society of Photo-Optical Instrumentation Engineers
  (SPIE) Conference Series. Vol. 7732 of Society of Photo-Optical
  Instrumentation Engineers (SPIE) Conference Series. p.~0.

\bibitem[{{Harding} and {Lai}(2006)}]{2006RPPh...69.2631H}
{Harding}, A.~K., {Lai}, D., Sep. 2006. {Physics of strongly magnetized neutron
  stars}. Reports on Progress in Physics 69, 2631--2708.

\bibitem[{{Harrison} et~al.(2013){Harrison}, {Craig}, {Christensen}, {Hailey},
  {Zhang}, {Boggs}, {Stern}, {Cook}, {Forster}, {Giommi}, {Grefenstette},
  {Kim}, {Kitaguchi}, {Koglin}, {Madsen}, {Mao}, {Miyasaka}, {Mori}, {Perri},
  {Pivovaroff}, {Puccetti}, {Rana}, {Westergaard}, {Willis}, {Zoglauer}, {An},
  {Bachetti}, {Barri{\`e}re}, {Bellm}, {Bhalerao}, {Brejnholt}, {Fuerst},
  {Liebe}, {Markwardt}, {Nynka}, {Vogel}, {Walton}, {Wik}, {Alexander},
  {Cominsky}, {Hornschemeier}, {Hornstrup}, {Kaspi}, {Madejski}, {Matt},
  {Molendi}, {Smith}, {Tomsick}, {Ajello}, {Ballantyne}, {Balokovi{\'c}},
  {Barret}, {Bauer}, {Blandford}, {Brandt}, {Brenneman}, {Chiang},
  {Chakrabarty}, {Chenevez}, {Comastri}, {Dufour}, {Elvis}, {Fabian}, {Farrah},
  {Fryer}, {Gotthelf}, {Grindlay}, {Helfand}, {Krivonos}, {Meier}, {Miller},
  {Natalucci}, {Ogle}, {Ofek}, {Ptak}, {Reynolds}, {Rigby}, {Tagliaferri},
  {Thorsett}, {Treister}, and {Urry}}]{2013ApJ...770..103H}
{Harrison}, F.~A., {Craig}, W.~W., {Christensen}, F.~E., et~al., Jun. 2013.
  {The Nuclear Spectroscopic Telescope Array (NuSTAR) High-energy X-Ray
  Mission}. \apj 770, 103.

\bibitem[{{Hasinger} and {van der Klis}(1989)}]{zatoll}
{Hasinger}, G., {van der Klis}, M., Nov. 1989. {Two patterns of correlated
  X-ray timing and spectral behaviour in low-mass X-ray binaries}. \aap 225,
  79--96.

\bibitem[{{Heil} et~al.(2015){Heil}, {Uttley}, and
  {Klein-Wolt}}]{2015MNRAS.448.3348H}
{Heil}, L.~M., {Uttley}, P., {Klein-Wolt}, M., Apr. 2015.
  {Inclination-dependent spectral and timing properties in transient black hole
  X-ray binaries}. \mnras 448, 3348--3353.

\bibitem[{{Hill} et~al.(2014){Hill}, {Black}, {Emmett}, {Enoto}, {Jahoda},
  {Kaaret}, {Nolan}, and {Tamagawa}}]{2014SPIE.9144E..1NH}
{Hill}, J.~E., {Black}, J.~K., {Emmett}, T.~J., et~al., Jul. 2014. {Design
  improvements and x-ray performance of a time projection chamber polarimeter
  for persistent astronomical sources}. In: Society of Photo-Optical
  Instrumentation Engineers (SPIE) Conference Series. Vol. 9144 of Society of
  Photo-Optical Instrumentation Engineers (SPIE) Conference Series. p.~1.

\bibitem[{{Hughes} et~al.(1984){Hughes}, {Long}, and
  {Novick}}]{1984ApJ...280..255H}
{Hughes}, J.~P., {Long}, K.~S., {Novick}, R., May 1984. {A search for X-ray
  polarization in cosmic X-ray sources}. \apj 280, 255--258.

\bibitem[{{Ingram} et~al.(2009){Ingram}, {Done}, and
  {Fragile}}]{2009MNRAS.397L.101I}
{Ingram}, A., {Done}, C., {Fragile}, P.~C., Jul. 2009. {Low-frequency
  quasi-periodic oscillations spectra and Lense-Thirring precession}. \mnras
  397, L101--L105.

\bibitem[{{Ingram} et~al.(2015){Ingram}, {Maccarone}, {Poutanen}, and
  {Krawczynski}}]{2015arXiv150500015I}
{Ingram}, A., {Maccarone}, T., {Poutanen}, J., {Krawczynski}, H., Apr. 2015.
  {Polarization modulation from Lense-Thirring precession in X-ray binaries}.
  ArXiv e-prints.

\bibitem[{{Ingram} and {van der Klis}(2013)}]{2013MNRAS.434.1476I}
{Ingram}, A., {van der Klis}, M., Sep. 2013. {An exact analytic treatment of
  propagating mass accretion rate fluctuations in X-ray binaries}. \mnras 434,
  1476--1485.

\bibitem[{{Jahoda} et~al.(2015){Jahoda}, {Kouveliotou}, {Kallman}, and {Praxys
  Team}}]{2015AAS...22533840J}
{Jahoda}, K., {Kouveliotou}, C., {Kallman}, T.~R., {Praxys Team}, Jan. 2015.
  {Polarization from Relativistic Astrophysical X-raY Sourses: The PRAXYS Small
  Explorer Observatory}. In: American Astronomical Society Meeting Abstracts.
  Vol. 225 of American Astronomical Society Meeting Abstracts. p. \#338.40.

\bibitem[{{Jahoda} et~al.(1996){Jahoda}, {Swank}, {Giles}, {Stark},
  {Strohmayer}, {Zhang}, and {Morgan}}]{1996SPIE.2808...59J}
{Jahoda}, K., {Swank}, J.~H., {Giles}, A.~B., et~al., Oct. 1996. {In-orbit
  performance and calibration of the Rossi X-ray Timing Explorer (RXTE)
  Proportional Counter Array (PCA)}. In: {Siegmund}, O.~H., {Gummin}, M.~A.
  (Eds.), EUV, X-Ray, and Gamma-Ray Instrumentation for Astronomy VII. Vol.
  2808 of Society of Photo-Optical Instrumentation Engineers (SPIE) Conference
  Series. pp. 59--70.

\bibitem[{{Jahoda} et~al.(2014){Jahoda}, {Black}, {Hill}, {Kallman}, {Kaaret},
  {Markwardt}, {Okajima}, {Petre}, {Soong}, {Strohmayer}, {Tamagawa}, and
  {Tawara}}]{2014SPIE.9144E..0NJ}
{Jahoda}, K.~M., {Black}, J.~K., {Hill}, J.~E., et~al., Jul. 2014. {X-ray
  polarization capabilities of a small explorer mission}. In: Society of
  Photo-Optical Instrumentation Engineers (SPIE) Conference Series. Vol. 9144
  of Society of Photo-Optical Instrumentation Engineers (SPIE) Conference
  Series. p.~0.

\bibitem[{{Jiang} et~al.(2014){Jiang}, {Stone}, and
  {Davis}}]{2014ApJ...796..106J}
{Jiang}, Y.-F., {Stone}, J.~M., {Davis}, S.~W., Dec. 2014. {A Global
  Three-dimensional Radiation Magneto-hydrodynamic Simulation of
  Super-Eddington Accretion Disks}. \apj 796, 106.

\bibitem[{{Kalemci} et~al.(2007){Kalemci}, {Boggs}, {Kouveliotou}, {Finger},
  and {Baring}}]{2007ApJS..169...75K}
{Kalemci}, E., {Boggs}, S.~E., {Kouveliotou}, C., {Finger}, M., {Baring},
  M.~G., Mar. 2007. {Search for Polarization from the Prompt Gamma-Ray Emission
  of GRB 041219a with SPI on INTEGRAL}. \apjs 169, 75--82.

\bibitem[{{Kara} et~al.(2013){Kara}, {Fabian}, {Cackett}, {Uttley}, {Wilkins},
  and {Zoghbi}}]{kara13}
{Kara}, E., {Fabian}, A.~C., {Cackett}, E.~M., et~al., Sep. 2013. {Discovery of
  high-frequency iron K lags in Ark 564 and Mrk 335}. \mnras 434, 1129--1137.

\bibitem[{{Kara} et~al.(2014){Kara}, {Fabian}, {Marinucci}, {Matt}, {Parker},
  {Alston}, {Brenneman}, {Cackett}, and {Miniutti}}]{kara14}
{Kara}, E., {Fabian}, A.~C., {Marinucci}, A., et~al., Nov. 2014. {The changing
  X-ray time lag in MCG-6-30-15}. \mnras 445, 56--65.

\bibitem[{{Kara} et~al.(2015){Kara}, {Zoghbi}, {Marinucci}, {Walton}, {Fabian},
  {Risaliti}, {Boggs}, {Christensen}, {Fuerst}, {Hailey}, {Harrison}, {Matt},
  {Parker}, {Reynolds}, {Stern}, and {Zhang}}]{kara15}
{Kara}, E., {Zoghbi}, A., {Marinucci}, A., et~al., Jan. 2015. {Iron K and
  Compton hump reverberation in SWIFT J2127.4+5654 and NGC 1365 revealed by
  NuSTAR and XMM-Newton}. \mnras 446, 737--749.

\bibitem[{{Katz}(1976)}]{1976ApJ...206..910K}
{Katz}, J.~I., Jun. 1976. {Nonrelativistic Compton scattering and models of
  quasars}. \apj 206, 910--916.

\bibitem[{{Keck} et~al.(2015){Keck}, {Brenneman}, {Ballantyne}, {Bauer},
  {Boggs}, {Christensen}, {Craig}, {Dauser}, {Elvis}, {Fabian}, {Fuerst},
  {Garc{\'{\i}}a}, {Grefenstette}, {Hailey}, {Harrison}, {Madejski},
  {Marinucci}, {Matt}, {Reynolds}, {Stern}, {Walton}, and {Zoghbi}}]{keck15}
{Keck}, M.~L., {Brenneman}, L.~W., {Ballantyne}, D.~R., et~al., Apr. 2015.
  {NuSTAR and Suzaku X-ray Spectroscopy of NGC 4151: Evidence for Reflection
  from the Inner Accretion Disk}. ArXiv e-prints.

\bibitem[{{Kellogg}(1971)}]{1971IAUS...46...42K}
{Kellogg}, E.~M., 1971. {X-Ray Observations of the Crab Nebula}. In: {Davies},
  R.~D., {Graham-Smith}, F. (Eds.), The Crab Nebula. Vol.~46 of IAU Symposium.
  p.~42.

\bibitem[{{Kestenbaum} et~al.(1976){Kestenbaum}, {Cohen}, {Long}, {Novick},
  {Silver}, {Weisskopf}, and {Wolff}}]{1976ApJ...210..805K}
{Kestenbaum}, H.~L., {Cohen}, G.~G., {Long}, K.~S., et~al., Dec. 1976. {The
  graphite crystal X-ray spectrometer on OSO-8}. \apj 210, 805--809.

\bibitem[{{Kii}(1987)}]{1987PASJ...39..781K}
{Kii}, T., 1987. {X-ray polarizations from accreting strongly magnetized
  neutron stars - Case studies for the X-ray pulsars 4U 1626-67 and Hercules
  X-1}. \pasj 39, 781--800.

\bibitem[{{Kislat} et~al.(2015){Kislat}, {Clark}, {Beilicke}, and
  {Krawczynski}}]{kislat_etal_2015a}
{Kislat}, F., {Clark}, B., {Beilicke}, M., {Krawczynski}, H., Aug. 2015.
  {Analyzing the data from X-ray polarimeters with Stokes parameters}.
  Astroparticle Physics 68, 45--51.

\bibitem[{{Kitaguchi} et~al.(2014){Kitaguchi}, {Tamagawa}, {Hayato}, {Enoto},
  {Yoshikawa}, {Kaneko}, {Takeuchi}, {Black}, {Hill}, {Jahoda}, {Krizmanic},
  {Sturner}, {Griffiths}, {Kaaret}, and {Marlowe}}]{2014SPIE.9144E..4LK}
{Kitaguchi}, T., {Tamagawa}, T., {Hayato}, A., et~al., Jul. 2014. {Monte-Carlo
  estimation of the inflight performance of the GEMS satellite x-ray
  polarimeter}. In: Society of Photo-Optical Instrumentation Engineers (SPIE)
  Conference Series. Vol. 9144 of Society of Photo-Optical Instrumentation
  Engineers (SPIE) Conference Series. p.~4.

\bibitem[{{Kosteleck{\'y}} and {Mewes}(2013)}]{2013PhRvL.110t1601K}
{Kosteleck{\'y}}, V.~A., {Mewes}, M., May 2013. {Constraints on Relativity
  Violations from Gamma-Ray Bursts}. Physical Review Letters 110~(20), 201601.

\bibitem[{Krawczynski(2011)}]{krawczynski_h_2011a}
Krawczynski, H., 2011. {Analysis of the data from Compton X-ray polarimeters
  which measure the azimuthal and polar scattering angles}. Astropart. Phys.
  34, 784--788.

\bibitem[{{Krawczynski}(2012)}]{2012ApJ...744...30K}
{Krawczynski}, H., Jan. 2012. {The Polarization Properties of Inverse Compton
  Emission and Implications for Blazar Observations with the GEMS X-Ray
  Polarimeter}. \apj 744, 30.

\bibitem[{{Krawczynski} et~al.(2013){Krawczynski}, {Angelini}, {Baring},
  {Baumgartner}, {Black}, {Dotson}, {Ghosh}, {Harding}, {Hill}, {Jahoda},
  {Kaaret}, {Kallman}, {Krolik}, {Lai}, {Markwardt}, {Marshall}, {Martoff},
  {Morris}, {Okajima}, {Petre}, {Poutanen}, {Reynolds}, {Scargle},
  {Schnittman}, {Serlemitsos}, {Soong}, {Strohmayer}, {Swank}, {Tawara}, and
  {Tamagawa}}]{2013arXiv1303.7158K}
{Krawczynski}, H., {Angelini}, L., {Baring}, M., et~al., Mar. 2013. {White
  Paper for Blazar Observations with a GEMS-like X-ray Polarimetry Mission}.
  ArXiv e-prints.

\bibitem[{{Krawczynski} et~al.(2011){Krawczynski}, {Garson}, {Guo}, {Baring},
  {Ghosh}, {Beilicke}, and {Lee}}]{2011APh....34..550K}
{Krawczynski}, H., {Garson}, A., {Guo}, Q., et~al., Feb. 2011. {Scientific
  prospects for hard X-ray polarimetry}. Astroparticle Physics 34, 550--567.

\bibitem[{{Kuiper} et~al.(2004){Kuiper}, {Hermsen}, and
  {Mendez}}]{2004ApJ...613.1173K}
{Kuiper}, L., {Hermsen}, W., {Mendez}, M., Oct. 2004. {Discovery of Hard
  Nonthermal Pulsed X-Ray Emission from the Anomalous X-Ray Pulsar 1E
  1841-045}. \apj 613, 1173--1178.

\bibitem[{{Lai} and {Ho}(2003)}]{lai2003}
{Lai}, D., {Ho}, W.~C.~G., May 2003. {Transfer of Polarized Radiation in
  Strongly Magnetized Plasmas and Thermal Emission from Magnetars: Effect of
  Vacuum Polarization}. \apj 588, 962--974.

\bibitem[{{Laurent} et~al.(2011){Laurent}, {Rodriguez}, {Wilms}, {Cadolle Bel},
  {Pottschmidt}, and {Grinberg}}]{2011Sci...332..438L}
{Laurent}, P., {Rodriguez}, J., {Wilms}, J., et~al., Apr. 2011. {Polarized
  Gamma-Ray Emission from the Galactic Black Hole Cygnus X-1}. Science 332,
  438--.

\bibitem[{{Lei} et~al.(1997){Lei}, {Dean}, and {Hills}}]{1997SSRv...82..309L}
{Lei}, F., {Dean}, A.~J., {Hills}, G.~L., Nov. 1997. {Compton Polarimetry in
  Gamma-Ray Astronomy}. \ssr 82, 309--388.

\bibitem[{{Levine} et~al.(1996){Levine}, {Bradt}, {Cui}, {Jernigan}, {Morgan},
  {Remillard}, {Shirey}, and {Smith}}]{1996ApJ...469L..33L}
{Levine}, A.~M., {Bradt}, H., {Cui}, W., et~al., Sep. 1996. {First Results from
  the All-Sky Monitor on the Rossi X-Ray Timing Explorer}. \apjl 469, L33.

\bibitem[{{Li} et~al.(2009){Li}, {Narayan}, and
  {McClintock}}]{2009ApJ...691..847L}
{Li}, L.-X., {Narayan}, R., {McClintock}, J.~E., Jan. 2009. {Inferring the
  Inclination of a Black Hole Accretion Disk from Observations of its Polarized
  Continuum Radiation}. \apj 691, 847--865.

\bibitem[{{Liebe} et~al.(2012){Liebe}, {Craig}, {Kim}, {McLean}, {Meras},
  {Raffanti}, and {Scholz}}]{2012OptEn..51d3605L}
{Liebe}, C.~C., {Craig}, W., {Kim}, Y., et~al., Apr. 2012. {Calibration and
  alignment of metrology system for the Nuclear Spectroscopic Telescope Array
  mission}. Optical Engineering 51~(4), 043605.

\bibitem[{{Lodato} and {Price}(2010)}]{2010MNRAS.405.1212L}
{Lodato}, G., {Price}, D.~J., Jun. 2010. {On the diffusive propagation of warps
  in thin accretion discs}. \mnras 405, 1212--1226.

\bibitem[{{Lyubarsky}(2012)}]{2012MNRAS.427.1497L}
{Lyubarsky}, Y.~E., Dec. 2012. {Highly magnetized region in pulsar wind nebulae
  and origin of the Crab gamma-ray flares}. \mnras 427, 1497--1502.

\bibitem[{{Madsen} et~al.(2015){Madsen}, {Harrison}, {Markwardt}, {An},
  {Grefenstette}, {Bachetti}, {Miyasaka}, {Kitaguchi}, {Bhalerao},
  {Christensen}, {Craig}, {Fuerst}, {Walton}, {Hailey}, {Rana}, {Stern},
  {Westergaard}, and {Zhang}}]{2015arXiv150401672M}
{Madsen}, K.~K., {Harrison}, F.~A., {Markwardt}, C., et~al., Apr. 2015.
  {Calibration of the NuSTAR High Energy Focusing X-ray Telescope}. ArXiv
  e-prints.

\bibitem[{{Marinucci} et~al.(2014{\natexlab{a}}){Marinucci}, {Matt}, {Kara},
  {Miniutti}, {Elvis}, {Arevalo}, {Ballantyne}, {Balokovi{\'c}}, {Bauer},
  {Brenneman}, {Boggs}, {Cappi}, {Christensen}, {Craig}, {Fabian}, {Fuerst},
  {Hailey}, {Harrison}, {Risaliti}, {Reynolds}, {Stern}, {Walton}, and
  {Zhang}}]{marinucci14}
{Marinucci}, A., {Matt}, G., {Kara}, E., et~al., May 2014{\natexlab{a}}.
  {Simultaneous NuSTAR and XMM-Newton 0.5-80 keV spectroscopy of the
  narrow-line Seyfert 1 galaxy SWIFT J2127.4+5654}. \mnras 440, 2347--2356.

\bibitem[{{Marinucci} et~al.(2014{\natexlab{b}}){Marinucci}, {Matt},
  {Miniutti}, {Guainazzi}, {Parker}, {Brenneman}, {Fabian}, {Kara}, {Arevalo},
  {Ballantyne}, {Boggs}, {Cappi}, {Christensen}, {Craig}, {Elvis}, {Hailey},
  {Harrison}, {Reynolds}, {Risaliti}, {Stern}, {Walton}, and
  {Zhang}}]{marinucci14b}
{Marinucci}, A., {Matt}, G., {Miniutti}, G., et~al., May 2014{\natexlab{b}}.
  {The Broadband Spectral Variability of MCG-6-30-15 Observed by NuSTAR and
  XMM-Newton}. \apj 787, 83.

\bibitem[{{Marscher} et~al.(2008){Marscher}, {Jorstad}, {D'Arcangelo}, {Smith},
  {Williams}, {Larionov}, {Oh}, {Olmstead}, {Aller}, {Aller}, {McHardy},
  {L{\"a}hteenm{\"a}ki}, {Tornikoski}, {Valtaoja}, {Hagen-Thorn}, {Kopatskaya},
  {Gear}, {Tosti}, {Kurtanidze}, {Nikolashvili}, {Sigua}, {Miller}, and
  {Ryle}}]{2008Natur.452..966M}
{Marscher}, A.~P., {Jorstad}, S.~G., {D'Arcangelo}, F.~D., et~al., Apr. 2008.
  {The inner jet of an active galactic nucleus as revealed by a
  radio-to-{$\gamma$}-ray outburst}. \nat 452, 966--969.

\bibitem[{{McGlynn} et~al.(2007){McGlynn}, {Clark}, {Dean}, {Hanlon},
  {McBreen}, {Willis}, {McBreen}, {Bird}, and {Foley}}]{2007A&A...466..895M}
{McGlynn}, S., {Clark}, D.~J., {Dean}, A.~J., et~al., May 2007. {Polarisation
  studies of the prompt gamma-ray emission from GRB 041219a using the
  spectrometer aboard INTEGRAL}. \aap 466, 895--904.

\bibitem[{{McKinney}(2006)}]{2006MNRAS.368.1561M}
{McKinney}, J.~C., Jun. 2006. {General relativistic magnetohydrodynamic
  simulations of the jet formation and large-scale propagation from black hole
  accretion systems}. \mnras 368, 1561--1582.

\bibitem[{{McKinney} and {Blandford}(2009)}]{2009MNRAS.394L.126M}
{McKinney}, J.~C., {Blandford}, R.~D., Mar. 2009. {Stability of relativistic
  jets from rotating, accreting black holes via fully three-dimensional
  magnetohydrodynamic simulations}. \mnras 394, L126--L130.

\bibitem[{{McKinney} et~al.(2013){McKinney}, {Tchekhovskoy}, and
  {Blandford}}]{2013Sci...339...49M}
{McKinney}, J.~C., {Tchekhovskoy}, A., {Blandford}, R.~D., Jan. 2013.
  {Alignment of Magnetized Accretion Disks and Relativistic Jets with Spinning
  Black Holes}. Science 339, 49--.

\bibitem[{{McKinney} et~al.(2014){McKinney}, {Tchekhovskoy}, {Sadowski}, and
  {Narayan}}]{2014MNRAS.441.3177M}
{McKinney}, J.~C., {Tchekhovskoy}, A., {Sadowski}, A., {Narayan}, R., Jul.
  2014. {Three-dimensional general relativistic radiation magnetohydrodynamical
  simulation of super-Eddington accretion, using a new code HARMRAD with M1
  closure}. \mnras 441, 3177--3208.

\bibitem[{{M{\'e}sz{\'a}ros}(1992)}]{1992hrfm.book.....M}
{M{\'e}sz{\'a}ros}, P., 1992. {High-energy radiation from magnetized neutron
  stars.} The University of Chicago Press.

\bibitem[{{Meszaros} et~al.(1988){Meszaros}, {Novick}, {Szentgyorgyi},
  {Chanan}, and {Weisskopf}}]{1988ApJ...324.1056M}
{Meszaros}, P., {Novick}, R., {Szentgyorgyi}, A., {Chanan}, G.~A., {Weisskopf},
  M.~C., Jan. 1988. {Astrophysical implications and observational prospects of
  X-ray polarimetry}. \apj 324, 1056--1067.

\bibitem[{{Migliari} and {Fender}(2006)}]{migliari}
{Migliari}, S., {Fender}, R.~P., Feb. 2006. {Jets in neutron star X-ray
  binaries: a comparison with black holes}. \mnras 366, 79--91.

\bibitem[{{Miniutti} et~al.(2003){Miniutti}, {Fabian}, {Goyder}, and
  {Lasenby}}]{miniutti03}
{Miniutti}, G., {Fabian}, A.~C., {Goyder}, R., {Lasenby}, A.~N., Sep. 2003.
  {The lack of variability of the iron line in MCG-6-30-15: general
  relativistic effects}. \mnras 344, L22--L26.

\bibitem[{{Moran} et~al.(2013){Moran}, {Shearer}, {Mignani}, {S{\l}owikowska},
  {De Luca}, {Gouiff{\`e}s}, and {Laurent}}]{2013MNRAS.433.2564M}
{Moran}, P., {Shearer}, A., {Mignani}, R.~P., et~al., Aug. 2013. {Optical
  polarimetry of the inner Crab nebula and pulsar}. \mnras 433, 2564--2575.

\bibitem[{{Morgan} et~al.(1997){Morgan}, {Remillard}, and
  {Greiner}}]{1997ApJ...482..993M}
{Morgan}, E.~H., {Remillard}, R.~A., {Greiner}, J., Jun. 1997. {RXTE
  Observations of QPOs in the Black Hole Candidate GRS 1915+105}. \apj 482,
  993--1010.

\bibitem[{{Mori} et~al.(2013){Mori}, {Gotthelf}, {Zhang}, {An}, {Baganoff},
  {Barri{\`e}re}, {Beloborodov}, {Boggs}, {Christensen}, {Craig}, {Dufour},
  {Grefenstette}, {Hailey}, {Harrison}, {Hong}, {Kaspi}, {Kennea}, {Madsen},
  {Markwardt}, {Nynka}, {Stern}, {Tomsick}, and {Zhang}}]{2013ApJ...770L..23M}
{Mori}, K., {Gotthelf}, E.~V., {Zhang}, S., et~al., Jun. 2013. {NuSTAR
  Discovery of a 3.76 s Transient Magnetar Near Sagittarius A*}. \apjl 770,
  L23.

\bibitem[{{Motta} et~al.(2015){Motta}, {Casella}, {Henze}, {Mu{\~n}oz-Darias},
  {Sanna}, {Fender}, and {Belloni}}]{2015MNRAS.447.2059M}
{Motta}, S.~E., {Casella}, P., {Henze}, M., et~al., Feb. 2015. {Geometrical
  constraints on the origin of timing signals from black holes}. \mnras 447,
  2059--2072.

\bibitem[{{National Aeronautics and Space Administration}(2014)}]{SMEX2014}
{National Aeronautics and Space Administration}, 2014. Announcement of
  opportunity for astrophysics small explorer mission, solicitation number:
  Nnh14zda013o.
\newline\urlprefix\url{https://www.fbo.gov/}

\bibitem[{{Novick}(1975)}]{1975SSRv...18..389N}
{Novick}, R., Dec. 1975. {Stellar and Solar X-Ray Polarimetry}. \ssr 18,
  389--408.

\bibitem[{{{\"O}zel}(2001)}]{2001ApJ...563..276O}
{{\"O}zel}, F., Dec. 2001. {Surface Emission Properties of Strongly Magnetic
  Neutron Stars}. \apj 563, 276--288.

\bibitem[{{Papadakis} et~al.(2005){Papadakis}, {Kazanas}, and
  {Akylas}}]{papadakis05}
{Papadakis}, I.~E., {Kazanas}, D., {Akylas}, A., Oct. 2005. {Fourier-Resolved
  Spectroscopy of the XMM-Newton Observations of MCG -06-30-15}. \apj 631,
  727--732.

\bibitem[{{Parker} et~al.(2014){Parker}, {Wilkins}, {Fabian}, {Grupe},
  {Dauser}, {Matt}, {Harrison}, {Brenneman}, {Boggs}, {Christensen}, {Craig},
  {Gallo}, {Hailey}, {Kara}, {Komossa}, {Marinucci}, {Miller}, {Risaliti},
  {Stern}, {Walton}, and {Zhang}}]{parker14}
{Parker}, M.~L., {Wilkins}, D.~R., {Fabian}, A.~C., et~al., Sep. 2014. {The
  NuSTAR spectrum of Mrk 335: extreme relativistic effects within two
  gravitational radii of the event horizon?} \mnras 443, 1723--1732.

\bibitem[{{Ramsey}(2014)}]{2014Ramsey}
{Ramsey}, B.~f., 2014. {IXPE - The Imaging X-ray Polarimetry Explorer}.

\bibitem[{{Rana} et~al.(2009){Rana}, {Cook}, {Harrison}, {Mao}, and
  {Miyasaka}}]{2009SPIE.7435E..03R}
{Rana}, V.~R., {Cook}, III, W.~R., {Harrison}, F.~A., {Mao}, P.~H., {Miyasaka},
  H., Aug. 2009. {Development of focal plane detectors for the Nuclear
  Spectroscopic Telescope Array (NuSTAR) mission}. In: Society of Photo-Optical
  Instrumentation Engineers (SPIE) Conference Series. Vol. 7435 of Society of
  Photo-Optical Instrumentation Engineers (SPIE) Conference Series. p.~3.

\bibitem[{{Reid} et~al.(2014){Reid}, {McClintock}, {Steiner}, {Steeghs},
  {Remillard}, {Dhawan}, and {Narayan}}]{2014ApJ...796....2R}
{Reid}, M.~J., {McClintock}, J.~E., {Steiner}, J.~F., et~al., Nov. 2014. {A
  Parallax Distance to the Microquasar GRS 1915+105 and a Revised Estimate of
  its Black Hole Mass}. \apj 796, 2.

\bibitem[{{Remillard} and {McClintock}(2006)}]{2006ARA&A..44...49R}
{Remillard}, R.~A., {McClintock}, J.~E., Sep. 2006. {X-Ray Properties of
  Black-Hole Binaries}. \araa 44, 49--92.

\bibitem[{{Reynolds}(2014)}]{2014SSRv..183..277R}
{Reynolds}, C.~S., Sep. 2014. {Measuring Black Hole Spin Using X-Ray Reflection
  Spectroscopy}. \ssr 183, 277--294.

\bibitem[{{Risaliti} et~al.(2013){Risaliti}, {Harrison}, {Madsen}, {Walton},
  {Boggs}, {Christensen}, {Craig}, {Grefenstette}, {Hailey}, {Nardini},
  {Stern}, and {Zhang}}]{2013Natur.494..449R}
{Risaliti}, G., {Harrison}, F.~A., {Madsen}, K.~K., et~al., Feb. 2013. {A
  rapidly spinning supermassive black hole at the centre of NGC 1365}. \nat
  494, 449--451.

\bibitem[{{Rodriguez} et~al.(2015){Rodriguez}, {Grinberg}, {Laurent}, {Cadolle
  Bel}, {Pottschmidt}, {Pooley}, {Bodaghee}, {Wilms}, and
  {Gouiff{\`e}s}}]{2015ApJ...807...17R}
{Rodriguez}, J., {Grinberg}, V., {Laurent}, P., et~al., Jul. 2015. {Spectral
  State Dependence of the 0.4-2 MeV Polarized Emission in Cygnus X-1 Seen with
  INTEGRAL/IBIS, and Links with the AMI Radio Data}. \apj 807, 17.

\bibitem[{{Ross} and {Fabian}(1993)}]{1993MNRAS.261...74R}
{Ross}, R.~R., {Fabian}, A.~C., Mar. 1993. {The effects of photoionization on
  X-ray reflection spectra in active galactic nuclei}. \mnras 261, 74--82.

\bibitem[{{Rudy} et~al.(2015){Rudy}, {Horns}, {DeLuca}, {Kolodziejczak},
  {Tennant}, {Yuan}, {Buehler}, {Arons}, {Blandford}, {Caraveo}, {Costa},
  {Funk}, {Hays}, {Lobanov}, {Max}, {Mayer}, {Mignani}, {O'Dell}, {Romani},
  {Tavani}, and {Weisskopf}}]{2015arXiv150404613R}
{Rudy}, A., {Horns}, D., {DeLuca}, A., et~al., Apr. 2015. {Characterization of
  the Inner Knot of the Crab: The Site of the Gamma-ray Flares?} ArXiv
  e-prints.

\bibitem[{{Ryan} et~al.(2015){Ryan}, {Dolence}, and
  {Gammie}}]{2015ApJ...807...31R}
{Ryan}, B.~R., {Dolence}, J.~C., {Gammie}, C.~F., Jul. 2015. {bhlight: General
  Relativistic Radiation Magnetohydrodynamics with Monte Carlo Transport}. \apj
  807, 31.

\bibitem[{{Sadowski} and {Narayan}(2015{\natexlab{a}})}]{2015arXiv150804980S}
{Sadowski}, A., {Narayan}, R., Aug. 2015{\natexlab{a}}. {Photon-conserving
  Comptonization in simulations of accretion disks around black holes}. ArXiv
  e-prints.

\bibitem[{{Sadowski} and {Narayan}(2015{\natexlab{b}})}]{2015arXiv150300654S}
{Sadowski}, A., {Narayan}, R., Mar. 2015{\natexlab{b}}. {Powerful radiative
  jets in super-critical accretion disks around non-spinning black holes}.
  ArXiv e-prints.

\bibitem[{{Sanchez Almeida} and {Martinez Pillet}(1993)}]{1993ApOpt..32.4231S}
{Sanchez Almeida}, J., {Martinez Pillet}, V., Aug. 1993. {Polarizing properties
  of grazing-incidence X-ray mirrors - Comment}. Applied Optics 32, 4231--4235.

\bibitem[{{S{\c a}dowski} et~al.(2013){S{\c a}dowski}, {Narayan},
  {Tchekhovskoy}, and {Zhu}}]{2013MNRAS.429.3533S}
{S{\c a}dowski}, A., {Narayan}, R., {Tchekhovskoy}, A., {Zhu}, Y., Mar. 2013.
  {Semi-implicit scheme for treating radiation under M1 closure in general
  relativistic conservative fluid dynamics codes}. \mnras 429, 3533--3550.

\bibitem[{{Schnittman} and {Krolik}(2009)}]{2009ApJ...701.1175S}
{Schnittman}, J.~D., {Krolik}, J.~H., Aug. 2009. {X-ray Polarization from
  Accreting Black Holes: The Thermal State}. \apj 701, 1175--1187.

\bibitem[{{Schnittman} and {Krolik}(2010)}]{2010ApJ...712..908S}
{Schnittman}, J.~D., {Krolik}, J.~H., Apr. 2010. {X-ray Polarization from
  Accreting Black Holes: Coronal Emission}. \apj 712, 908--924.

\bibitem[{{Schnittman} and {Krolik}(2013)}]{2013ApJ...777...11S}
{Schnittman}, J.~D., {Krolik}, J.~H., Nov. 2013. {A Monte Carlo Code for
  Relativistic Radiation Transport around Kerr Black Holes}. \apj 777, 11.

\bibitem[{{Schnittman} et~al.(2013){Schnittman}, {Krolik}, and
  {Noble}}]{2013ApJ...769..156S}
{Schnittman}, J.~D., {Krolik}, J.~H., {Noble}, S.~C., Jun. 2013. {X-Ray Spectra
  from Magnetohydrodynamic Simulations of Accreting Black Holes}. \apj 769,
  156.

\bibitem[{{Shapiro} et~al.(1976){Shapiro}, {Lightman}, and
  {Eardley}}]{1976ApJ...204..187S}
{Shapiro}, S.~L., {Lightman}, A.~P., {Eardley}, D.~M., Feb. 1976. {A
  two-temperature accretion disk model for Cygnus X-1 - Structure and
  spectrum}. \apj 204, 187--199.

\bibitem[{{Shaviv} et~al.(1999){Shaviv}, {Heyl}, and
  {Lithwick}}]{1999MNRAS.306..333S}
{Shaviv}, N.~J., {Heyl}, J.~S., {Lithwick}, Y., Jun. 1999. {Magnetic lensing
  near ultramagnetized neutron stars}. \mnras 306, 333--347.

\bibitem[{{Shcherbakov} et~al.(2012){Shcherbakov}, {Penna}, and
  {McKinney}}]{2012ApJ...755..133S}
{Shcherbakov}, R.~V., {Penna}, R.~F., {McKinney}, J.~C., Aug. 2012.
  {Sagittarius A* Accretion Flow and Black Hole Parameters from General
  Relativistic Dynamical and Polarized Radiative Modeling}. \apj 755, 133.

\bibitem[{{Silver} et~al.(1979){Silver}, {Weisskopf}, {Kestenbaum}, {Long},
  {Novick}, and {Wolff}}]{1979ApJ...232..248S}
{Silver}, E.~H., {Weisskopf}, M.~C., {Kestenbaum}, H.~L., et~al., Aug. 1979.
  {The first search for X-ray polarization in the Centaurus X-3 and Hercules
  X-1 pulsars}. \apj 232, 248--254.

\bibitem[{{Soffitta} et~al.(2013){Soffitta}, {Barcons}, {Bellazzini}, {Braga},
  {Costa}, {Fraser}, {Gburek}, {Huovelin}, {Matt}, {Pearce}, {Poutanen},
  {Reglero}, {Santangelo}, {Sunyaev}, {Tagliaferri}, {Weisskopf}, {Aloisio},
  {Amato}, {Attin{\'a}}, {Axelsson}, {Baldini}, {Basso}, {Bianchi}, {Blasi},
  {Bregeon}, {Brez}, {Bucciantini}, {Burderi}, {Burwitz}, {Casella},
  {Churazov}, {Civitani}, {Covino}, {Curado da Silva}, {Cusumano}, {Dadina},
  {D'Amico}, {De Rosa}, {Di Cosimo}, {Di Persio}, {Di Salvo}, {Dovciak},
  {Elsner}, {Eyles}, {Fabian}, {Fabiani}, {Feng}, {Giarrusso}, {Goosmann},
  {Grandi}, {Grosso}, {Israel}, {Jackson}, {Kaaret}, {Karas}, {Kuss}, {Lai},
  {Rosa}, {Larsson}, {Larsson}, {Latronico}, {Maggio}, {Maia}, {Marin},
  {Massai}, {Mineo}, {Minuti}, {Moretti}, {Muleri}, {O'Dell}, {Pareschi},
  {Peres}, {Pesce}, {Petrucci}, {Pinchera}, {Porquet}, {Ramsey}, {Rea},
  {Reale}, {Rodrigo}, {R{\'o}{\.z}a{\'n}ska}, {Rubini}, {Rudawy}, {Ryde},
  {Salvati}, {de Santiago}, {Sazonov}, {Sgr{\'o}}, {Silver}, {Spandre},
  {Spiga}, {Stella}, {Tamagawa}, {Tamborra}, {Tavecchio}, {Teixeira Dias}, {van
  Adelsberg}, {Wu}, and {Zane}}]{2013ExA....36..523S}
{Soffitta}, P., {Barcons}, X., {Bellazzini}, R., et~al., Dec. 2013. {XIPE: the
  X-ray imaging polarimetry explorer}. Experimental Astronomy 36, 523--567.

\bibitem[{{Spruit}(2011)}]{2011AIPC.1381..227S}
{Spruit}, H.~C., Sep. 2011. {Magnetically powered jets}. In: {Aharonian},
  F.~A., {Hofmann}, W., {Rieger}, F.~M. (Eds.), American Institute of Physics
  Conference Series. Vol. 1381 of American Institute of Physics Conference
  Series. pp. 227--246.

\bibitem[{Stokes(1852)}]{stokes_gg_1852}
Stokes, G.~G., 1852. {Composition and resolution of streams of polarized light
  from multiple sources}. Trans. Cambridge Philos. Soc. 9, 399--416, reprinted
  in Mathematical and Physical Papers, Vol.~3, Cambridge University Press,
  London, 1901.

\bibitem[{{Sunyaev} and {Thorne}(1973)}]{1973Sunyaev}
{Sunyaev}, R.~A., {Thorne}, K.~S., 1973. {Discussion in Moscow, see Thorne \&
  Price (1973).}

\bibitem[{{Sunyaev} and {Titarchuk}(1980)}]{1980A&A....86..121S}
{Sunyaev}, R.~A., {Titarchuk}, L.~G., Jun. 1980. {Comptonization of X-rays in
  plasma clouds - Typical radiation spectra}. \aap 86, 121--138.

\bibitem[{{Sunyaev} and {Titarchuk}(1985)}]{1985A&A...143..374S}
{Sunyaev}, R.~A., {Titarchuk}, L.~G., Feb. 1985. {Comptonization of
  low-frequency radiation in accretion disks Angular distribution and
  polarization of hard radiation}. \aap 143, 374--388.

\bibitem[{{Takeuchi} et~al.(2014){Takeuchi}, {Kitaguchi}, {Hayato}, {Tamagawa},
  {Iwakiri}, {Asami}, {Yoshikawa}, {Kaneko}, {Enoto}, {Black}, {Hill}, and
  {Jahoda}}]{2014SPIE.9144E..4NT}
{Takeuchi}, Y., {Kitaguchi}, T., {Hayato}, A., et~al., Jul. 2014. {Properties
  of the flight model gas electron multiplier for the GEMS mission}. In:
  Society of Photo-Optical Instrumentation Engineers (SPIE) Conference Series.
  Vol. 9144 of Society of Photo-Optical Instrumentation Engineers (SPIE)
  Conference Series. p.~4.

\bibitem[{{Tavani} et~al.(2011){Tavani}, {Bulgarelli}, {Vittorini},
  {Pellizzoni}, {Striani}, {Caraveo}, {Weisskopf}, {Tennant}, {Pucella},
  {Trois}, {Costa}, {Evangelista}, {Pittori}, {Verrecchia}, {Del Monte},
  {Campana}, {Pilia}, {De Luca}, {Donnarumma}, {Horns}, {Ferrigno}, {Heinke},
  {Trifoglio}, {Gianotti}, {Vercellone}, {Argan}, {Barbiellini}, {Cattaneo},
  {Chen}, {Contessi}, {D'Ammando}, {DeParis}, {Di Cocco}, {Di Persio},
  {Feroci}, {Ferrari}, {Galli}, {Giuliani}, {Giusti}, {Labanti}, {Lapshov},
  {Lazzarotto}, {Lipari}, {Longo}, {Fuschino}, {Marisaldi}, {Mereghetti},
  {Morelli}, {Moretti}, {Morselli}, {Pacciani}, {Perotti}, {Piano}, {Picozza},
  {Prest}, {Rapisarda}, {Rappoldi}, {Rubini}, {Sabatini}, {Soffitta},
  {Vallazza}, {Zambra}, {Zanello}, {Lucarelli}, {Santolamazza}, {Giommi},
  {Salotti}, and {Bignami}}]{2011Sci...331..736T}
{Tavani}, M., {Bulgarelli}, A., {Vittorini}, V., et~al., Feb. 2011. {Discovery
  of Powerful Gamma-Ray Flares from the Crab Nebula}. Science 331, 736--.

\bibitem[{{Taverna} et~al.(2014){Taverna}, {Muleri}, {Turolla}, {Soffitta},
  {Fabiani}, and {Nobili}}]{2014MNRAS.438.1686T}
{Taverna}, R., {Muleri}, F., {Turolla}, R., et~al., Feb. 2014. {Probing
  magnetar magnetosphere through X-ray polarization measurements}. \mnras 438,
  1686--1697.

\bibitem[{{Tchekhovskoy} et~al.(2011){Tchekhovskoy}, {Narayan}, and
  {McKinney}}]{2011MNRAS.418L..79T}
{Tchekhovskoy}, A., {Narayan}, R., {McKinney}, J.~C., Nov. 2011. {Efficient
  generation of jets from magnetically arrested accretion on a rapidly spinning
  black hole}. \mnras 418, L79--L83.

\bibitem[{{Tendulkar} et~al.(2014){Tendulkar}, {F{\"u}rst}, {Pottschmidt},
  {Bachetti}, {Bhalerao}, {Boggs}, {Christensen}, {Craig}, {Hailey},
  {Harrison}, {Stern}, {Tomsick}, {Walton}, and {Zhang}}]{2014ApJ...795..154T}
{Tendulkar}, S.~P., {F{\"u}rst}, F., {Pottschmidt}, K., et~al., Nov. 2014.
  {NuSTAR Discovery of a Cyclotron Line in the Be/X-Ray Binary RX J0520.5-6932
  during Outburst}. \apj 795, 154.

\bibitem[{{The European Space Agency}(2015)}]{m4}
{The European Space Agency}, 2015. {Three Candidates for ESA's Next Medium
  Class Science Mission}.
\newline\urlprefix\url{http://www.esa.int/Our\_Activities
  /Space\_Science/Three\_candidates\_for\_ESA\_s\_next
  \_medium-class\_science\_mission}

\bibitem[{{The National Aeronautics and Space Administration}(2015)}]{SMEXD}
{The National Aeronautics and Space Administration}, 2015. {Three Candidates
  for ESA's Next Medium Class Science Mission}.
\newline\urlprefix\url{http://www.nasa.gov/press-release
  /nasa-selects-proposals-to-study-neutron- stars-black-holes-and-more}

\bibitem[{{Thompson} and {Duncan}(1995)}]{thompson1995}
{Thompson}, C., {Duncan}, R.~C., Jul. 1995. {The soft gamma repeaters as very
  strongly magnetized neutron stars - I. Radiative mechanism for outbursts}.
  MNRAS 275, 255--300.

\bibitem[{{Thorne} and {Price}(1975)}]{1975ApJ...195L.101T}
{Thorne}, K.~S., {Price}, R.~H., Feb. 1975. {Cygnus X-1 - an interpretation of
  the spectrum and its variability}. \apjl 195, L101--L105.

\bibitem[{{Tomsick} and {Kaaret}(2001)}]{2001ApJ...548..401T}
{Tomsick}, J.~A., {Kaaret}, P., Feb. 2001. {The X-Ray Properties of
  Low-Frequency Quasi-periodic Oscillations from GRS 1915+105 Up to 120 KEV}.
  \apj 548, 401--409.

\bibitem[{{Tomsick} et~al.(2014){Tomsick}, {Nowak}, {Parker}, {Miller},
  {Fabian}, {Harrison}, {Bachetti}, {Barret}, {Boggs}, {Christensen}, {Craig},
  {Forster}, {F{\"u}rst}, {Grefenstette}, {Hailey}, {King}, {Madsen},
  {Natalucci}, {Pottschmidt}, {Ross}, {Stern}, {Walton}, {Wilms}, and
  {Zhang}}]{2014ApJ...780...78T}
{Tomsick}, J.~A., {Nowak}, M.~A., {Parker}, M., et~al., Jan. 2014. {The
  Reflection Component from Cygnus X-1 in the Soft State Measured by NuSTAR and
  Suzaku}. \apj 780, 78.

\bibitem[{{Tosti} et~al.(1998){Tosti}, {Fiorucci}, {Luciani}, {Efimov},
  {Shakhovskoy}, {Valtaoja}, {Teraesranta}, {Sillanpaeae}, {Takalo}, {Villata},
  {Raiteri}, {de Francesco}, and {Sobrito}}]{1998A&A...339...41T}
{Tosti}, G., {Fiorucci}, M., {Luciani}, M., et~al., Nov. 1998. {Radio, optical
  and photopolarimetric observations of Markarian 421 around the great 1996-97
  outburst}. \aap 339, 41--51.

\bibitem[{{Tr{\"u}mper} et~al.(2013){Tr{\"u}mper}, {Dennerl}, {Kylafis},
  {Ertan}, and {Zezas}}]{2013ApJ...764...49T}
{Tr{\"u}mper}, J.~E., {Dennerl}, K., {Kylafis}, N.~D., {Ertan}, {\"U}.,
  {Zezas}, A., Feb. 2013. {An Accretion Model for the Anomalous X-Ray Pulsar 4U
  0142+61}. \apj 764, 49.

\bibitem[{{Tr{\"u}mper} et~al.(2010){Tr{\"u}mper}, {Zezas}, {Ertan}, and
  {Kylafis}}]{2010A&A...518A..46T}
{Tr{\"u}mper}, J.~E., {Zezas}, A., {Ertan}, {\"U}., {Kylafis}, N.~D., Jul.
  2010. {The energy spectrum of anomalous X-ray pulsars and soft gamma-ray
  repeaters}. \aap 518, A46.

\bibitem[{{Ubertini} et~al.(2003){Ubertini}, {Lebrun}, {Di Cocco}, {Bazzano},
  {Bird}, {Broenstad}, {Goldwurm}, {La Rosa}, {Labanti}, {Laurent}, {Mirabel},
  {Quadrini}, {Ramsey}, {Reglero}, {Sabau}, {Sacco}, {Staubert}, {Vigroux},
  {Weisskopf}, and {Zdziarski}}]{2003A&A...411L.131U}
{Ubertini}, P., {Lebrun}, F., {Di Cocco}, G., et~al., Nov. 2003. {IBIS: The
  Imager on-board INTEGRAL}. \aap 411, L131--L139.

\bibitem[{{Ueda} et~al.(2010){Ueda}, {Honda}, {Takahashi}, {Done}, {Shirai},
  {Fukazawa}, {Yamaoka}, {Naik}, {Awaki}, {Ebisawa}, {Rodriguez}, and
  {Chaty}}]{2010ApJ...713..257U}
{Ueda}, Y., {Honda}, K., {Takahashi}, H., et~al., Apr. 2010. {Suzaku
  Observation of GRS 1915+105: Evolution of Accretion Disk Structure during
  Limit-cycle Oscillation}. \apj 713, 257--268.

\bibitem[{{Uttley} et~al.(2014){Uttley}, {Cackett}, {Fabian}, {Kara}, and
  {Wilkins}}]{2014A&ARv..22...72U}
{Uttley}, P., {Cackett}, E.~M., {Fabian}, A.~C., {Kara}, E., {Wilkins}, D.~R.,
  Aug. 2014. {X-ray reverberation around accreting black holes}. \aapr 22, 72.

\bibitem[{{Uttley} and {McHardy}(2001)}]{2001MNRAS.323L..26U}
{Uttley}, P., {McHardy}, I.~M., May 2001. {The flux-dependent amplitude of
  broadband noise variability in X-ray binaries and active galaxies}. \mnras
  323, L26--L30.

\bibitem[{{Uttley} et~al.(2005){Uttley}, {McHardy}, and
  {Vaughan}}]{2005MNRAS.359..345U}
{Uttley}, P., {McHardy}, I.~M., {Vaughan}, S., May 2005. {Non-linear X-ray
  variability in X-ray binaries and active galaxies}. \mnras 359, 345--362.

\bibitem[{{Vaughan} et~al.(2003){Vaughan}, {Fabian}, and {Nandra}}]{vaughan03}
{Vaughan}, S., {Fabian}, A.~C., {Nandra}, K., Mar. 2003. {X-ray continuum
  variability of MCG-6-30-15}. \mnras 339, 1237--1255.

\bibitem[{{Vedrenne} et~al.(2003){Vedrenne}, {Roques}, {Sch{\"o}nfelder},
  {Mandrou}, {Lichti}, {von Kienlin}, {Cordier}, {Schanne}, {Kn{\"o}dlseder},
  {Skinner}, {Jean}, {Sanchez}, {Caraveo}, {Teegarden}, {von Ballmoos},
  {Bouchet}, {Paul}, {Matteson}, {Boggs}, {Wunderer}, {Leleux},
  {Weidenspointner}, {Durouchoux}, {Diehl}, {Strong}, {Cass{\'e}}, {Clair}, and
  {Andr{\'e}}}]{2003A&A...411L..63V}
{Vedrenne}, G., {Roques}, J.-P., {Sch{\"o}nfelder}, V., et~al., Nov. 2003.
  {SPI: The spectrometer aboard INTEGRAL}. \aap 411, L63--L70.

\bibitem[{Vinokur(1965)}]{vinokur_m_1965}
Vinokur, M., 1965. {Optimisation dans la Recherche d'une Sinusoïde de Période
  Connue en Présence de Bruit. Application a la Radioastronomie}. Annales
  d'Astrophysique 28, 412--445.

\bibitem[{{Walton} et~al.(2014){Walton}, {Risaliti}, {Harrison}, {Fabian},
  {Miller}, {Arevalo}, {Ballantyne}, {Boggs}, {Brenneman}, {Christensen},
  {Craig}, {Elvis}, {Fuerst}, {Gandhi}, {Grefenstette}, {Hailey}, {Kara},
  {Luo}, {Madsen}, {Marinucci}, {Matt}, {Parker}, {Reynolds}, {Rivers}, {Ross},
  {Stern}, and {Zhang}}]{walton14}
{Walton}, D.~J., {Risaliti}, G., {Harrison}, F.~A., et~al., Jun. 2014. {NuSTAR
  and XMM-NEWTON Observations of NGC 1365: Extreme Absorption Variability and a
  Constant Inner Accretion Disk}. \apj 788, 76.

\bibitem[{{Weisskopf} et~al.(2006){Weisskopf}, {Elsner}, {Hanna}, {Kaspi},
  {O'Dell}, {Pavlov}, and {Ramsey}}]{2006astro.ph.11483W}
{Weisskopf}, M.~C., {Elsner}, R.~F., {Hanna}, D., et~al., Nov. 2006. {The
  prospects for X-ray polarimetry and its potential use for understanding
  neutron stars}. ArXiv Astrophysics e-prints.

\bibitem[{{Weisskopf} et~al.(1978){Weisskopf}, {Silver}, {Kestenbaum}, {Long},
  and {Novick}}]{1978ApJ...220L.117W}
{Weisskopf}, M.~C., {Silver}, E.~H., {Kestenbaum}, H.~L., {Long}, K.~S.,
  {Novick}, R., Mar. 1978. {A precision measurement of the X-ray polarization
  of the Crab Nebula without pulsar contamination}. \apjl 220, L117--L121.

\bibitem[{Weisskopf et~al.(2006)}]{weisskopf_mc_2006}
Weisskopf, M.~C., et~al., 2006. {The prospects for X-ray polarimetry and its
  potential use for understanding neutron stars}. Paper presented at the 363rd
  Heraeus Seminar in Bad Honnef, Germany, 2006. arXiv: astro-ph/0611483v1.

\bibitem[{{Wijnands} and {van der Klis}(1999)}]{1999ApJ...514..939W}
{Wijnands}, R., {van der Klis}, M., Apr. 1999. {The Broadband Power Spectra of
  X-Ray Binaries}. \apj 514, 939--944.

\bibitem[{{Willis} et~al.(2005){Willis}, {Barlow}, {Bird}, {Clark}, {Dean},
  {McConnell}, {Moran}, {Shaw}, and {Sguera}}]{2005A&A...439..245W}
{Willis}, D.~R., {Barlow}, E.~J., {Bird}, A.~J., et~al., Aug. 2005. {Evidence
  of polarisation in the prompt gamma-ray emission from GRB 930131 and GRB
  960924}. \aap 439, 245--253.

\bibitem[{{Wilms} et~al.(2014){Wilms}, {Brand}, {Barret}, {Beuchert}, {den
  Herder}, {Kreykenbohm}, {Lotti}, {Meidinger}, {Nandra}, {Peille}, {Piro},
  {Rau}, {Schmid}, {Smith}, {Tenzer}, {Wille}, and
  {Willingale}}]{2014SPIE.9144E..5XW}
{Wilms}, J., {Brand}, T., {Barret}, D., et~al., Jul. 2014. {ATHENA end-to-end
  simulations}. In: Society of Photo-Optical Instrumentation Engineers (SPIE)
  Conference Series. Vol. 9144 of Society of Photo-Optical Instrumentation
  Engineers (SPIE) Conference Series. p.~5.

\bibitem[{{Wolter}(1952)}]{1952AnP...445...94W}
{Wolter}, H., 1952. {Spiegelsysteme streifenden Einfalls als abbildende Optiken
  f{\"u}r R{\"o}ntgenstrahlen}. Annalen der Physik 445, 94--114.

\bibitem[{{Yonetoku} et~al.(2012){Yonetoku}, {Murakami}, {Gunji}, {Mihara},
  {Toma}, {Morihara}, {Takahashi}, {Wakashima}, {Yonemochi}, {Sakashita},
  {Toukairin}, {Fujimoto}, and {Kodama}}]{2012ApJ...758L...1Y}
{Yonetoku}, D., {Murakami}, T., {Gunji}, S., et~al., Oct. 2012. {Magnetic
  Structures in Gamma-Ray Burst Jets Probed by Gamma-Ray Polarization}. \apjl
  758, L1.

\bibitem[{{Yonetoku} et~al.(2011){Yonetoku}, {Murakami}, {Gunji}, {Mihara},
  {Toma}, {Sakashita}, {Morihara}, {Takahashi}, {Toukairin}, {Fujimoto},
  {Kodama}, {Kubo}, and {IKAROS Demonstration Team}}]{2011ApJ...743L..30Y}
{Yonetoku}, D., {Murakami}, T., {Gunji}, S., et~al., Dec. 2011. {Detection of
  Gamma-Ray Polarization in Prompt Emission of
  2011Sci...332..438L2011Sci...332..438L 100826A}. \apjl 743, L30.

\bibitem[{{Zhang} and {B{\"o}ttcher}(2013)}]{2013ApJ...774...18Z}
{Zhang}, H., {B{\"o}ttcher}, M., Sep. 2013. {X-Ray and Gamma-Ray Polarization
  in Leptonic and Hadronic Jet Models of Blazars}. \apj 774, 18.

\bibitem[{{Zhang} et~al.(2014){Zhang}, {Chen}, and
  {B{\"o}ttcher}}]{2014ApJ...789...66Z}
{Zhang}, H., {Chen}, X., {B{\"o}ttcher}, M., Jul. 2014. {Synchrotron
  Polarization in Blazars}. \apj 789, 66.

\bibitem[{{Zhang} et~al.(2015{\natexlab{a}}){Zhang}, {Chen}, {B{\"o}ttcher},
  {Guo}, and {Li}}]{2015ApJ...804...58Z}
{Zhang}, H., {Chen}, X., {B{\"o}ttcher}, M., {Guo}, F., {Li}, H., May
  2015{\natexlab{a}}. {Polarization Swings Reveal Magnetic Energy Dissipation
  in Blazars}. \apj 804, 58.

\bibitem[{{Zhang} et~al.(2015{\natexlab{b}}){Zhang}, {Chen}, {Qu}, and
  {Bu}}]{2015AJ....149...82Z}
{Zhang}, L., {Chen}, L., {Qu}, J.-l., {Bu}, Q.-c., Feb. 2015{\natexlab{b}}.
  {The NuSTAR View of a QPO Evolution of GRS 1915+105}. \aj 149, 82.

\bibitem[{{Zhang}(2013)}]{2013FrPhy...8..630Z}
{Zhang}, S.-N., Dec. 2013. {Black hole binaries and microquasars}. Frontiers of
  Physics 8, 630--660.

\bibitem[{{Zhang}(2009)}]{2009SPIE.7437E..0NZ}
{Zhang}, W.~W., Aug. 2009. {Manufacture of mirror glass substrates for the
  NuSTAR mission}. In: Society of Photo-Optical Instrumentation Engineers
  (SPIE) Conference Series. Vol. 7437 of Society of Photo-Optical
  Instrumentation Engineers (SPIE) Conference Series. p.~0.

\bibitem[{{Zhu} et~al.(2015){Zhu}, {Narayan}, {Sadowski}, and
  {Psaltis}}]{2015MNRAS.451.1661Z}
{Zhu}, Y., {Narayan}, R., {Sadowski}, A., {Psaltis}, D., Aug. 2015. {HERO - A
  3D general relativistic radiative post-processor for accretion discs around
  black holes}. \mnras 451, 1661--1681.

\bibitem[{{Zoghbi} et~al.(2014){Zoghbi}, {Cackett}, {Reynolds}, {Kara},
  {Harrison}, {Fabian}, {Lohfink}, {Matt}, {Balokovic}, {Boggs}, {Christensen},
  {Craig}, {Hailey}, {Stern}, and {Zhang}}]{zoghbi14}
{Zoghbi}, A., {Cackett}, E.~M., {Reynolds}, C., et~al., Jul. 2014.
  {Observations of MCG-5-23-16 with Suzaku, XMM-Newton and NuSTAR: Disk
  Tomography and Compton Hump Reverberation}. \apj 789, 56.

\bibitem[{{Zoghbi} et~al.(2012){Zoghbi}, {Fabian}, {Reynolds}, and
  {Cackett}}]{zoghbi12}
{Zoghbi}, A., {Fabian}, A.~C., {Reynolds}, C.~S., {Cackett}, E.~M., May 2012.
  {Relativistic iron K X-ray reverberation in NGC 4151}. \mnras 422, 129--134.

\end{thebibliography}
\end{document}